\documentclass{amsarte}

\usepackage{latexsym,pifont}
\usepackage{epsfig}
\usepackage{rotating}

\usepackage{ifpdf}
\ifpdf
\usepackage{epstopdf}
\usepackage{hyperref}
\else
\usepackage[hypertex]{hyperref}
\fi

\usepackage{extarrows}
\usepackage{amsfonts,amsmath,amssymb,mathrsfs,extarrows,MnSymbol}
\usepackage[all]{xy}

\usepackage{tikz}

\usepackage{cancel}

\usetikzlibrary{matrix,arrows}

\newcommand{\alxydim}[2]{\begin{aligned}\xymatrix#1{#2}\end{aligned}}

\newcommand{\brem}{\begin{Rem}}%%%%%%%%%%%% open-close
\newcommand{\erem}{\end{Rem}\medskip}
\newcommand{\beg}{\begin{Eg}}
\newcommand{\eeg}{\end{Eg}}
\newcommand{\bedef}{\begin{Def}}
\newcommand{\exdef}{\begin{flushright}$\diamond$\end{flushright}
\end{Def}\vskip0.1cm}
\newcommand{\berop}{\begin{Prop}}
\newcommand{\eerop}{\end{Prop}}
\newcommand{\belem}{\begin{Lem}}
\newcommand{\elem}{\end{Lem}}
\newcommand{\bethe}{\begin{Thm}}
\newcommand{\ethe}{\end{Thm}}
\newcommand{\becor}{\begin{Cor}}
\newcommand{\ecor}{\end{Cor}}
\newcommand{\beroof}{\noindent\begin{proof}}
\newcommand{\eroof}{\end{proof}}
\newcommand{\becon}{\begin{Conv}}
\newcommand{\econ}{\begin{flushright}$\checkmark$\end{flushright}\end{Conv}}
\newcommand{\befact}{\begin{Fact}}
\newcommand{\efact}{\begin{flushright}$\checkmark$\end{flushright}\end{Fact}}
\newcommand{\bequest}{\begin{Quest}}
\newcommand{\equest}{\end{Quest}}
\newcommand{\brob}{\begin{Prob}}
\newcommand{\erob}{\end{Prob}}

\newcommand{\barr}{\begin{array}}
\newcommand{\earr}{\end{array}}
\newcommand{\ben}{\begin{enumerate}}
\newcommand{\een}{\end{enumerate}}
\newcommand{\bit}{\begin{itemize}}
\newcommand{\eit}{\end{itemize}}

\newcommand{\qq}{\begin{eqnarray}}
\newcommand{\qqq}{\end{eqnarray}}

\newcommand{\nn}{\nonumber}%%%%%%%%%%% control

\newcommand{\ovl}[1]{\overline{#1}}
\newcommand{\unl}[1]{\underline{#1}}

\newcommand{\Reqref}[1]{Eq.\,\eqref{#1}}
\newcommand{\Rcite}[1]{Ref.\,\cite{#1}}

\newcommand\void[1]{}

\newcommand{\tx}[1]{\textrm{#1}} %%%%%%%%%%%%% style
\newcommand{\ciut}[1]{\tiny$#1$}

\newcommand{\gt}[1]{\mathfrak{#1}}

\def\cA{\mathcal{A}}%%%%%%%%%% calligraphy

\def\cC{\mathcal{C}}

\def\cG{\mathcal{G}}

\def\cK{\mathcal{K}}
\def\ceL{\mathcal{L}}
\def\cM{\mathcal{M}}

\def\cO{\mathcal{O}}

\def\cR{\mathcal{R}}

\def\xcL{\mathscr{L}}
\def\xcM{\mathscr{M}}

\def\xcW{\mathscr{W}}

\def\t{\mathbf{t}}

\def\bC{{\mathbb{C}}}

\def\bN{{\mathbb{N}}}

\def\bR{{\mathbb{R}}}
\def\bS{{\mathbb{S}}}

\def\bX{{\mathbb{X}}}

\def\a{\alpha}%%%%%%%%%% Greek
\def\b{\beta}
\def\g{\gamma}
\def\G{\Gamma}
\def\d{\delta}
\def\D{\Delta}
\def\ep{\epsilon}
\def\vep{\varepsilon}

\def\k{\kappa}

\def\la{\lambda}
\def\La{\Lambda}
\def\om{\omega}
\def\Om{\Omega}

\def\si{\sigma}

\def\t{\tau}

\def\z{\zeta}

\def\agt{\gt{a}}%%%%%%%%%%%%%%%%% Gothic

\def\Cgt{\gt{C}}

\def\dgt{\gt{d}}

\def\egt{\gt{e}}

\def\fgt{\gt{f}}

\def\ggt{\gt{g}}

\def\hgt{\gt{h}}

\def\kgt{\gt{k}}

\def\lgt{\gt{l}}

\def\tgt{\gt{t}}

\def\zgt{\gt{z}}

\newcommand{\sfd}{{\mathsf d}}

\newcommand{\sfi}{{\mathsf i}}

\newcommand{\sfT}{{\mathsf T}}

\newcommand{\sfY}{{\mathsf Y}}

\newcommand{\txA}{{\rm A}}

\newcommand{\txB}{{\rm B}}

\newcommand{\ee}{{\rm e}}

\newcommand{\txg}{{\rm g}}
\newcommand{\txG}{{\rm G}}

\newcommand{\txH}{{\rm H}}

\newcommand{\txK}{{\rm K}}

\newcommand{\txP}{{\rm P}}

\newcommand{\txT}{{\rm T}}

\newcommand{\txZ}{{\rm Z}}

\def\Cv{\v{C}}

\def\vH{\check{H}}

\def\exp{{\rm exp}}
\def\id{{\rm id}}
\newcommand{\pr}{{\rm pr}}
\def\sign{{\rm sign}}

\def\too{\longrightarrow}

\def\1morf{1{\rm -Mor}}
\def\2morf{2{\rm -Mor}}
\def\dim{{\rm dim}}
\def\im{{\rm im}}

\def\Vol{{\rm Vol}}%%%%%%%% diff-geom

\newcommand{\pLie}[1]{\,{-\hspace{-8pt}\xcL}_{#1}}
\def\p{\partial}

\def\con{\righthalfcup}

\newcommand{\sG}{\mathcal{sG}}

\def\bd1{{\boldsymbol{1}}}
\def\brd0{{\boldsymbol{0}}}

\def\det{{\rm det}}
\def\tr{{\rm tr}}
\def\diag{\textrm{diag}}

\def\ad{{\rm ad}}
\def\Ad{{\rm Ad}}

\def\Cliff{{\rm Cliff}}

\newcommand{\uj}{{\rm U}(1)}

\def\x{\times}%%%%%%%%%% operations, equalities
\def\ox{\otimes}

\def\rx{\rtimes}

\def\must{\stackrel{!}{=}}

\def\rstr{\mathord{\restriction}}

\newcommand{\corr}[1]{\left\langle #1 \right\rangle}

\newtheorem{Thm}{Theorem}
\newtheorem{Prop}[Thm]{Proposition}
\newtheorem{Lem}[Thm]{Lemma}
\newtheorem{Cor}[Thm]{Corollary}
\theoremstyle{definition}
\newtheorem{Rem}[Thm]{Remark}
\newtheorem{Def}[Thm]{Definition}
\newtheorem{Eg}[Thm]{Example}
\newtheorem{Conv}[Thm]{Convention}
\newtheorem{Fact}[Thm]{Fact}
\newtheorem{Quest}[Thm]{Question}
\newtheorem{Prob}[Thm]{Problem}

\numberwithin{equation}{section} \numberwithin{Thm}{section}

\newcount\hour\newcount\minute
        \hour=\time \divide\hour by60 \minute=\time
        {\multiply\hour by60 \global\advance\minute by-\hour}
        \edef\militarytime{\number\hour:\ifnum\minute<10 0\fi\number\minute}

\begin{document}

\title{Equivariant prequantisation of the super-0-brane in ${\rm AdS}_2\x\bS^2$\\ -- a toy model for supergerbe theory on curved spaces}

\author{Rafa\l ~R. ~Suszek}
\address{R.R.S.:\ Katedra Metod Matematycznych Fizyki, Wydzia\l ~Fizyki
Uniwersytetu Warszawskiego, ul.\ Pasteura 5, PL-02-093 Warszawa,
Poland} \email{suszek@fuw.edu.pl}
%\thanks{}

\begin{abstract}
The paper is another step towards a realisation of the goal, advanced in articles 1706.05682 [hep-th] and 1808.04470 [hep-th], of a systematic supersymmetry-equivariant geometrisation of physically distinguished Green--Schwarz super-$(p+2)$-cocycles defining classes in the supersymmetry-invariant refinement of the de Rham cohomology of homogeneous spaces of (supersymmetry) Lie supergroups, associated with reductive decompositions of their Lie superalgebras. It deals with a correlated geometrisation of a \emph{pair} of such super-$(p+2)$-cocycles on spaces in correspondence under a blow-up transformation dual to the \.In\"on\"u--Wigner contraction that relates the respective (supersymmetry) Lie superalgebras, the latter correspondence being taken as the organising principle of the geometrisation procedure that exploits the link between the Cartan--Eilenberg cohomology of the supersymmetry group and the Chevalley--Eilenberg cohomology of its Lie superalgebra, alongside a cohomological classification of central extensions thereof. A general scheme of a correlated geometrisation compatible with the contraction is laid out and illustrated on the nontrivial example of a pair of consistent super-0-brane backgrounds: the super-Minkowski space $\,{\rm sMink}^{3,1\,\vert\,2\cdot 4}\,$ with the standard $\,N=2\,$ Green--Schwarz super-2-cocycle on it and the super-${\rm AdS}_2\x\bS^2\,$ space with Zhou's super-2-cocycle on it, asymptoting to the former in the limit of an infinite common radius of the $\,{\rm AdS}_2\,$ and the $\,\bS^2\,$ in the body $\,{\rm AdS}_2\x\bS^2\,$ of the supertarget. The geometrisation yields the respective supersymmetry-equivariant super-0-gerbes, {\it i.e.}, the prequantum bundles of the associated Green--Schwarz super-$\si$-models in the Nambu--Goto formulation. Upon passing to the equivalent Hughes--Polchinski formulation of the models, the relevant extended super-0-gerbes are verified to possess a weak $\k$-equivariant structure.
\end{abstract}

%\date{\today, \militarytime\,(GMT+1)}

\maketitle

\tableofcontents

\section{Introduction}\label{sec:intro0}

The fundamental r\^ole of higher geometry and various incarnations of cohomology in the lagrangean formulation, in identification and the gauging of symmetries as well as in geometric quantisation and, finally, in classification of the simple geometric dynamics of the $p$-branes of string theory, captured by the so-called nonlinear $\si$-models, with an action functional $\,S_\si$,\ for (patchwise) smooth embeddings
\qq\nn
\bX\ :\ \Om_{p+1}\too\xcM
\qqq
of the $(p+1)$-dimensional closed worldvolume $\,\Om_{p+1},\ \p\Om_{p+1}=\emptyset\,$ of the $p$-brane in a target space $\,\xcM\,$ with a metric body $\,(|\xcM|,\txg)\,$ and a de Rham $(p+2)$-cocycle $\,\underset{\tx{\ciut{(p+2)}}}{\chi}\in Z^{p+2}_{\rm dR}(\xcM)\,$ on it, is, by now, well established and amply documented in both the standard ({\it i.e.}, non-graded) and super-geometric settings -- {\it cp} Refs.\,\cite{Gawedzki:1987ak,Gawedzki:1999bq,Gawedzki:2002se,Gawedzki:2003pm,Gawedzki:2004tu,Schreiber:2005mi,Recknagel:2006hp,Gawedzki:2007uz,Fuchs:2007fw,Runkel:2008gr,Gawedzki:2008um,Gawedzki:2010rn,Suszek:2011hg,Suszek:2012ddg,Gawedzki:2012fu,Suszek:2013,Suszek:2017xlw,Suszek:2018bvx}. Higher geometry enters the picture by associating a (super-)geometric object, termed the (super-)$p$-gerbe $\,\underset{\tx{\ciut{(p)}}}{\cG}$,\ with the $(p+2)$-cocycle $\,\underset{\tx{\ciut{(p+2)}}}{\chi}\,$ in a manner structually analogous to that in which a principal $\bC^\x$-bundle with a compatible connection is associated with a de Rham 2-cocycle (with integer periods), ultimately identified as its curvature. In the non-graded setting, the $p$-gerbe, representable by a class in the real Deligne--Beilinson cohomology of $\,\xcM\,$ in degree $p+2$,\ determines a differential character whose evaluation on the embedded worldvolume yields a rigorous definition of the topological (Wess--Zumino) term $\,\cA_{\rm DF(WZ)}[\bX]\,$ in the Dirac--Feynman amplitude 
\qq\nn
\cA_{\rm DF}[\bX]=\ee^{\sfi\,S_\si[\bX]}\,,
\qqq
locally (that is for $\,\bX\,$ that factors through a contractible open subset $\,\cO\subset\xcM$) given by the exponent of the integral of the pullback of a primitive $\,\sfd^{-1}\underset{\tx{\ciut{(p+2)}}}{\chi}\,$ of $\,\underset{\tx{\ciut{(p+2)}}}{\chi}$,\ 
\qq\nn
\cA_{\rm DF(WZ)}[\bX]\equiv\ee^{\sfi\,S^{(p+1)}_{\rm WZ}[\bX]}\,,\qquad\qquad S^{(p+1)}_{\rm WZ}[\bX]=\int_{\Om_{p+1}}\,\bX^*\sfd^{-1}\underset{\tx{\ciut{(p+2)}}}{\chi}\,.
\qqq
The differential character can be understood as a generalisation of the line holonomy of a principal $\bC^\x$-bundle, termed the $(p+1)$-surface holonomy of the $p$-gerbe $\,\underset{\tx{\ciut{(p)}}}{\cG}\,$ along $\,\bX$,
\qq\nn
\cA_{\rm DF(WZ)}\equiv{\rm Hol}_{\underset{\tx{\ciut{(p)}}}{\cG}}(\cdot)\ :\ C^\infty(\Om_{p+1},\xcM)\too\uj\,,
\qqq
and assigning to $\,\underset{\tx{\ciut{(p)}}}{\cG}\,$ the image of the class
\qq\nn
[\bX^*\underset{\tx{\ciut{(p)}}}{\cG}]\in\vH^{p+1}\bigl(\Om_{p+1},\unl\uj\bigr)
\qqq
of its pullback along $\,\bX\,$ in the \Cv ech cohomology group $\,\vH^{p+1}\bigl(\Om_{p+1},\unl\uj\bigr)\,$ with values in the sheaf $\,\unl\uj\,$ of locally constant maps $\,\Om_{p+1}\too\uj\,$ under the isomorphism
\qq\nn
\vH^{p+1}\bigl(\Om_{p+1},\unl\uj\bigr)\cong\uj\,.
\qqq
But most importantly, the (super-)$p$-gerbe determines a geometric quantisation scheme for the classical theory $\,S_\si\,$ as it transgresses to a principal $\bC^\x$-bundle over the configuration ($p$-loop) space of the theory over $\,\xcM\,$ with a compatible connection of curvature $\,\underset{\tx{\ciut{(p+2)}}}{\chi}$,\ that is (the nontrivial component of) the prequantum bundle of the $\si$-model. Its square-integrable sections become, upon polarisation, the wave functionals of the field theory under consideration. 

In supersymmetric field theories, the de Rham cohomology $\,H^\bullet_{\rm dR}(\xcM)\,$ is replaced in our considerations by its refinement $\,H^\bullet_{\rm dR}(\xcM)^\txG\,$ spanned on classes of closed superdifferential forms invariant under the natural action of the supersymmetry group $\,\txG$.\ Given the non-compactness of the latter, there arises, generically, a discrepancy between the two cohomologies, $\,H^\bullet_{\rm dR}(\xcM)\,$ and $\,H^\bullet_{\rm dR}(\xcM)^\txG$,\ which apparently quantifies the topology enhancement\footnote{{\it I.e.}, the appearance of new non-contractible cycles.} accompanying the descent from the original supermanifold $\,\xcM\,$ to its Rabin--Crane(--Kosteleck\'y) orbifold, constructed explicitly for the super-Minkowski space in Refs.\,\cite{Rabin:1984rm,Rabin:1985tv}, subsequently used, for the same supertarget, in \Rcite{Suszek:2017xlw}, and finally explored in the field-theoretic context on both the super-Minkowski space $\,{\rm sMink}^{9,1\,\vert\,32}\,$ and the super-${\rm AdS}_5\x\bS^5\,$ space $\,{\rm s}({\rm AdS}_5\x\bS^5)\,$ in \Rcite{Suszek:2018bvx}. The encrypted topological content of $\,H^\bullet_{\rm dR}(\xcM)^\txG\,$ motivates and justifies the attempts, initiated in the super-Minkowskian setting in \Rcite{Suszek:2017xlw} and continued in the super-${\rm AdS}_5\x\bS^5\,$ setting in \Rcite{Suszek:2018bvx}, at a geometrisation, analogous to that known from the non-graded geometry ({\it cp} Refs.\,\cite{Murray:1994db,Murray:1999ew}), of physically distinguished super-$(p+2)$-cocycles representing classes in the supersymmetry-invariant cohomology of supertargets. Superstring theory further suggests, as natural candidates, supermanifolds with the structure of a homogeneous space of a Lie supergroup, such as, {\it e.g.},  the super-Minkowski space and the supermanifolds $\,{\rm s}({\rm AdS}_2\x\bS^2),{\rm s}({\rm AdS}_3\x\bS^3),{\rm s}({\rm AdS}_4\x\bS^7),{\rm s}({\rm AdS}_7\x\bS^4)\,$ and $\,{\rm s}({\rm AdS}_5\x\bS^5)\,$ with curved and topologically nontrivial bodies. The study of these distinguished supermanifolds has an additional advantage, to wit, it enables us to use the Cartan calculus of supersymmetric ({\it i.e.}, left-invariant) superdifferential forms on $\,\txG\,$ in the construction and analysis of the super-$\si$-models on homogeneous spaces $\,\xcM\equiv\txG/\txH\,$ (here, $\,\txH\,$ is the isotropy group of an arbitrary point $\,m\in\xcM$), and this, in turn, paves the way to the application of the Cartan--Eilenberg cohomology $\,{\rm CaE}^\bullet(\txG)\equiv H^\bullet_{\rm dR}(\txG)^\txG\,$ of the supersymmetry group $\,\txG$.\ The latter cohomology is isomorphic with the Chevalley--Eilenberg cohomology $\,{\rm CE}^\bullet(\ggt,\bR)\,$ of the Lie superalgebra $\,\ggt\,$ of $\,\txG\,$ with values in its trivial module $\,\bR$,\ and so the concept of geometrisation acquires a purely Lie-superalgebraic reformulation. In particular, cohomological trivialisation of the Cartan--Eilenberg super-$(p+2)$-cocycles upon pullback to supermanifolds that surjectively submerse their supports $\,\xcM$,\ at the root of the standard geometrisation scheme, is transcribed -- with the help of the one-to-one correspondence between classes in the 2nd cohomology group $\,{\rm CE}^2(\ggt,\agt)\,$ of $\,\ggt\,$ with values in an abelian module $\,\agt\,$ of $\,\ggt\,$ and equivalence classes of central extensions of $\,\ggt\,$ by $\,\agt\,$ ({\it cp} \Rcite{Suszek:2017xlw}) -- into a construction of suitable extensions of the supersymmetry algebra $\,\ggt\,$ whose integration to Lie supergroups, whenever possible, yields the desired surjective submersions. This is the basic idea underlying the geometrisation programme proposed and tested in \Rcite{Suszek:2017xlw} and elaborated, in the metrically and topologically nontrivial supergeometry of the supermanifold $\,{\rm s}({\rm AdS}_5\x\bS^5)$,\ in \Rcite{Suszek:2018bvx}. The supergeometric objects associated, as a result of the geometrisation, with the physically relevant super-$(p+2)$-cocycles were termed (Green--Schwarz) super-$p$-gerbes in \Rcite{Suszek:2017xlw}. 

A crucial property of any such super-$p$-gerbe is the existence of a weak $\k$-equivariant structure, as defined in  \Rcite{Suszek:2017xlw}, on its extension by a trivial super-$p$-gerbe determined by the Hughes--Polchinski formulation of the associated super-$\si$-model. The formulation was originally discovered in \Rcite{Hughes:1986dn} and analysed at great length, also in the present context, in \Rcite{Suszek:2017xlw}. In it, the metric term of the super-$\si$-model in the standard Nambu--Goto (or Polyakov) formulation goes over to an extra topological term, of the same nature as the original Wess--Zumino term, and thus gives rise to a trivial super-$p$-gerbe that extends the super-$p$-gerbe associated with the super-$(p+2)$-cocycle $\,\underset{\tx{\ciut{(p+2)}}}{\chi}$.\ The ensuing extended super-$p$-gerbe is then required to carry an incomplete equivariant structure with respect to \emph{infinitesimal} (\emph{tangential}) \emph{right} Gra\ss mann-odd translations on $\,\txG\,$ taken from the image of a projector acting on the Gra\ss mann-odd component $\,\ggt^{(1)}\,$ of $\,\ggt\,$ and annihilating exactly half of the Gra\ss mann-odd degrees of freedom present. Such distinguished linearised supertranslations preserve the super-$\si$-model action functional in the Hughes--Polchinski formulation and the presence of this gauge supersymmetry serves to establish actual supersymmetric balance between the bosonic and fermionic degrees of freedom in the field theories under consideration, which is why we insist upon geometrising the symmetry.

\medskip

The realisation of the programme delineated above was initiated in \Rcite{Suszek:2017xlw} in the topologically and metrically trivial setting of the super-Minkowski space. The long-known Green--Schwarz super-$(p+2)$-cocycles for the super-$p$-branes were geometrised explicitly for $\,p\in\{0,1,2\}\,$ and the ensuing super-$p$-gerbes were subsequently shown to possess the expected supersymmetry-($\Ad_\cdot$-)equi\-vari\-ant structure, in perfect analogy with their bosonic counterpart for $\,p=1$,\ {\it cp} Refs.\,\cite{Gawedzki:2010rn,Gawedzki:2012fu,Suszek:2011,Suszek:2012ddg,Suszek:2013}, whose appearance in this picture follows from the identification of the GS super-$3$-cocycle on the super-Minkowski space as a super-variant of the canonical Cartan 3-form on a Lie group, and that of the associated super-$\si$-model in the Polyakov formulation as the super-variant of the well-known Wess--Zumino--Witten $\si$-model of Refs.\,\cite{Witten:1983ar,Gawedzki:1990jc,Gawedzki:1999bq,Gawedzki:2001rm}. The associated extended super-$p$-gerbes were demonstrated to possess a weak $\k$-equivariant structure.

The intuitions and tools gathered in the manageable super-Minkowskian setting were next adapted to the  highly nontrivial, both topologically and metrically, superbackground of the two-dimensional Metsaev--Tseytlin super-$\si$-model of \Rcite{Metsaev:1998it}, of much physical relevance in the context of the recent studies, founded on the ${\rm AdS}$/CFT correspondence, of QCD-type systems at strong coupling. The corresponding supertarget of the super-$\si$-model has the desired structure of a homogeneous space
\qq\nn
{\rm s}\bigl({\rm AdS}_5\x\bS^5\bigr)\equiv{\rm SU}(2,2\,\vert\,4)/\bigl({\rm SO}(4,1)\x{\rm SO}(5)\bigr)
\qqq
of the supersymmetry group $\,{\rm SU}(2,2\,\vert\,4)\,$ and supports a Green--Schwarz super-3-cocycle descended from a left-invariant closed super-3-form $\,\underset{\tx{\ciut{(3)}}}{\chi}^{\rm MT}\,$ on $\,{\rm SU}(2,2\,\vert\,4)\,$ given by a linear combination, with ${\rm SO}(4,1)\x{\rm SO}(5)$-invariant tensors as coefficients, of wedge products of components of the $\gt{su}(2,2\,\vert\,4)$-valued Maurer--Cartan super-1-form on $\,{\rm SU}(2,2\,\vert\,4)\,$ along the direct-sum completion of the isotropy algebra $\,\gt{so}(4,1)\oplus\gt{so}(5)\,$ within $\,\gt{su}(2,2\,\vert\,4)\,$ (which ensures that it descends to the quotient $\,{\rm SU}(2,2\,\vert\,4)/({\rm SO}(4,1)\x{\rm SO}(5))$). The super-3-cocycle admits a \emph{global} supersymmetric primitive, found by Roiban and Siegel in \Rcite{Roiban:2000yy}, of the same structure as the super-3-cocycle which -- consequently -- also descends to $\,{\rm s}({\rm AdS}_5\x\bS^5)$.\ These observations led to the construction of a \emph{trivial} gerbe for the Metsaev--Tseytlin super-3-cocycle in \Rcite{Suszek:2017xlw}, but that turned out \emph{not} to be the end of the story.

Indeed, the Metsaev--Tseytlin superbackground 
\qq\nn
\bigl({\rm s}\bigl({\rm AdS}_5\x\bS^5\bigr),\underset{\tx{\ciut{(3)}}}{\chi}^{\rm MT}\bigr)
\qqq
is related to a flat superbackground
\qq\nn
\bigl({\rm sMink}^{9,1\,\vert\,32},\underset{\tx{\ciut{(3)}}}{\chi}^{\rm GS}\bigr)\,,
\qqq
with a certain Green--Schwarz super-3-cocycle $\,\underset{\tx{\ciut{(3)}}}{\chi}^{\rm GS}\in Z^3_{\rm dR}({\rm sMink}^{9,1\,\vert\,32})^{{\rm sISO}(9,1\,\vert\,32)}$,\ through the flattening limit
\qq\nn
R\to\infty
\qqq
of the \emph{common} radius $\,R\,$ of the generating 1-cycle of $\,{\rm AdS}_5\cong\bS^1\x\bR^{\x 4}\,$ and that of the 5-sphere in the body of the supertarget. The limit is the Lie-supergroup dual of the \.In\"on\"u--Wigner contraction
\qq\label{eq:su224contrsmink91}
\gt{su}(2,2\,\vert\,4)\xrightarrow[{\rm rescaling}]{\ R-{\rm dependent}\ }\gt{su}(2,2\,\vert\,4)_R\xrightarrow{\ R\to\infty\ }\gt{smink}^{d,1\,\vert\,D_{d,1}}\,.
\qqq
In fact, the asymptotic relation between the two superbackgrounds is one of the basic guiding principles on which Metsaev and Tseytlin based their construction of the two-dimensional super-$\si$-model for $\,{\rm s}({\rm AdS}_5\x\bS^5)$.\ Therefore, it seems natural to demand that a geometrisation of the super-3-cocycle defining the super-$\si$-model with the supersymmetry group $\,{\rm SU}(2,2\,\vert\,4)\,$ should be compatible with the blow-up transformation
\qq
\bigl({\rm s}\bigl({\rm AdS}_5\x\bS^5\bigr),\underset{\tx{\ciut{(3)}}}{\chi}^{\rm MT}\bigr)\xrightarrow[{\rm rescaling}]{\ R-{\rm dependent}\ }\bigl({\rm s}\bigl({\rm AdS}_5(R)\x\bS^5(R)\bigr),\underset{\tx{\ciut{(3)}}}{\chi}^{\rm MT}(R)\bigr)\xrightarrow{\ R\to\infty\ }\bigl({\rm sMink}^{9,1\,\vert\,32},\underset{\tx{\ciut{(3)}}}{\chi}^{\rm GS}\bigr)\,.\cr \label{eq:AdSbacktozMinkback}
\qqq
This is \emph{not} the case for the trivial super-1-gerbe associated with the Roiban--Siegel primitive of the Metsaev--Tseytlin super-3-cocycle, as the said primitive asymptotes to an exact super-2-form on $\,{\rm sMink}^{9,1\,\vert\,32}$.\ The last constatation prompted a systematic search, with partial results reported in \Rcite{Suszek:2018bvx}, for an alternative geometrisation of $\,\underset{\tx{\ciut{(3)}}}{\chi}^{\rm MT}$.\ In conformity with the fomerly presented logic, the task in hand was reformulated in Lie-superalgebraic terms in which it boiled down to finding an extension of the supersymmetry algebra $\,\gt{su}(2,2\,\vert\,4)\,$ (integrable to a Lie supergroup surjectively submersing $\,{\rm SU}(2,2\,\vert\,4)$) that would contract to the central extension of $\,\gt{smink}^{9,1\,\vert\,32}\,$ determined by the Green--Schwarz super-3-cocycle $\,\underset{\tx{\ciut{(3)}}}{\chi}^{\rm GS}$.\ Analysis of two parametric families of natural deformations of $\,\gt{su}(2,2\,\vert\,4)$,\ motivated by the study of (the asymptotics of) the wrapping-charge anomalies (including those of the Kosteleck\'y--Rabin type) in the Poisson algebra of Noether charges of supersymmetry in the canonical description of the Metsaev--Tseytlin super-$\si$-model, yielded negative results, leaving us with the non-contractible trivial super-1-gerbe as the only consistent geometrisation of the Metsaev--Tseytlin super-3-cocycle to date.

\medskip

In the failed attempt at constructing a contractible geometrisation of the Metsaev--Tseytlin super-3-cocycle on $\,{\rm s}({\rm AdS}_5\x\bS^5)$,\ the asymptotic relation \eqref{eq:AdSbacktozMinkback} was assumed fundamental in that we worked with a \emph{fixed pair} of super-3-cocycles: $\,(\underset{\tx{\ciut{(3)}}}{\chi}^{\rm MT},\underset{\tx{\ciut{(3)}}}{\chi}^{\rm GS})\,$ and merely sought a trivialisation of the former one subject to the asymptotic constraint defined by the known trivialisation of the latter one. But clearly, this is not a unique approach, and the general principle upon which the construction of the Metsaev--Tseytlin super-$\si$-model was founded, to wit, that the super-$\si$-model and the super-3-cocycle asymptote to their respective super-Minkowskian counterparts and exhibit $\k$-symmetry, leaves much room for an alternative. Indeed, upon taking, instead, the contraction \eqref{eq:su224contrsmink91} to be fundamental and, consequently, demanding that the latter lift to the respective extensions, with the extension of the super-Minkowskian algebra fixed by the known trivialisation of the Green--Schwarz super-3-cocycle (taken as the fixed reference datum of the geometrisation), we arrive at a geometrisation scheme in which the point of departure is a pair $\,(\widetilde\ggt_2,\widetilde\ggt_1)\,$ of Lie superalgebras extending a given pair of geometric Lie superalgebras $\,(\ggt_2,\ggt_1)\,$ with distinguished (isotropy) Lie subalgebras $\,(\hgt_2,\hgt_1)$,\ unchanged by the deformation, in such a manner that an \.In\"on\"u--Wigner contraction ($\vec R\,$ is the scaling parameter with the relevant limiting value $\,\vec R_*$)
\qq\nn
\widetilde\ggt_2\xrightarrow[{\rm rescaling}]{\ \vec R-{\rm dependent}\ }\widetilde\ggt_{2,\vec R}\xrightarrow{\ \vec R\to\vec R_*}\widetilde\ggt_1/\hgt_1
\qqq
can be devised that projects to the original \.In\"on\"u--Wigner contraction
\qq\nn
\ggt_2\xrightarrow[{\rm rescaling}]{\ \vec R-{\rm dependent}\ }\ggt_{2,\vec R}\xrightarrow{\ \vec R\to\vec R_*}\ggt_1/\hgt_1\,.
\qqq
Under certain circumstances, to be specified in due course, the extension $\,\widetilde\ggt_2\,$ \emph{determines} a super-$(p+1)$-form (a super-2-form in the original example) that descends to the homogeneous space $\,\widetilde\txG_2/\txH_2\,$ and asymptotes to the predetermined primitive of the super-$(p+2)$-cocycle on $\,\widetilde\txG_1/\txH_1$,\ a {\it sine qua non} condition for the contractiblity of the associated geometrisation. In this scenario, it is the contractible primitive on $\,\widetilde\txG_2/\txH_2\,$ that \emph{defines} the underlying super-$(p+2)$-cocycle (through exterior differentiation), and so also the actual supertarget of the associated super-$\si$-model, the latter being identified with the (minimal) homogeneous space to which the super-$(p+2)$-cocycle descends.

\medskip

A pilot study of the alternative (correlated) geometrisation scheme outlined above was carried out already in \Rcite{Suszek:2018bvx} where a non-standard super-extension of the ${\rm AdS}$-algebra, contractible to the standard (Kosteleck\'y--Rabin) superstring extension of the super-Minkowskian algebra, was shown to yield a super-3-cocycle which would \emph{not} descend from the homogeneous space $\,\widetilde\txG_2/\txH_2\,$ of the corresponding extended supersymmetry group $\,\widetilde\txG_2$,\ a fact to be interpreted as evidence of the potential necessity of taking the extended homogeneous space $\,\widetilde\txG_2/\txH_2\,$ as the supertarget of the relevant super-$\si$-model. In the present paper, we test the validity of the geometrisation scheme thus constrained on the pair of consistent $\,N=2\,$ super-0-brane superbackgrounds related by a blow-up transformation, namely: the super-Minkowski space
\qq\nn
{\rm sMink}^{3,1\,\vert\,2\cdot 4}\equiv{\rm sISO}(3,1\,\vert\,2\cdot 4)/{\rm SO}(3,1)
\qqq 
with the standard $\,N=2\,$ Green--Schwarz super-2-cocycle on it and the super-${\rm AdS}_2\x\bS^2\,$ space 
\qq\nn
{\rm s}\bigl({\rm AdS}_2\x\bS^2\bigr)\equiv{\rm SU}(1,1\,\vert\,2)_2/\bigl({\rm SO}(1,1)\x{\rm SO}(2)\bigr)
\qqq
with a particular super-2-cocycle from the 2-parameter family of super-2-cocycles on it derived by Zhou in \Rcite{Zhou:1999sm}, asymptoting to the former in the limit of an infinite common radius of the $\,{\rm AdS}_2\cong\bS^1\x\bR\,$ and the $\,\bS^2\,$ in the body $\,{\rm AdS}_2\x\bS^2\,$ of the supertarget. Our analysis yields a super-0-gerbe over $\,{\rm s}({\rm AdS}_2\x\bS^2)\,$ (in particular, its curvature \emph{is} the one found by Zhou) that contracts to its super-Minkowskian counterpart. The super-0-gerbe is shown to admit an extension, determined by the Hughes--Polchinski reformulation of Zhou's super-$\si$-model, that exhibits a manifest weak $\k$-equivariance. It is to be emphasised that Zhou's super-2-cocycle admits, upon pullback to the supersymmetry group $\,{\rm SU}(1,1\,\vert\,2)_2$,\ a global primitive that \emph{does not} descend to the quotient $\,{\rm SU}(1,1\,\vert\,2)_2/({\rm SO}(1,1)\x{\rm SO}(2))\,$ but \emph{does} asymptote to the non-supersymmetric primitive of the Green--Schwarz super-2-cocycle on $\,{\rm sMink}^{3,1\,\vert\,2\cdot 4}$.\ Furthermore, the low dimensionality of the super-$\si$-model has critical impact on the complexity of the construction reported hereunder. Altogether, then, our study accounts but for a very particular case from a rich variety of supergeometric circumstances, known to be realised in the field-theoretic context under consideration, to which we are planning to return in the future work. 

The paper is organised as follows:
\bit
\item in Section \ref{sec:Carthomsp}, we recapitulate the construction of a lagrangean field theory on a homogeneous space $\,\txG/\txH\,$ of a (super)symmetry Lie (super)group $\,\txG\,$ (corresponding to a reductive decomposition $\,\ggt=\tgt\oplus\hgt\,$ of the supersymmetry Lie superalgebra $\,\ggt$) using suitable elements of the Cartan differential calculus on $\,\txG$,\ and the ensuing induction of a non-linear realisation of supersymmetry; subsequently, we specialise the general discussion to the setting of interest, {\it i.e.}, the Nambu--Goto and the Hughes--Polchinski formulations of the (Green--Schwarz) super-$\si$-model of super-$p$-brane dynamics on $\,\txG/\txH$;\ we also discuss the mechanism of induction of a Lie-superalgebra extension from a super-$(p+2)$-cocycle representing a class in $\,{\rm CaE}^{p+2}(\txG)\,$ and give, after Tuynman and Wiegenrinck, the necessary and sufficient conditions for a central extension of a Lie (super)algebra to integrate to a central extension of the corresponding Lie supergroup;
\item in Section \ref{sec:scheme}, we lay down the general scheme of a correlated geometrisation of a pair of super-$(p+2)$-cocycles on homogeneous spaces in correspondence determined by a blow-up transformation dual to an \.In\"on\"u--Wigner contraction relating the respective Lie superalgebras; 
\item in Section \ref{sec:ints0bext}, we derive an extension of the $\,N=2\,$ supersymmetry algebra $\,\gt{siso}(3,1\,\vert\,2\cdot 4)\,$ on the basis of the field-theoretic realisation of the supersymmetry in the associated Green--Schwarz super-$\si$-model and verify its integrability to a Lie-supergroup extension; subsequently, we construct an extension of the supersymmetry algebra $\,\gt{su}(1,1\,\vert\,2)_2\,$ of Zhou's super-$\si$-model that admits an \.In\"on\"u--Wigner contraction projecting to the one that relates the underlying geometric Lie superalgebras; the latter extension is checked to integrate to a Lie-supergroup extension of the supersymmetry group $\,{\rm SU}(1,1\,\vert\,2)_2$; 
\item in Section \ref{sec:geometrise}, we use the Lie-supergroup extensions derived in the preceding section to induce surjective submersions over the respective supertargets: $\,{\rm sMink}^{3,1\,\vert\,2\cdot 4}\,$ and $\,{\rm s}({\rm AdS}_2\x\bS^2)\,$ and endow them with supersymmetric primitives for the pullbacks of the Green--Schwarz super-2-cocycles, thereby completing their correlated geometrisation compatible with the contraction which yields a pair of super-0-gerbes in correspondence under the blow-up transformation;
\item in Section \ref{sec:kappa}, we define Hughes--Polchinski extensions of the super-0-gerbes obtained in the preceding section and verify the existence of a suitably defined weak $\k$-equivariant structure thereupon;
\item in Section \ref{sec:C&O}, we recapitulate the work reported in the present paper and indicate possible directions of its future continuation;
\item in the Appendix, we give the conventions on and facts regarding the Clifford algebra relevant for the Cartan-geometric description of the super-${\rm AdS}_2\x\bS^2\,$ space.
\eit

\section{The Cartan geometry of homogeneous superspaces and super-$\si$-models thereon}\label{sec:Carthomsp}

Let $\,\txG\,$ be a Lie supergroup, to be referred to as {\bf the supersymmetry group} in what follows, and let $\,\txH\,$ be a closed Lie subgroup of $\,\txG$,\ to be termed {\bf the isotropy group}, with a distinguished closed Lie subgroup
\qq\label{eq:HvacH}
\txH_{\rm vac}\subseteq\txH\,,
\qqq 
to be termed {\bf the vacuum isotropy group}. Let the corresponding Lie (super)algebras be: $\,\ggt\,$ for $\,\txG$,\ to be called {\bf the supersymmetry algebra}, and $\,\hgt\supset[\hgt,\hgt]\,$ (resp.\ $\,\hgt_{\rm vac}\supset[\hgt_{\rm vac},\hgt_{\rm vac}]$) for $\,\txH\,$ (resp.\ $\,\txH_{\rm vac}$),\ to be called {\bf the isotropy algebra} (resp.\ {\bf the vacuum isotropy algebra}). We shall denote the direct-sum complement of $\,\hgt\,$ within $\,\ggt\,$ as $\,\tgt$,
\qq\label{eq:ggtgthgt}
\ggt=\tgt\oplus\hgt\,,
\qqq
further assuming it to be an ${\rm ad}$-module of the isotropy algebra,
\qq\nn
[\hgt,\tgt]\subset\tgt\,,
\qqq
which qualifies decomposition \eqref{eq:ggtgthgt} as {\bf reductive}. The Lie superalgebra $\,\ggt\,$ admits a super-grading
\qq\nn
\ggt=\ggt^{(0)}\oplus\ggt^{(1)}
\qqq
in which $\,\ggt^{(0)}\,$ is the Gra\ss mann-even Lie subalgebra of $\,\ggt$,
\qq\nn 
[\ggt^{(0)},\ggt^{(0)}]\subset\ggt^{(0)}\,,
\qqq
containing $\,\hgt$, 
\qq\nn
\hgt\subset\ggt^{(0)}\,,
\qqq
and $\,\ggt^{(1)}\,$ is the Gra\ss mann-odd $\ad$-module thereof,
\qq\nn
[\ggt^{(0)},\ggt^{(1)}]\subset\ggt^{(1)}\,.
\qqq
The super-grading is inherited by the subspace $\,\tgt$,
\qq\nn
\tgt=\tgt^{(0)}\oplus\tgt^{(1)}\,.
\qqq
The direct-sum complement of $\,\hgt_{\rm vac}\,$ within $\,\hgt\,$ shall be denoted as $\,\dgt$,
\qq\nn
\hgt=\dgt\oplus\hgt_{\rm vac}\,.
\qqq
Finally, we distinguish a subspace 
\qq\nn
\tgt^{(0)}_{\rm vac}\subseteq\tgt^{(0)}
\qqq
closed under the ${\rm ad}$-action of the vacuum isotropy algebra,
\qq\nn
[\hgt_{\rm vac},\tgt^{(0)}_{\rm vac}]\subset\tgt^{(0)}_{\rm vac}\,.
\qqq
Its direct-sum complement within $\,\tgt^{(0)}\,$ shall be denoted as $\,\egt$,
\qq\nn
\tgt^{(0)}=\tgt^{(0)}_{\rm vac}\oplus\egt^{(0)}\,.
\qqq
We assume the decomposition
\qq\nn
\ggt=\fgt\oplus\hgt_{\rm vac}\,,\qquad\qquad\fgt=\tgt\oplus\dgt
\qqq
to be reductive as well,
\qq\nn
[\hgt_{\rm vac},\fgt]\subset\fgt\,.
\qqq
The adjoint action of $\,\hgt_{\rm vac}\,$ on $\,\tgt^{(0)}_{\rm vac}\,$ is taken to integrate to a {\bf unimodular} (adjoint) action of the Lie group $\,\txH_{\rm vac}\,$ on the same space, {\it i.e.},
\qq\label{eq:Hvacunimod}
\forall_{h\in\txH_{\rm vac}}\ :\ \det\,\bigl(\sfT_e\Ad_h\rstr_{\tgt^{(0)}_{\rm vac}}\bigr)=1\,.
\qqq
We set 
\qq\nn
(D,\d,\unl\d,d,p):=(\dim\,\ggt-1,\dim\,\tgt-1,\dim\,\fgt-1,\dim\,\tgt^{(0)}-1,\dim\,\tgt^{(0)}_{\rm vac}-1)
\qqq
and denote the respective basis vectors (generators) of the various (complexified) subalgebras and subspaces as 
\qq\nn
&\ggt=\bigoplus_{A=0}^D\,\corr{t_A}_\bC\,,\qquad\qquad\tgt=\bigoplus_{\unl A=0}^\d\,\corr{t_{\unl A}}_\bC\,,\qquad\qquad\hgt=\bigoplus_{S=1}^{D-\d}\,\corr{J_S}_\bC\,,&\cr\cr
&\tgt^{(0)}=\bigoplus_{\mu=0}^d\,\corr{P_\mu}_\bC\,,\qquad\qquad\tgt^{(1)}=\bigoplus_{\a=1}^{\d-d}\,\corr{Q_\a}_\bC\,,\qquad\qquad\tgt_{\rm vac}^{(0)}=\bigoplus_{\unl a=0}^p\,\corr{P_{\unl a}}_\bC\,,&\cr\cr
&\egt^{(0)}=\bigoplus_{\widehat a=p+1}^d\,\corr{P_{\widehat a}}_\bC\,,\qquad\qquad\hgt_{\rm vac}=\bigoplus_{\unl S=1}^{D-\unl\d}\,\corr{J_{\unl S}}_\bC\,,\qquad\qquad\dgt=\bigoplus_{\widehat S=D-\unl\d+1}^{D-\d}\,\corr{J_{\widehat S}}_\bC\,.&
\qqq
These satisfy structure relations
\qq\nn
[t_A,t_B\}=f_{AB}^{\ \ C}\,t_C
\qqq
in which the $\,f_{AB}^{\ \ C}\,$ are structure constants with symmetry properties, expressed in terms of the Gra\ss mann parities $\,|A|\equiv|t_A|\,$ and $\,|B|\equiv|t_B|\,$ of the respective (homogeneous) generators $\,t_A\,$ and $\,t_B$,
\qq\nn
f_{AB}^{\ \ \ C}=(-1)^{|A|\cdot|B|+1}\,f_{BA}^{\ \ \ C}\in\bC\,.
\qqq

The homogeneous space $\,\txG/\txK,\ \txK\in\{\txH,\txH_{\rm vac}\}\,$ can be realised locally as a section of the principal bundle\footnote{The first arrow denotes the free and transitive action of the structure group on the fibre.}
\qq\nn
\alxydim{@C=1cm@R=1cm}{\txK \ar[r] & \txG \ar[d]^{\pi_{\txG/\txK}} \\ & \txG/\txK}
\qqq
with the structure group $\,\txK$.\ Thus, we shall work with a family of submanifolds embedded in $\,\txG\,$ by the respective (local) sections\footnote{That is, equivalently, by a collection of local trivialisations.}
\qq\nn
\si_i^\txK\ :\ \cO_i^\txK\too\txG\ :\ g\txK\longmapsto g\cdot h_i^\txK(g)\,,\quad i\in I^\txK\,,
\qqq
of the submersive projection on the base $\,\pi_{\txG/\txK}$,\ associated with a trivialising cover $\,\cO^\txK=\{\cO_i^\txK\}_{i\in I^\txK}\,$ of the latter,
\qq\nn
\txG/\txK=\bigcup_{i\in I^\txK}\,\cO_i^\txK\,.
\qqq
The redundancy of such a realisation over any non-empty intersection, $\,\cO_{ij}\hspace{-6pt}{}^\txK\equiv\cO_i^\txK\cap\cO_j^\txK\neq\emptyset$,\ is accounted for by a collection of locally smooth (transition) maps
\qq\nn
h^\txK_{ij}\ :\ \cO^\txK_{ij}\too\txK\subset\txG
\qqq
fixed by the condition
\qq\nn
\forall_{x\in\cO^\txK_{ij}}\ :\ \si^\txK_j(x)=\si_i^\txK(x)\cdot h^\txK_{ij}(x)\,.
\qqq
The homogeneous space admits a natural action of the supersymmetry group induced by the left regular action
\qq\nn
\ell_\cdot\ :\ \txG\x\txG\too\txG\ :\ (g',g)\longmapsto g'\cdot g\equiv\ell_{g'}(g)\,,
\qqq
namely,
\qq\label{eq:cosetlact}
[\ell^\txK]_\cdot\ :\ \txG\x\txG/\txK\too\txG/\txK\ :\ (g',g\txK)\longmapsto(g'\cdot g)\txK\,.
\qqq
The latter is transcribed, through the $\,\si_i^\txK$,\ into a geometric realisation of $\,\txG\,$ on the image of $\,\txG/\txK\,$ within $\,\txG$,\ with the same obvious redundancy. Indeed, consider a point $\,x\in\cO^\txK_i\,$ and an element $\,g\in\txG$.\ Upon choosing an \emph{arbitrary} index $\,j\in I^\txK\,$ with the property
\qq\nn
\widetilde x(x;g'):=\pi_{\txG/\txK}\bigl(g'\cdot\si^\txK_i(x)\bigr)\in\cO^\txK_j\,,
\qqq
we find a unique $\,\unl h^\txK_{ij}(x;g')\in\txK\,$ defined (on some open neighbourhood of $\,(x,g')$) by the condition
\qq\nn
g'\cdot\si^\txK_i(x)=\si^\txK_j\bigl(\widetilde x(x;g')\bigr)\cdot\unl h^\txK_{ij}(x;g')^{-1}\,.
\qqq
Note that for $\,\widetilde x(x;g')\in\cO^\txK_{jk}\,$ we have
\qq\nn
\unl h^\txK_{ik}(x;g')=\unl h^\txK_{ij}(x;g')\cdot h^\txK_{jk}\bigl(\widetilde x(x;g')\bigr)\,,
\qqq
so that the two realisations of the action are related by a compensating transformation from the structure group $\,\txK$.

The realisation of the homogeneous space $\,\txG/\txK\,$ within $\,\txG\,$ described above enables us to reconstruct the differential calculus on the former space from that on the latter. In particular, the tangent sheaf of $\,\txG/\txK\,$ over $\,\cO_i^\txK\ni x\,$ is spanned on (restrictions of) the fundamental vector fields $\,\Xi_X^\txK,\ X\in\ggt\,$ of $\,[\ell^\txK]_\cdot$.\ Over $\,\cO^\txK_i$,\ these push forward to the functional-linear combinations $\,\cK_X^\txK\,$ of the right- and left-invariant vector fields on the Lie supergroup
\qq\nn
\cK_X^\txK\bigl(\si^\txK_i(x)\bigr)=\sfT_x\si_i\bigl(\Xi_X(x)\bigr)\,,
\qqq
described in detail in \cite[Sec.\,2]{Suszek:2018bvx}. In the dual sheaf of (super)differential forms on $\,\txG/\txK$,\ on the other hand, we find global sections descended from the Lie supergroup $\,\txG$.\ Denote the relevant direct-sum decomposition of the supersymmetry algebra as 
\qq\nn
\ggt=\lgt\oplus\kgt\,,\qquad\qquad(\lgt,\kgt)\in\{(\tgt,\hgt),(\fgt,\hgt_{\rm vac})\}
\qqq
and the corresponding generators as
\qq\nn
\lgt=\bigoplus_{\z=0}^{\dim\,\lgt-1}\,\corr{L_\z}_\bC\,,\qquad\qquad\kgt=\bigoplus_{Z=1}^{\dim\,\kgt}\,\corr{J_Z}\,.
\qqq
Among the global sections of the latter sheaf, there are superdifferential forms on the homogeneous space with pullbacks along $\,\pi_{\txG/\txK}\,$ given by linear combinations of wedge products of the components of the left-invariant $\ggt$-valued Maurer--Cartan super-1-form (the $\,E^A_{\ B}\,$ are components of the standard Vielbein in the local coordinates $\,\{X^A\}^{A\in\ovl{0,D}}\,$ on $\,\txG$)
\qq\nn
\theta_{\rm L}=\theta_{\rm L}^A\ox t_A\equiv E^A_{\ B}(\cdot)\,\sfd X^B\ox t_A
\qqq
on $\,\txG\,$ along $\,\lgt$,\ with arbitrary $\txK$-invariant tensors as coefficients. Indeed, the said components transform tensorially under right regular $\txK$-translations on $\,\txG$,\ and so the combinations are manifestly $\txK$-basic. Consequently, pullbacks, along the local sections $\,\si^\txK_i\,$ over $\,\cO^\txK_i$,\ of super-$k$-forms
\qq\nn
\underset{\tx{\ciut{(k)}}}{\om}=\om_{\z_1\z_2\ldots\z_k}\,\theta_{\rm L}^{\z_1}\wedge\theta_{\rm L}^{\z_2}\wedge\cdots\wedge\theta_{\rm L}^{\z_k}\,,\qquad\z_i\in\ovl{0,\dim\,\lgt-1}\,,\quad i\in\ovl{1,k}\,,
\qqq
with -- for any left-invariant vector field $\,\ceL_{L_\z}\,$ associated with $\,L_\z\in\lgt\,$ in the standard manner --
\qq\nn
\ceL_{L_\z}\con\theta_{\rm L}^{\z'}=\d^\z_{\ \z'}
\qqq
and
\qq\nn
\om_{\z_1\z_2\ldots\z_k}\,\bigl(\sfT_e\Ad_h\bigr)^{\z_1}_{\ \z_1'}\,\bigl(\sfT_e\Ad_h\bigr)^{\z_2}_{\ \z_2'}\,\cdots\,\bigl(\sfT_e\Ad_h\bigr)^{\z_k}_{\ \z_k'}=\om_{\z_1'\z_2'\ldots\z_k'}\,,\qquad h\in\txK\,,
\qqq
do \emph{not} depend on the choice of the local section and hence glue smoothly over non-empty intersections $\,\cO^\txK_{ij}\,$ to \emph{global} superdifferential forms on $\,\txG/\txK$,\ mentioned earlier. This is the point of departure in the construction of a supersymmetric lagrangean field theory with the fibre of the covariant configuration bundle given by (or, to put it differently, with fields in the lagrangean density taking values in) $\,\txG/\txK$.

Among the lagrangean field theories of the type indicated, we find two classes of particular interest to us, and intimately related to one another, to wit, the Nambu--Goto super-$\si$-model of smooth embeddings of a $(p+1)$-dimensional worldvolume $\,\Om_p\,$ of a super-$p$-brane in $\,\txG/\txH\,$ and the Hughes--Polchinski model of smooth embeddings of $\,\Om_p\,$ in $\,\txG/\txH_{\rm vac}$,\ put forward in \Rcite{Hughes:1986dn}, elaborated in \Rcite{Gauntlett:1989qe} and recently revived in \Rcite{Suszek:2018bvx}. The main supergeometric datum that enters the definition of \emph{both} models (in correspondence) is a distinguished Cartan--Eilenberg super-$(p+2)$-cocycle $\,\underset{\tx{\ciut{(p+2)}}}{\chi}\in Z^{p+2}_{\rm dR}(\txG)^\txG\,$ on $\,\txG\,$ given by a linear combination 
\qq\nn
\underset{\tx{\ciut{(p+2)}}}{\chi}=\tfrac{1}{(p+2)!}\,\chi_{\unl A_1\unl A_2\ldots\unl A_{p+2}}\,\theta^{\unl A_1}_{\rm L}\wedge\theta^{\unl A_2}_{\rm L}\wedge\cdots\wedge\theta^{\unl A_{p+2}}_{\rm L}
\qqq
of $(p+2)$-fold wedge products of the components $\,\theta^{\unl A}_{\rm L},\ \unl A\in\ovl{0,\d}\,$ of the Maurer--Cartan super-1-form $\,\theta_{\rm L}\,$ along $\,\tgt\,$ with $\txH$-invariant tensors $\,\chi_{\unl A_1\unl A_2\ldots\unl A_{p+2}}\,$ as coefficients. Clearly, the super-$(p+2)$-cocycle descends to $\,\txG/\txH\,$ (and so also to $\,\txG/\txH_{\rm vac}$), that is, there exists a (unique) {\bf Green--Schwarz super-$(p+2)$-cocycle} $\,\underset{\tx{\ciut{(p+2)}}}{\txH}\in Z^{p+2}_{\rm dR}(\txG/\txH)^\txG\,$ with the property
\qq\nn
\underset{\tx{\ciut{(p+2)}}}{\chi}=\pi_{\txG/\txH}^*\underset{\tx{\ciut{(p+2)}}}{\txH}\,,
\qqq
or, equivalently, 
\qq\nn
\underset{\tx{\ciut{(p+2)}}}{\txH_i}\equiv\underset{\tx{\ciut{(p+2)}}}{\txH}\rstr_{\cO^\txH_i}=\si^\txH_i{}^*\underset{\tx{\ciut{(p+2)}}}{\chi}\,,
\qqq
and it is further assumed that the restrictions $\,\underset{\tx{\ciut{(p+2)}}}{\txH_i}\,$ are de Rham coboundaries with primitives $\,\underset{\tx{\ciut{(p+1)}}}{\txB_i}$,
\qq\nn
\underset{\tx{\ciut{(p+2)}}}{\txH_i}=\sfd\underset{\tx{\ciut{(p+1)}}}{\txB_i}\,,
\qqq
forming, under the induced action $\,[\ell^\txH]_\cdot\,$ of the supersymmetry group, a pseudo-invariant family in the sense of the relation
\qq\nn
[\ell]_g^*\underset{\tx{\ciut{(p+1)}}}{\txB_j}(x)=\underset{\tx{\ciut{(p+1)}}}{\txB_i}(x)+\sfd\underset{\tx{\ciut{(p)}}}{\D^g_{ij}}(x)
\qqq
valid for all $\,(g,x)\in\txG\x\cO^\txH_i$,\ for $\,j\in I\,$ such that $\,[\ell]_g(x)\in\cO^\txH_j$,\ and for some $\,\underset{\tx{\ciut{(p)}}}{\D^g_{ij}}\in\Om^p(\cO^\txH_i)$.\ In fact, in the most studied examples, $\,\underset{\tx{\ciut{(p+2)}}}{\chi}\,$ is de Rham-exact, and so it is the behaviour of its globally smooth primitive under left $\txG$-translations and right $\txK$-translations that determines its cohomological status on $\,\txG/\txK$,\ and -- through the latter -- the well-definedness of the corresponding field theory.

The Nambu--Goto super-$\si$-model requires yet another datum: a metric tensor $\,\unl\txg\,$ on $\,\txG/\txH\,$ descended from a left-$\txG$-invariant and right-$\txH$-basic metric tensor on $\,\txG\,$ as
\qq\nn
\pi_{\txG/\txH}^*\unl\txg=\txg_{\unl A\unl B}\,\theta_{\rm L}^{\unl A}\ox\theta_{\rm L}^{\unl B}\equiv\txg\,,
\qqq
where the $\,\txg_{\unl A\unl B}\,$ are components of an $\txH$-invariant tensor,
\qq\nn
\txg_{\unl A\unl B}\,\bigl(\sfT_e\Ad_h\bigr)^{\unl A}_{\ \unl A'}\,\bigl(\sfT_e\Ad_h\bigr)^{\unl B}_{\ \unl B'}=\txg_{\unl A'\unl B'}\,,\qquad h\in\txH\,.
\qqq
Given the pair $\,(\underset{\tx{\ciut{(p+2)}}}{\chi},\txg)$,\ we define the super-$\si$-model as the theory of smooth embeddings 
\qq\nn
\xi\ :\ \Om_p\too\txG/\txH
\qqq
of the $(p+1)$-dimensional worldvolume $\,\Om_p\,$ in the homogeneous space $\,\txG/\txH\,$ of the  supersymmetry group $\,\txG\,$ determined by (the principle of least action for) an action functional constructed in the following fashion. Let $\,(\theta_i^\a,X_i^\mu)\,$ be local coordinates on $\,\cO^\txH_i$,\ centred on the reference point $\,\unl g_i\,\txH\in\cO^\txH_i\,$ (with $\,(\theta_i^\a,X_i^\mu)(\unl g_i\,\txH)=(0,0)$), and consider the corresponding local sections of the principal $\txH$-bundle $\,\txG\too\txG/\txH\,$ of the form
\qq\nn
\si^\txH_i\ :\ \cO^\txH_i\too\txG\ :\ Z_i\equiv\bigl(\theta^\a_i,X^\mu_i\bigr)\longmapsto\unl g_i\cdot g_i(X_i)\cdot\ee^{\Theta_i(Z_i)}\,,\qquad i\in I^\txH\,,
\qqq
with
\qq\nn
g_i(X_i)=\ee^{X_i^\mu\,P_\mu}\in\vert\txG\vert
\qqq
and
\qq\nn
\Theta_i^\a(Z_i)=\theta_i^\b\,f_{i\,\b}^{\ \ \a}(X_i)
\qqq
in general depending upon the Gra\ss mann-even coordinate (through some functions $\,f_{i\,\b}^{\ \ \a}$). Next, take an arbitrary tesselation $\,\triangle(\Om_p)\,$ of $\,\Om_p\,$ subordinate, for a given map $\,\xi$,\ to the open cover $\,\cO^\txH$,\ as reflected by the existence of a map $\,i_\cdot\ :\ \triangle(\Om_p)\too I^\txH\,$ with the property
\qq\nn
\forall_{\z\in\triangle(\Om_p)}\ :\ \xi(\z)\subset\cO^\txH_{i_\z}\,.
\qqq
Let $\,\Cgt\subset\triangle(\Om_p)\,$ be the set of $(p+1)$-cells of the tesselation,
\qq\nn
\Om_p=\bigcup_{\t\in\Cgt}\,\t\,.
\qqq
The Nambu--Goto action functional is now given by the sum
\qq\label{eq:NGGS}
S^{({\rm NG})}_{\rm GS}[\xi]=S^{({\rm NG})}_{\rm GS,metr}[\xi]+S^{({\rm NG})}_{\rm GS,top}[\xi]
\qqq
of the metric term
\qq
S^{({\rm NG})}_{\rm GS,metr}[\xi]&:=&\sum_{\t\in\Cgt}\,S^{(\t)}_{{\rm GS,metr}}[\xi_\t]\,,\qquad\qquad\xi_\t:=\xi\rstr_\t\cr\label{eq:SmetrNG}&&\\ \cr
S^{(\t)}_{\rm GS,metr}[\xi_\t]&=&-\tfrac{1}{2}\,\int_\t\,\Vol(\Om)\,\sqrt{\det_{(p+1)}\,\bigl(\txg_{\unl A\unl B}\,\bigl(\p_a\con(\si^\txH_{i_\t}\circ\xi_\t)^*\theta^{\unl A}_{\rm L}\bigr)\,\bigl(\p_b\con(\si^\txH_{i_\t}\circ\xi_\t)^*\theta^{\unl B}_{\rm L}\bigr)\bigr)}\,,\nonumber
\qqq
expressed in terms of the (local) coordinate vector fields $\,\p_a\equiv\frac{\p\ }{\p\si^a},\ a\in\ovl{0,p}\,$ on $\,\t\subset\Om_p$,\ and of the Wess--Zumino term which may be formally written as the integral
\qq\label{eq:StopNG}
S^{({\rm NG})}_{\rm WZ,top}[\xi]=\int_{\Om_p}\,\sfd^{-1}\xi^*\underset{\tx{\ciut{(p+2)}}}{\txH}
\qqq
of a primitive of (the pullback of) the GS super-$(p+2)$-cocycle. Generically, the latter is \emph{not} a globally smooth supersymmetric super-$(p+1)$-form, and so either we restrict to a class of embeddings $\,\xi\,$ with $\,\xi(\Om_p)\subset\cO^\txH_i\,$ for some $\,i\in I^\txH\,$ and supersymmetry transformations from a vicinity of the identity (resp.\ cut the worldvolume open, in which case we may sometimes define the action functional as above but lose the possibility to compare values taken by the functional on maps with cobordant images in $\,\txG/\txH$), or we write it out in terms of worldvolume (de Rham) currents associated with the tesselation $\,\triangle(\Om_p)\,$ (or a suitable refinement thereof), whereupon it sums up to (a local presentation of the logarithm of) the volume holonomy, along $\,\xi(\Om_p)$,\ of the geometrisation of the GS super-$(p+2)$-cocycle. The geometrisation allows for a rigorous definition of the topological WZ term of the super-$\si$-model and is the subject of the present report for $\,p=0\,$ and a both mathematically and physically motivated choice of the supertarget $\,\txG/\txH$.\ We shall give geometrisations for $\,p=0\,$ over two supertargets: $\,{\rm sMink}^{3,1\,\vert\,2\cdot 4}\,$ and $\,{\rm s}({\rm AdS}_2\x\bS^2)\,$ in Sec.\,\ref{sec:geometrise}. These supertargets are related by a blow-up transformartion dual to the \.In\"on\"u--Wigner contraction that sets the associated supersymmetry algebras in correspondence, and the geometrisations may be chosen compatible with the correspondence in the sense rendered precise, in all generality, in Sec.\,\ref{sec:scheme}. The first step in the geometrisation of the GS super-$(p+2)$-cocycle is the construction of a surjective submersion $\,\pi_{\sfY(\txG/\txH)}\ :\ \sfY(\txG/\txH)\too\txG/\txH\,$ alongside a lift of $\,[\ell^\txH]_\cdot\,$ to its total space $\,\sfY(\txG/\txH)\,$ such that there exists a globally smooth super-$(p+1)$-form $\,\underset{\tx{\ciut{(p+1)}}}{\txB}\,$ on $\,\sfY(\txG/\txH)\,$ invariant under the lifted supersymmetry and with the property
\qq\nn
\pi_{\sfY(\txG/\txH)}^*\underset{\tx{\ciut{(p+2)}}}{\txH}=\sfd\underset{\tx{\ciut{(p+1)}}}{\txB}\,.
\qqq
In what follows, we contemplate a more restrictive scenario in which it is the super-$(p+2)$-cocycle $\,\underset{\tx{\ciut{(p+2)}}}{\chi}$,\ defining a possibly non-trivial class in $\,{\rm CaE}^{p+2}(\txG)$,\ that can be trivialised, also in the Cartan--Eilenberg cohomology, through pullback along a surjective submersion 
\qq\label{eq:Gext}
\pi_{\widetilde\txG}\ :\ \widetilde\txG\too\txG
\qqq
to a Lie supergroup $\,\widetilde\txG$,\ to be termed {\bf the extended supersymmetry group} in the present context, integrating a Lie-superalgebra extension $\,\widetilde\ggt\,$ (with the Lie superbracket $\,[\cdot,\cdot\}^\sim$) of $\,\ggt$, 
\qq\label{eq:gext}
\brd0\too\zgt\too\widetilde\ggt\too\ggt\too\brd0
\qqq
by a Lie superalgebra 
\qq\nn
\zgt=\bigoplus_{i=1}^N\,\corr{Z_i}_\bC 
\qqq
embedded in $\,\widetilde\ggt\,$ as a normal subalgebra,
\qq\label{eq:znormsub}
[\widetilde\ggt,\zgt\}^\sim\subset\zgt\,,
\qqq
so that -- in particular -- the extension has, again, a reductive decomposition 
\qq\nn
\widetilde\ggt=\widetilde\tgt\oplus\hgt\,,\qquad\qquad[\hgt,\widetilde\tgt\}^\sim\subset\widetilde\tgt\,,
\qqq
with
\qq\nn
\widetilde\tgt\equiv\zgt\oplus\tgt\,.
\qqq
Thus, the extension is a principal bundle 
\qq\label{eq:Gextasbun}
\alxydim{@C=1cm@R=1cm}{\txZ \ar[r] & \widetilde\txG \ar[d]^{\pi_{\widetilde\txG}} \\ & \txG}
\qqq
with the structure group
\qq\nn
\txZ\equiv\exp\,\zgt\,.
\qqq
The extension is determined by $\,\underset{\tx{\ciut{(p+2)}}}{\chi}\,$ in the following manner. Denote the generators of $\,\widetilde\ggt\,$ as $\,\{\widetilde t_{\widetilde A}\}_{\widetilde A\in\ovl{1,\dim\,\widetilde\ggt}}\,$ and, among them, those that generate $\,\widetilde\tgt\,$ as $\,\{\widetilde t_{\unl{\widetilde A}}\}_{\unl{\widetilde A}\in\ovl{1,\dim\,\widetilde\tgt}}$,\ and, within this subset, those that generate $\,\tgt\,$ as $\,\{t_{\unl A}\}_{\unl A\in\ovl{1,\dim\,\tgt}}$,\ and write the structure equations of the extension as
\qq\nn
\bigl[\widetilde t_{\widetilde A},\widetilde t_{\widetilde B}\bigr\}^\sim=\widetilde f_{\widetilde A\widetilde B}^{\ \ \ \widetilde C}\,\widetilde t_{\widetilde C}\,.
\qqq
With every generator $\,Z_i,\ i\in\ovl{1,\dim\,\zgt}$,\ we may then associate a trivialisation, in the Cartan--Eilenberg cohomology of $\,\widetilde\txG$,\ of a super-2-cocycle
\qq\nn
\underset{\tx{\ciut{(2)}}}{\varpi}{}^i:=(-1)^{|\widetilde A|\cdot|\widetilde B|}\,\widetilde f_{\widetilde A\widetilde B}^{\ \ \ i}\,\widetilde\theta_{\rm L}^{\widetilde A}\wedge\widetilde\theta_{\rm L}^{\widetilde B}\,,
\qqq
written in terms of the components $\,\widetilde\theta_{\rm L}^{\widetilde A}\,$ of the $\widetilde\ggt$-valued left-invariant Maurer--Cartan super-1-form 
\qq\nn
\widetilde\theta_{\rm L}=\widetilde\theta_{\rm L}^{\widetilde A}\ox\widetilde t_{\widetilde A}
\qqq
of the respective Gra\ss mann parities $\,\vert\widetilde\theta_{\rm L}^{\widetilde A}\vert\equiv\vert\widetilde t_{\widetilde A}\vert=:|\widetilde A|$.\ We take the extension $\,\widetilde\ggt\,$ to be \emph{constructed}, whenever possible, according to a (systematic yet non-algorithmic) sequential scheme\footnote{In some concrete applications, the super-2-cocycles $\,\underset{\tx{\ciut{(2)}}}{\varpi}{}^j\,$ are obtained, {\it e.g.}, by contracting the super-$(p+2)$-cocycle with suitably chosen right-invariant vector fields.} illustrated amply in Refs.\,\cite{Chryssomalakos:2000xd,Suszek:2017xlw}, so that there exists, on $\,\widetilde\txG$,\ a left-invariant primitive $\,\underset{\tx{\ciut{(p+1)}}}{\widetilde\b}\,$ of the pullback of $\,\underset{\tx{\ciut{(p+2)}}}{\chi}\,$ along $\,\pi_{\widetilde\txG}$,
\qq\nn
\pi_{\widetilde\txG}^*\underset{\tx{\ciut{(p+2)}}}{\chi}=\sfd\underset{\tx{\ciut{(p+1)}}}{\widetilde\b}\,,
\qqq
with the structure of a linear combination, with $\txH$-invariant tensors $\,\mu_{\widetilde{\unl A}_1\widetilde{\unl A}_2\ldots\widetilde{\unl A}_{p+1}}\,$ as coefficients, of wedge products of components of $\,\widetilde\theta_{\rm L}\,$ along $\,\widetilde\tgt$, 
\qq\nn
\underset{\tx{\ciut{(p+1)}}}{\widetilde\b}=\mu_{\widetilde{\unl A}_1\widetilde{\unl A}_2\ldots\widetilde{\unl A}_{p+1}}\,\theta^{\widetilde{\unl A}_1}_{\rm L}\wedge\theta^{\widetilde{\unl A}_2}_{\rm L}\wedge\cdots\wedge\theta^{\widetilde{\unl A}_{p+1}}_{\rm L}\,,
\qqq
in which \emph{all} the components $\,\widetilde\theta_{\rm L}^i,\ i\in\ovl{1,\dim\,\zgt}\,$ are present. Thus, the extension defines (or, indeed, is defined by) a supersymmetric trivialisation of $\,\underset{\tx{\ciut{(p+2)}}}{\chi}\,$ that descends to the homogeneous space $\,\widetilde\txG/\txH$.\ The latter constitutes the base of the principal $\txH$-bundle 
\qq\nn
\alxydim{@C=1cm@R=1cm}{\txH \ar[r] & \widetilde\txG \ar[d]^{\pi_{\widetilde\txG/\txH}} \\ & \widetilde\txG/\txH}
\qqq
and we assume it to surjectively submerse over the original homogeneous space
\qq\nn
\widetilde\pi\ :\ \widetilde\txG/\txH\equiv\sfY(\txG/\txH)\too\txG/\txH\,,
\qqq
for which it is only natural to presuppose that the following diagram of principal bundles commute
\qq\label{diag:buncomm}
\alxydim{@C=2cm@R=2cm}{ & \txZ \ar[d] \ar@{=}[rr] & & \txZ \ar[d] \\ \txH \ar[r] \ar@{=}[d] & \widetilde\txG \ar[rr]^{\pi_{\widetilde\txG/\txH}} \ar[d]^{\pi_{\widetilde\txG}} & & \widetilde\txG/\txH \ar[d]^{[\pi_{\widetilde\txG}]} \\ \txH \ar[r] & \txG \ar[rr]_{\pi_{\txG/\txH}} & & \txG/\txH }\,.
\qqq
The original action $\,[\ell]_\cdot\,$ of the supersymmetry group on the homogeneous space $\,\txG/\txH\,$ lifts to a projective action thereof
\qq\nn
\widetilde{\unl{[\ell]}}_\cdot\ :\ \txG\x\sfY\bigl(\txG/\txH\bigr)\too\sfY\bigl(\txG/\txH\bigr)\ :\ \bigl(g'.\widetilde g\,\txH\bigr)\longmapsto\bigl(g'\cdot\widetilde g\bigr)\,\txH
\qqq
on $\,\widetilde\txG/\txH$,\ with the homomorphicity 2-cochain ({\it cp} \Reqref{eq:znormsub})
\qq\nn
d^{(0)}_{\cdot,\cdot}\in C^2(\txG,\txZ)
\qqq
entering through the identity
\qq\nn
\widetilde{\unl{[\ell]}}_{g_1}\circ\widetilde{\unl{[\ell]}}_{g_2}\circ\widetilde{\unl{[\ell]}}_{(g_1\cdot g_2)^{-1}}=\widetilde{\unl{[\ell]}}_{d^{(0)}_{g_1,g_2}}
\qqq
that reflects the deformation \eqref{eq:gext} of the supersymmetry algebra. We obtain a standard (left) action only upon replacing $\,\txG\,$ with $\,\widetilde\txG$,
\qq\nn
\widetilde{[\ell]}_\cdot\ :\ \widetilde\txG\x\sfY\bigl(\txG/\txH\bigr)\too\sfY\bigl(\txG/\txH\bigr)\ :\ \bigl(\widetilde g'.\widetilde g\,\txH\bigr)\longmapsto\bigl(\widetilde g'\cdot\widetilde g\bigr)\,\txH\,,
\qqq
and so we are led to identify $\,\widetilde\txG\,$ as the supersymmetry group of the surjective submersion $\,\sfY(\txG/\txH)$,\ whence also the name given to it earlier.

In general (that is, for $p$ arbitrary), the construction of the surjective submersion $\,\sfY(\txG/\txH)\,$ is merely the first step towards a full-blown geometrisation of the GS super-$(p+2)$-cocycle. The situation simplifies for $\,p=0\,$ as $\,\zgt\,$ is then one-dimensional and supercentral in $\,\widetilde\ggt\,$ in virtue of \cite[Prop.\,C.5]{Suszek:2017xlw}, and the extension \eqref{eq:Gext}, whenever integrable, carries the structure of a principal $\bC^\x$-bundle with a natural action of the extended supersymmetry group that descends to the principal $\bC^\x$-bundle $\,\sfY(\txG/\txH)\too\txG/\txH\,$ along the horizontal arrows in Diag.\,\eqref{diag:buncomm}. The supersymmetric primitive $\,\underset{\tx{\ciut{(1)}}}{\widetilde\b}\,$ is the principal $\bC^\x$-connection 1-form on \eqref{eq:Gextasbun} and descends to the quotient $\,\widetilde\txG/\txH\,$ by construction. This mechanism for $\,p=0\,$ shall be illustrated on nontrivial examples in Secs.\,\ref{sec:ints0bext} and \ref{sec:geometrise}. Rather than contemplating the next steps of the geometrisation for $\,p>0$,\ necessarily in abstraction from concrete examples, we leave the construction at this preliminary stage and focus on its other physically significant aspects, accessible in the Hughes--Polchinski formulation.

Prior to passing to the latter, however, we pause to make a comment on the group extension \eqref{eq:Gext} used in the construction. {\it A priori}, we might restrict our discussion directly to the homogeneous space $\,\txG/\txH$,\ and -- indeed -- the case study to follow provides an example of a successful reconstruction of a (contractible) geometrisation of a GS super-2-cocycle without any reference to the extended Lie supergroup $\,\widetilde\txG$.\ The advantage of working with the Lie-supergroup extension becomes apparent only in the field-theoretic context when the issue of an accompanying (and coherent) geometrisation of the linearised gauged supersymmetry of the associated super-$\si$-model discovered in Refs.\,\cite{deAzcarraga:1982njd,Siegel:1983hh}, aka $\k$-symmetry, is raised in the setting particularly suited to its straightforward analysis, to wit, in the Hughes--Polchinski formulation of the super-$\si$-model. Access to the Cartan calculus on $\,\widetilde\txG\,$ will be demonstrated to facilitate the study of the (weak-)$\k$-equivariance of the geometrisation of an extension of the GS super-$(p+2)$-cocycle $\,\underset{\tx{\ciut{(p+2)}}}{\txH}$.

Thus, we are confronted with the question of integrability of the short exact sequence \eqref{eq:gext}. In concrete physical considerations, it is determined by the super-$(p+2)$-cocycle $\,\underset{\tx{\ciut{(p+2)}}}{\chi}$,\ and in general the said question has to be answered through direct computation. Whenever, though, the path from the original Lie superalgebra $\,\ggt\,$ to its extension $\,\widetilde\ggt\,$ leads through a sequence of super-central extensions, as was the case, {\it e.g.}, in \Rcite{Suszek:2017xlw}, the answer can be given, at each step, by an adaptation to the supergeometric setting of interest of the classical result 
\bethe\cite[Thm.\,5.4]{Tuynman:1987ij}\label{thm:TWextgrp} 
Let $\,\txG\,$ be a Lie group with Lie algebra $\,\ggt\,$ and suppose $\,\om\,$ is a Lie-algebra 2-cocycle with values in $\,\bR$,\ {\it i.e.}, $\,[\om]\in H^2(\ggt,\bR)\cong{\rm CaE}^2(\txG)$.\ Then, there exists, for $\,D\,$ a discrete subgroup of $\,\bR$,\ a Lie-group central extension $\,\widetilde\txG\,$ of $\,\txG\,$ by $\,\bR/D\,$ associated to the Lie-algebra extension $\,\widetilde\ggt\,$ of $\,\ggt\,$ by $\,\bR\,$ (defined by $\,\om\,$ through Prop.\,3.10 {\it ibidem}) iff the following two conditions are satisfied:
\bit
\item[(i)] the abelian group $\,{\rm Per}(\om)\subset\bR\,$ of the periods of $\,\om\,$ is contained in $\,D$;
\item[(ii)] there exists a moment map for the left regular action of $\,\txG\,$ on $\,(\txG,\om)$,\ {\it i.e.}, an $\bR$-linear map 
\qq\nn
\mu_\cdot\ :\ \ggt\too C^\infty(\txG,\bR)\ :\ X\longmapsto\mu_X
\qqq
which satisfies the relation
\qq\nn
\cR_X\con\om=-\sfd\mu_X
\qqq
for the right-invariant vector field $\,\cR_X\in\G(\sfT\txG)\,$ associated with an arbitrary $\,X\in\ggt$.
\eit
\ethe

We may finally take a closer look at the Hughes--Polchinski model. The model is a theory of smooth embeddings 
\qq\nn
\widetilde\xi\ :\ \Om_p\too\txG/\txH_{\rm vac}
\qqq
whose definition calls for a choice of the $(p+1)$-dimensional subspace $\,\tgt^{(0)}_{\rm vac}\subset\tgt\,$ with the following property, stated in terms of a tesselation $\,\widetilde\triangle(\Om_p)\,$ subordinate, for a given embedding $\,\widetilde\xi$,\ to a trivialising open cover $\,\cO^{\txH_{\rm vac}}\,$ of $\,\txG/\txH_{\rm vac}\,$ (with the set $\,\widetilde\Cgt\,$ of $(p+1)$-cells): the subspace $\,\sfT_e\ell_{\si^{\txH_{\rm vac}}_{i_{\widetilde\t}}\circ\widetilde\xi_*(\si)}\tgt^{(0)},\ \widetilde\t\ni\si$,\ spanned on the left-invariant vector fields $\,\ceL_{P_{\unl a}},\ \unl a\in\ovl{0,p}$,\ coincides with the projection of the tangent space $\,\sfT_{\si^{\txH_{\rm vac}}_{i_{\widetilde\t}}\circ\widetilde\xi_*(\si)}(\si_{i_{\widetilde\t}}^{\txH_{\rm vac}}\circ\widetilde\xi_*(\Om_p))\,$ of the embedded worldvolume to the tangent space of the body $\,|\txG|\,$ in the so-called static gauge $\,\widetilde\xi=\widetilde\xi_*$.\ This choice is accompanied by the identification of the Lie subalgebra $\,\hgt_{\rm vac}\subset\hgt\,$ that preserves $\,\tgt^{(0)}_{\rm vac}\,$ under the adjoint action. Given these, we write out the action functional in terms of the distinguished local sections
\qq\nn
\si^{\txH_{\rm vac}}_i\ :\ \cO^{\txH_{\rm vac}}_i\too\txG\ :\ \widehat Z_i\equiv\bigl(\theta^\a_i,X^\mu_i,\phi^{\widehat S}_i\bigr)\longmapsto\unl{\widetilde g_i}\cdot g_i(X_i)\cdot\ee^{\Theta_i(\theta_i,X_i)}\cdot\ee^{\phi_i^{\widehat S}\,J_{\widehat S}}\,,\qquad i\in I^{\txH_{\rm vac}}\,,
\qqq
expressed in terms of local coordinates $\,(\theta^\a_i,X^\mu_i,\phi^{\widehat S}_i)\,$ on $\,\cO^{\txH_{\rm vac}}_i\,$ centred on the reference point $\,\unl{\widetilde g_i}\,\txH_{\rm vac}\in\cO^{\txH_{\rm vac}}_i$,\ as the sum
\qq\label{eq:HPGS}
S^{({\rm HP})}_{\rm GS}\bigl[\widetilde\xi\bigr]=S^{({\rm HP})}_{\rm GS,metr}\bigl[\widetilde\xi\bigr]+S^{({\rm HP})}_{\rm GS,top}\bigl[\widetilde\xi\bigr]
\qqq
of the topological WZ term 
\qq\nn
S^{({\rm HP})}_{\rm WZ,top}\bigl[\widetilde\xi\bigr]=\int_{\Om_p}\,\sfd^{-1}\widetilde\xi^*\underset{\tx{\ciut{(p+2)}}}{\widetilde\txH}\,,\qquad\qquad\underset{\tx{\ciut{(p+2)}}}{\widetilde\txH}\rstr_{\cO^{\txH_{\rm vac}}_i}\equiv\si^{\txH_{\rm vac}}_i{}^*\underset{\tx{\ciut{(p+2)}}}{\chi}\,,
\qqq
to be understood as in the NG model, and of the complementary `metric' term
\qq\nn
S^{({\rm HP})}_{\rm GS,metr}[\widetilde\xi\bigr]&:=&\sum_{\widetilde\t\in\widetilde\Cgt}\,S^{(\widetilde\t)}_{{\rm GS,metr}}[\widetilde\xi_{\widetilde\t}\bigr]\,,\qquad\qquad\xi_{\widetilde\t}:=\xi\rstr_{\widetilde\t}\cr\cr\cr
S^{(\widetilde\t)}_{{\rm GS,metr}}\bigl[\widetilde\xi\bigr]&=&\int_{\widetilde\t}\,\bigl(\si^{\txH_{\rm vac}}_{i_{\widetilde\t}}\circ\widetilde\xi_{\widetilde\t}\bigr)^*\underset{\tx{\ciut{(p+1)}}}{\b}\hspace{-7pt}{}^{\rm (HP)}\,,
\qqq
with
\qq\label{eq:HPcurv}
\underset{\tx{\ciut{(p+1)}}}{\b}\hspace{-7pt}{}^{\rm (HP)}=\tfrac{1}{(p+1)!}\,\ep_{\unl a_0\unl a_1\ldots\unl a_p}\,\theta^{\unl a_0}_{\rm L}\wedge\theta^{\unl a_1}_{\rm L}\wedge\cdots\wedge\theta^{\unl a_p}_{\rm L}
\qqq
written in terms of the standard totally antisymmetric symbol
\qq\nn
\ep_{\unl a_0\unl a_1\ldots\unl a_p}=\left\{\barr{cl} \sign\left(\barr{cccc}0 & 1 &\ldots& p \\ \unl a_0 & \unl a_1 & \ldots & \unl a_p\earr\right) & \tx{ if } \{\unl a_0,\unl a_1,\ldots,\unl a_p\}=\ovl{0,p} \\ \\
0 & \tx{ otherwise}\earr\right.\,.
\qqq

The components $\,\phi^{\widehat S}\,$ of the lagrangean field $\,\widetilde\xi\,$ along $\,\dgt\,$ are non-dynamical -- in fact, they are the Goldstone bosons associated with the symmetry breakdown 
\qq\nn
\hgt\searrow\hgt_{\rm vac}
\qqq
effected by the `vacuum' of the theory, that is the classical embedding $\,\widetilde\xi\,$ of the super-$p$-brane worldvolume minimising the Hughes--Polchinski action functional. Under certain circumstances, they can be integrated out through the {\bf inverse Higgs mechanism} of \Rcite{Ivanov:1975zq}, whereupon a correspondence between the two field-theoretic models is established.
\bethe\cite[Prop.\,3.2]{Suszek:2017xlw}\label{prop:IHCartMink}
Adopt the notation defined between \Reqref{eq:HvacH} and \Reqref{eq:Hvacunimod}. If the following conditions are satisfied:
\bit
\item[(E1)] there exist non-degenerate bilinear symmetric forms: $\,\unl\g\,$ on $\,\tgt^{(0)}_{\rm vac}\,$ and $\,\widehat\g\,$ on $\,\egt^{(0)}\,$ with respective presentations 
\qq\nn
\unl\g=\unl\g_{\unl a\unl b}\,\tau^{\unl a}\ox\tau^{\unl b}\,,\qquad\qquad\widehat\g=\widehat\g_{\widehat a\widehat b}\,\tau^{\widehat a}\ox\tau^{\widehat b}
\qqq
in the basis $\,\{\tau^A\}^{A\in\ovl{0,D}}\,$ of $\,\ggt\,$ dual to $\,\{t_A\}_{A\in\ovl{0,D}}$,\
\qq\nn
\tau^A(t_B)=\d^A_{\ B}\,,\quad A,B\in\ovl{0,D}\,,
\qqq
for which the following identities hold true
\qq\label{eq:f2fIH}
\unl\g^{-1\,\unl c\unl a}\,f_{\widehat S\unl a}^{\ \ \widehat b}\,\widehat\g_{\widehat b\widehat d}=-f_{\widehat S\widehat d}^{\ \ \unl c}\,;
\qqq
\item[(E2)] $\,S^{({\rm HP})}_{{\rm GS},p}\,$ is restricted to field configurations satisfying the {\bf inverse Higgs constraint}
\qq\label{eq:IsHiggs}
\forall_{\widehat a\in\ovl{p+1,d}}\ :\ \bigl(\si^{\txH_{\rm vac}}_{i_\t}\circ\widetilde\xi\bigr)^*\theta^{\widehat a}_{\rm L}\must0
\qqq
whose solvability is ensured by the invertibility -- in an arbitrary (local) coordinate system $\,\{\si^a\}^{a\in\ovl{0,p}}\,$ on $\,\Om_p\,$ -- of the (tangent-transport) operator 
\qq\nn
E^{\unl a}_{\ A}\bigl(\si^{\txH_{\rm vac}}_{i_\t}\circ\widetilde\xi(\si)\bigr)\,\tfrac{\p\widetilde\xi^A}{\p\si^b}(\si)\equiv\unl\ep^{\unl a}_{\ b}(\si)\,,\quad\si\in\Om_p\,,
\qqq
\eit
the Hughes--Polchinski model on the homogeneous space $\,\txG/\txH_{\rm vac}$,\ determined by the action functional $\,S^{({\rm HP})}_{\rm GS}\,$ of \Reqref{eq:HPGS}, is equivalent to the Nambu--Goto super-$\si$-model on the homogeneous space $\,\txG/\txH$,\ defined by the action functional $\,S^{({\rm NG})}_{\rm GS}\,$ of \Reqref{eq:NGGS} with the metric term \eqref{eq:SmetrNG} for the metric $\,\txg=\unl\g\oplus\widehat\g\,$ and the same topological term \eqref{eq:StopNG}.

The inverse Higgs constraint is equivalent to the Euler--Lagrange equations of $\,S^{({\rm HP})}_{\rm GS}\,$ obtained by varying the functional in the direction of the Goldstone fields $\,\phi^{\widehat S},\ \widehat S\in\ovl{D-\unl\d,D-\d}$. 
\ethe

Whenever possible, the Hughes--Polchinski reformulation of the super-$\si$-model renders the field theory effectively topological upon encrypting the metric degrees of freedom in the Goldstone modes and extending the formerly postulated geometrisation \emph{trivially} to the whole lagrangean density through adjunction 
\qq\nn
\underset{\tx{\ciut{(p+2)}}}{\chi}\longmapsto\underset{\tx{\ciut{(p+2)}}}{\chi}+\sfd\underset{\tx{\ciut{(p+1)}}}{\b}\hspace{-7pt}{}^{\rm (HP)}
\qqq
of the trivial Cartan--Eilenberg super-$(p+2)$-cocycle $\,\sfd\underset{\tx{\ciut{(p+1)}}}{\b}\hspace{-7pt}{}^{\rm (HP)}$.\ An important advantage of working with the original field theory restructured thus is the simplification of the description of the gauged supersymmetry of the dynamics captured by both models, known under the name of the $\k$-symmetry. This is just invariance of the action functional under \emph{linearised} (functional) right translations on the Lie supergroup in the direction of a subspace 
\qq\nn
\tgt^{(1)}_{\rm vac}\subset\tgt^{(1)}\,,
\qqq
identified as the image
\qq\nn
\tgt^{(1)}_{\rm vac}\equiv\im\,\txP^{\tgt^{(1)}}_{\ \tgt^{(1)}_{\rm vac}}
\qqq
of a projector $\,\txP^{\tgt^{(1)}}_{\ \tgt^{(1)}_{\rm vac}}$,
\qq\nn
\txP^{\tgt^{(1)}}_{\ \tgt^{(1)}_{\rm vac}}\circ\txP^{\tgt^{(1)}}_{\ \tgt^{(1)}_{\rm vac}}=\txP^{\tgt^{(1)}}_{\ \tgt^{(1)}_{\rm vac}}\,,
\qqq
with the proprety 
\qq\nn
\dim\,\tgt^{(1)}_{\rm vac}=\tfrac{1}{2}\,\dim\,\tgt^{(1)}\,,
\qqq
reflected by the corresponding property of the projector
\qq\nn
\tr\,\txP^{\tgt^{(1)}}_{\ \tgt^{(1)}_{\rm vac}}=\tfrac{1}{2}\,\dim\,\tgt^{(1)}\,.
\qqq
It is straightforward to identify the symmetry by taking into account the variations, under tangential shifts $\,\k\in C^\infty(\Om_p,\tgt^{(1)}_{\rm vac})\,$ (with the postulated $\,\txP^{\tgt^{(1)}}_{\ \tgt^{(1)}_{\rm vac}}$,\ and so also $\,\tgt^{(1)}_{\rm vac}\,$ itself to be established in the process), of the various components $\,\theta_{\rm L}^{\unl A}\,$ of the Maurer--Cartan super-1-form on $\,\txG$.\ These read\footnote{The variations are to be understood as differential expressions valid over $\,\Om_p\x\txG$,\ with the obvious pullbacks along the canonical projections suppressed for the sake of transparency.} 
\qq\nn
\d_\k\theta_{\rm L}^{\unl A}\equiv\pLie{\kappa^\a\,\ceL_{Q_\a}}\,\theta_{\rm L}^{\unl A}=\d^{\unl A}_{\ \a}\,\sfd\k^\a-\tfrac{(-1)^{|B|\cdot|C|}}{2}\,f_{BC}^{\ \ \ \unl A}\,\k^\a\,\ceL_{Q_\a}\con\bigl(\theta_{\rm L}^B\wedge\theta_{\rm L}^C\bigr)=\d^{\unl A}_{\ \a}\,\sfd\k^\a+f_{B\a}^{\ \ \unl A}\,\k^\a\,\theta_{\rm L}^B
\qqq
and yield, formally,
\qq\nn
\d_\k\bigl(\underset{\tx{\ciut{(p+1)}}}{\b}\hspace{-7pt}{}^{\rm (HP)}+\sfd^{-1}\underset{\tx{\ciut{(p+2)}}}{\chi}\bigr)&=&\k^\a\,\bigl(\tfrac{1}{p!}\,f_{\a\b}^{\ \ \ \unl a}\,\ep_{\unl a\unl a_1\ldots\unl a_p}\,\theta^\b_{\rm L}\wedge\theta^{\unl a_1}_{\rm L}\wedge\cdots\wedge\theta^{\unl a_p}_{\rm L}+\ceL_{Q_\a}\con\underset{\tx{\ciut{(p+2)}}}{\chi}\bigr)+\sfd\bigl(\k^\a\,\ceL_{Q_\a}\con\sfd^{-1}\underset{\tx{\ciut{(p+2)}}}{\chi}\bigr)\cr\cr
&=&\tfrac{\k^\a}{p!}\,\bigl(f_{\a\b}^{\ \ \ \unl a}\,\ep_{\unl a\unl a_1\ldots\unl a_p}\,\theta^\b_{\rm L}\wedge\theta^{\unl a_1}_{\rm L}\wedge\cdots\wedge\theta^{\unl a_p}_{\rm L}+\tfrac{1}{p+1}\,\chi_{\a\unl A_1\unl A_2\ldots\unl A_{p+1}}\,\theta_{\rm L}^{\unl A_1}\wedge\theta_{\rm L}^{\unl A_2}\wedge\cdots\wedge\theta_{\rm L}^{\unl A_{p+1}}\bigr)\cr\cr
&&+\sfd\bigl(\k^\a\,\ceL_{Q_\a}\con\sfd^{-1}\underset{\tx{\ciut{(p+2)}}}{\chi}\bigr)\,,
\qqq
whence a natural {\bf $\k$-symmetry constraint}, to be imposed upon the geometric and algebraic data of the model (for some $\,\txP^{\tgt^{(1)}}_{\ \tgt^{(1)}_{\rm vac}}\,$ as above): there must exist constants $\,\la_{\b\,\unl A_1\unl A_2\ldots\unl A_{p+1}}\equiv\la_{\b\,[\unl A_1\unl A_2\ldots\unl A_{p+1}\}}\in\bC\,$ such that the following equality holds (on restriction to the image of $\,\txG/\txH_{\rm vac}\,$ within $\,\txG$)
\qq\nn
f_{\a\b}^{\ \ \ \unl a}\,\ep_{\unl a\unl a_1\ldots\unl a_p}\,\d^\b_{[\unl A_1}\,\d^{\unl a_1}_{\unl A_2}\cdots\d^{\unl a_p}_{\unl A_{p+1}\}}+\tfrac{1}{p+1}\,\chi_{\a\unl A_1\unl A_2\ldots\unl A_{p+1}}\must\bigl(\id_{\tgt^{(1)}}-\txP^{\tgt^{(1)}}_{\ \tgt^{(1)}_{\rm vac}}\bigr)^\b_{\ \a}\,\la_{\b\,\unl A_1\unl A_2\ldots\unl A_{p+1}}\,.
\qqq

Having identified the infinitesimal gauge symmetry of the field theory of interest, we may, finally, enquire as to the existence of a weak equivariant structure on the super-$p$-gerbe with respect to the distinguished linearised supertranslations from $\,\tgt^{(1)}_{\rm vac}$.\ By definition, this is equivalent to the existence of appropriate automorphisms of the supergeometric structures composing the geometrisation which preserve the associated connective structure, {\it cp} \cite[Sec.\,6]{Suszek:2017xlw}. In the setting in hand, that is for $\,p=0$,\ this boils down to the preservation of the principal connection 
\qq\nn
\underset{\tx{\ciut{(p+1)}}}{\widetilde\b}+\pi_{\widetilde\txG}^*\underset{\tx{\ciut{(p+1)}}}{\b^{\rm (HP)}}
\qqq 
on (the pullback, to the extended supersymmetry group, of) the extended super-0-gerbe along (infinitesimal) flows of left-invariant vector fields on $\,\widetilde\txG\,$ in the directions of $\,\tgt^{(1)}_{\rm vac}$.\ The weak $\k$-equivariance of each of the geometrisations given in Sec.\,\ref{sec:geometrise} shall be verified in Sec.\,\ref{sec:kappa}.

\section{Extension compatible with contraction -- the scheme}\label{sec:scheme}

Let $\,\txG_\a,\ \a\in\{1,2\}\,$ be a pair of (supersymmetry) Lie supergroups with the respective Lie superalgebras 
\qq\nn
\ggt_\a=\ggt_\a^{(0)}\oplus\ggt_\a^{(1)}
\qqq
(on which we have the respective Lie superbrackets $\,[\cdot,\cdot\}_\a$) with the Gra\ss mann-even subalgebra $\,\ggt_\a^{(0)}\,$ and its Gra\ss mann-odd $\ad$-module $\,\ggt_\a^{(1)}$,\ and with a fixed reductive direct-sum decomposition 
\qq\nn
\ggt_\a=\tgt_\a\oplus\hgt_\a
\qqq
into a Lie subalgebra $\,\hgt_\a\subset\ggt_\a^{(0)}\,$ with the corresponding Lie subgroup 
\qq\nn
\txH_\a\subset\txG_\a\,,
\qqq
assumed closed, and its $\ad$-module $\,\tgt_\a$.\ In what follows, we further assume $\,\tgt_1\,$ to be a Lie sub-superalgebra,
\qq\nn
[\tgt_1,\tgt_1\}_1\subset\tgt_1\,,
\qqq
associated with a Lie sub-supergroup 
\qq\nn
\txT_1\subset\txG_1\,,\qquad\qquad\txT_1\cong\txG_1/\txH_1\,.
\qqq
Let the Lie superalgebras $\,\ggt_2\,$ and $\,\tgt_1\,$ be related by an \.In\"on\"u--Wigner contraction ($\vec R\in\bC^{\x n},\ n\in\bN\,$ is a vector of scaling parameters)
\qq\label{eq:IWcontrsalg}
\ggt_2\xrightarrow[{\rm rescaling}]{\ \vec R-{\rm dependent}\ }\ggt_{2,\vec R}\xrightarrow{\ \vec R\to\vec R_*\ }\tgt_1\,,
\qqq
with
\qq\nn
\hgt_{2,\vec R}\equiv\hgt_2
\qqq
decoupling in the limit, to which there corresponds a (super)geometric transition -- to be referred to by the name of a blow-up transformation in the remainder of the paper -- between the homogeneous spaces
\qq\label{eq:IWcontrsgrp}
\txG_2/\txH_2\xrightarrow[{\rm dual\ rescaling}]{\ \vec R-{\rm dependent}\ }\txG_{2,\vec R}/\txH_2\xrightarrow{\ \vec R\to\vec R_*\ }\txG_1/\txH_1\cong\txT_1\,,
\qqq
effected by a dual rescaling of the local coordinates on $\,\txG_2\,$ associated with $\,\tgt_2\,$ and passing to the limit 
\qq\label{eq:IWlim}
\vec R\to\vec R_*
\qqq
for some $\,\vec R_*\in\ovl\bC^{\x n}$.\ The contraction is to be understood as follows: In the limit \eqref{eq:IWlim}, the Lie superbracket $\,[\cdot,\cdot\}_2\,$ closes on the $\ad$-module $\,\tgt_2$,\ turning the latter into a Lie superalgebra isomorphic with $\,\tgt_1$.

Let, next, $\,\underset{\tx{\ciut{(p+2)}}}{\chi^\a}\in Z^{p+2}_{\rm dR}(\txG_\a)^{\txG_\a}\,$ be Cartan--Eilenberg super-$(p+2)$-cocycles each of which is a linear combination, with $\txH_\a$-invariant tensors as coefficients, of wedge products of components of the $\ggt_\a$-valued left-invariant Maurer--Cartan super-1-form $\,\theta_{\a\,{\rm L}}\,$ on $\,\txG_\a\,$ along $\,\tgt_\a$.\ By the reasoning invoked in the previous section, the super-$(p+2)$-cocycles are pullbacks, along the projections $\,\pi_\a\equiv\pi_{\txG_\a/\txH_\a}\,$ to the base $\,\txG_\a/\txH_\a\,$ of the respective principal $\txH_\a$-bundles
\qq\nn
\alxydim{@C=1cm@R=1cm}{\txH_\a \ar[r] &\txG_\a \ar[d]^{\pi_\a} \\ & \txG_\a/\txH_\a}\,,
\qqq
of some $\txG_\a$-invariant de Rham super-$(p+2)$-cocycles $\,\underset{\tx{\ciut{(p+2)}}}{\txH^\a}\in Z^{p+2}_{\rm dR}(\txG_\a/\txH_\a)^{\txG_\a}$,
\qq\nn
\underset{\tx{\ciut{(p+2)}}}{\chi^\a}=\pi_\a^*\underset{\tx{\ciut{(p+2)}}}{\txH^\a}\,,
\qqq
invariant under the natural action 
\qq\nn
[\ell^\a]_\cdot\ :\ \txG_\a\x\bigl(\txG_\a/\txH_\a\bigr)\too\txG_\a/\txH_\a\ :\ (g_\a',g_\a\,\txH_\a)\longmapsto(g_\a'\cdot g_\a)\,\txH_\a
\qqq
of the supersymmetry group $\,\txG_\a$.

Each of the two homogeneous spaces admits a patchwise smooth realisation, detailed previously, within the total space of the bundle for which we merely fix the notation below. Thus, we have a family of local sections  
\qq\nn
\si^\a_{i_\a}\ :\ \cO^\a_{i_\a}\too\txG_\a\,,\quad i_\a\in I_\a
\qqq
defined over an open (trivialising) cover $\,\{\cO^\a_{i_\a}\}_{i_\a\in I_\a}\,$ of $\,\txG_\a/\txH_\a\,$ and related by locally smooth (transition) maps
\qq\nn
h^\a_{i_\a j_\a}\ :\ \cO^\a_{i_\a j_\a}\too\txH_\a\subset\txG_\a
\qqq
as 
\qq\nn
\forall_{x_\a\in\cO^\a_{i_\a j_\a}}\ :\ \si^\a_{j_\a}(x_\a)=\si^\a_{i_\a}(x_\a)\cdot h^\a_{i_\a j_\a}(x_\a)
\qqq
and composing a space 
\qq\nn
\si^\a(\txG_\a/\txH_\a):=\bigsqcup_{i_\a\in I_\a}\,\si^\a_{i_\a}(\cO_{i_\a})
\qqq
on which we have the quasi-action
\qq\label{eq:cosetactal}
g_\a'\cdot\si^\a_{i_\a}(x_\a)=\si^\a_{j_\a}\bigl(\widetilde x_\a(x_\a;g_\a')\bigr)\cdot\unl h^\a_{i_\a j_\a}(x_\a;g_\a')^{-1}\,,
\qqq
defined for every $\,(x_\a,g_\a)\in\cO^\a_{i_\a}\x\txG_\a\,$ and an arbitrary index $\,j_\a\in I_\a\,$ with the property
\qq\nn
\widetilde x_\a(x_\a;g_\a'):=\pi_\a\bigl(g_\a'\cdot\si^\a_{i_\a}(x_\a)\bigr)\in\cO^\a_{j_\a}\,,
\qqq
and with a unique $\,\unl h^\a_{i_\a j_\a}(x_\a,g_\a')\in\txH_\a\,$ defined on some open neighbourhood of $\,(x_\a,g_\a')$.\ The latter satisfies the gluing condition
\qq\label{eq:cosetactglue}
\unl h^\a_{i_\a k_\a}(x_\a;g_\a')=\unl h^\a_{i_\a j_\a}(x_\a;g_\a')\cdot h^\a_{j_\a k_\a}\bigl(\widetilde x_\a(x_\a;g_\a')\bigr)\,,\qquad\widetilde x_\a(x_\a;g_\a')\in\cO^\a_{j_\a k_\a}\,.
\qqq

Suppose, now, that the super-$(p+2)$-cocycle $\,\underset{\tx{\ciut{(p+2)}}}{\txH^2}{}_{\vec R}\,$ that pulls back to the super-$(p+2)$-cocycle $\,\underset{\tx{\ciut{(p+2)}}}{\chi^2}{}_{\vec R}\,$ on $\,\txG_{2,\vec R}\,$ given by the same linear combination of left-invariant super-1-forms as the one defining $\,\underset{\tx{\ciut{(p+2)}}}{\chi^2}\,$ satisfies the relation
\qq\nn
\lim_{\vec R\to\vec R_*}\,\underset{\tx{\ciut{(p+2)}}}{\txH^2}{}_{\vec R}=\underset{\tx{\ciut{(p+2)}}}{\txH^1}\,.
\qqq 
Assume, moreover, that $\,\underset{\tx{\ciut{(p+2)}}}{\chi^1}\,$ defines a possibly non-trivial class in $\,{\rm CaE}^{p+2}(\txG_1)\,$ but can be trivialised, also in the Cartan--Eilenberg cohomology, through pullback along a surjective submersion 
\qq\label{eq:G1ext}
\widetilde\pi_1\ :\ \widetilde\txG_1\too\txG_1
\qqq
to an (extended-supersymmetry) Lie supergroup $\,\widetilde\txG_1\,$ integrating a Lie-superalgebra extension $\,\widetilde\ggt_1\,$ (with the Lie superbracket $\,[\cdot,\cdot\}_1^\sim$) of $\,\ggt_1$, 
\qq\label{eq:g1ext}
\brd0\too\zgt_1\too\widetilde\ggt_1\too\ggt_1\too\brd0
\qqq
by a Lie superalgebra 
\qq\nn
\zgt_1=\bigoplus_{i=1}^{N_1}\,\corr{Z^1_i}_\bC 
\qqq
embedded in $\,\widetilde\ggt_1\,$ as a normal subalgebra,
\qq\nn
[\widetilde\ggt_1,\zgt_1\}_1^\sim\subset\zgt_1\,,
\qqq
so that -- in particular -- the extension has, again, a reductive decomposition 
\qq\nn
\widetilde\ggt_1=\widetilde\tgt_1\oplus\hgt_1\,,\qquad\qquad[\hgt_1,\widetilde\tgt_1\}^\sim_1\subset\widetilde\tgt_1\,,
\qqq
with
\qq\nn
\widetilde\tgt_1\equiv\zgt_1\oplus\tgt_1\,,
\qqq
and restricts to an extension of the Lie superalgebra $\,\tgt_1$,
\qq\nn
\brd0\too\zgt_1\too\widetilde\tgt_1\too\tgt_1\too\brd0\,,
\qqq
integrating to a Lie-supergroup extension
\qq\nn
\bd1\too\exp\,\zgt_1\too\widetilde T_1\xrightarrow{\ \widetilde{\unl\pi}_1\ }T_1\too\bd1\,.
\qqq
Denote the generators of $\,\widetilde\ggt_1\,$ as $\,\{\widetilde t^1_{\widetilde A}\}_{\widetilde A\in\ovl{1,\dim\,\widetilde\ggt_1}}\,$ and, among them, those that generate $\,\widetilde\tgt_1\,$ as $\,\{\widetilde t^1_{\unl{\widetilde A}}\}_{\unl{\widetilde A}\in\ovl{1,\dim\,\widetilde\tgt_1}}$,\ and, within this subset, those that generate $\,\tgt_1\,$ as $\,\{t^1_{\unl A}\}_{\unl A\in\ovl{1,\dim\,\tgt_1}}$,\ and write the structure equations of the extension as
\qq\nn
\bigl[\widetilde t^1_{\widetilde A},\widetilde t^1_{\widetilde B}\bigr\}^\sim_1=\widetilde f_{\widetilde A\widetilde B}^{1\ \ \widetilde C}\,\widetilde t^1_{\widetilde C}\,.
\qqq
We assume the extension to be engendered by $\,\underset{\tx{\ciut{(p+2)}}}{\chi^1}\,$ through the trivialisation mechanism described in the previous section, that is we take $\,\widetilde\txG_1\,$ to support a left-invariant primitive $\,\underset{\tx{\ciut{(p+1)}}}{\widetilde\b^1}\,$ of the pullback of $\,\underset{\tx{\ciut{(p+2)}}}{\chi^1}\,$ along $\,\widetilde\pi_1$,
\qq\nn
\widetilde\pi_1^*\underset{\tx{\ciut{(p+2)}}}{\chi^1}=\sfd\underset{\tx{\ciut{(p+1)}}}{\widetilde\b^1}\,,
\qqq
with the structure of a linear combination, with $\txH_1$-invariant tensors $\,\mu^1_{\widetilde{\unl A}_1\widetilde{\unl A}_2\ldots\widetilde{\unl A}_{p+1}}\,$ as coefficients, of wedge products of components $\,\widetilde\theta_{1\,{\rm L}}^{\widetilde A}\,$ (of the respective Gra\ss mann parities $\,\vert\widetilde\theta_{1\,{\rm L}}^{\widetilde A}\vert\equiv\vert\widetilde t^1_{\widetilde A}\vert=:|\widetilde A|$) of the $\widetilde\ggt_1$-valued left-invariant Maurer--Cartan super-1-form 
\qq\nn
\widetilde\theta_{1\,{\rm L}}=\widetilde\theta_{1\,{\rm L}}^{\widetilde A}\ox\widetilde t^1_{\widetilde A}
\qqq 
along $\,\widetilde\tgt_1$, 
\qq\nn
\underset{\tx{\ciut{(p+1)}}}{\widetilde\b^1}=\mu^1_{\widetilde{\unl A}_1\widetilde{\unl A}_2\ldots\widetilde{\unl A}_{p+1}}\,\theta^{\widetilde{\unl A}_1}_{1\,{\rm L}}\wedge\theta^{\widetilde{\unl A}_2}_{1\,{\rm L}}\wedge\cdots\wedge\theta^{\widetilde{\unl A}_{p+1}}_{1\,{\rm L}}\,,
\qqq
in which the components $\,\widetilde\theta_{1\,{\rm L}}^i,\ i\in\ovl{1,\dim\,\zgt_1}\,$ associated with \emph{every} generator $\,Z^1_i,\ i\in\ovl{1,\dim\,\zgt_1}\,$ are present. The latter are used to trivialise super-2-cocycles
\qq\nn
\underset{\tx{\ciut{(2)}}}{\varpi}{}_1^i:=(-1)^{|\widetilde A|\cdot|\widetilde B|}\,\widetilde f_{\widetilde A\widetilde B}^{1\ \ i}\,\widetilde\theta_{1\,{\rm L}}^{\widetilde A}\wedge\widetilde\theta_{1\,{\rm L}}^{\widetilde B}
\qqq
along the way.

Treating the asymptotic relations \eqref{eq:IWcontrsalg} and \eqref{eq:IWcontrsgrp} as fundamental, we may, next, search for a supersymmetric trivialisation of the other super-$(p+2)$-cocycle $\,\underset{\tx{\ciut{(p+2)}}}{\chi^2}\,$ on (pullback to) a Lie-supergroup extension $\,\widetilde\txG_2\,$ of $\,\txG_2$, 
a surjective submersion 
\qq\label{eq:G2ext}
\widetilde\pi_2\ :\ \widetilde\txG_2\too\txG_2
\qqq
that integrates a Lie-superalgebra extension $\,\widetilde\ggt_2\,$ (with the Lie superbracket $\,[\cdot,\cdot\}_2^\sim$) of $\,\ggt_2$,
\qq\label{eq:g2ext}
\brd0\too\zgt_2\too\widetilde\ggt_2\too\ggt_2\too\brd0
\qqq
by a Lie superalgebra 
\qq\nn
\zgt_2=\bigoplus_{\iota=1}^{N_2}\,\corr{Z^2_\iota}_\bC 
\qqq
embedded in $\,\widetilde\ggt_2\,$ as a normal subalgebra,
\qq\nn
[\widetilde\ggt_2,\zgt_2\}_2^\sim\subset\zgt_2\,,
\qqq
and defining its reductive decomposition
\qq\nn
\widetilde\ggt_2=\widetilde\tgt_2\oplus\hgt_2\,,\qquad\qquad[\hgt_2,\widetilde\tgt_2\}^\sim_2\subset\widetilde\tgt_2
\qqq
with
\qq\nn
\widetilde\tgt_2\equiv\zgt_2\oplus\tgt_2
\qqq
that contracts -- \`a la \.In\"on\"u \& Wigner, for an appropriate choice of the scaling (weights) of the new generators $\,Z^2_\iota,\ \iota\in\ovl{1,N_2}\,$ -- to the formerly obtained extension $\,\widetilde\tgt_1$,\ so that relation \eqref{eq:IWcontrsalg}, and with it also \eqref{eq:IWcontrsgrp}, lifts to the extension,
\qq\label{eq:IWcontrext}
\alxydim{@C=2.5cm@R=1cm}{\widetilde\ggt_2 \ar[r]^{{\rm rescaling}}_{\vec R-{\rm dependent}} \ar[d]_{\sfT_{\widetilde e_2}\widetilde\pi_2} & \widetilde\ggt_{2,\vec R} \ar[r]^{\vec R\to\vec R_*} \ar[r]^{\vec R\to\vec R_*} \ar[d] & \widetilde\tgt_1 \ar[d]^{\sfT_{\widetilde e_1}\widetilde{\unl\pi}_1} \\ \ggt_2 \ar[r]^{{\rm rescaling}}_{\vec R-{\rm dependent}} & \ggt_{2,\vec R} \ar[r]^{\vec R\to\vec R_*} & \tgt_1}\,.
\qqq
This entails distinguished linear combinations of the generators $\,\widetilde t^2_{\widetilde{\unl\a}},\ \widetilde{\unl\a}\in\ovl{1,\dim\,\widetilde\tgt_2}\,$ of $\,\widetilde\tgt_2\,$ asymptoting to those of $\,\widetilde\tgt_1$,\ $\,\widetilde t^1_{\widetilde{\unl A}},\ \widetilde{\unl A}\in\ovl{1,\dim\,\widetilde\tgt_1}$,\ which we shall write symbolically as
\qq\nn
\widehat t^2_{\widetilde{\unl A}}:=\La_{\widetilde{\unl A}}^{\ \widetilde{\unl\a}}\,\widetilde t^2_{\widetilde{\unl\a}}\xrightarrow{\ \vec R\to\vec R_*\ }\widetilde t^1_{\widetilde{\unl A}}\,,\qquad\La_{\widetilde{\unl A}}^{\ \widetilde{\unl\a}}\in\bC\,,\qquad\qquad \widetilde{\unl A}\in\ovl{1,\dim\,\widetilde\tgt_1}\,.
\qqq
The linear combinations $\,\widehat t^2_{\widetilde{\unl A}}\,$ surviving in the limit \eqref{eq:IWlim} are linearly independent, and so we may complete the set $\,\{\widehat t^2_{\widetilde{\unl A}}\}_{\widetilde{\unl A}\in\ovl{1,\dim\,\widetilde\tgt_1}}\,$ to a new basis of $\,\widetilde\tgt_2\,$ and associate with each of its elements a component $\,\widetilde\theta^{\widetilde{\unl A}}_{2\,{\rm L}}\,$ of the $\widetilde\ggt_2$-valued left-invariant Maurer--Cartan super-1-form $\,\widetilde\theta_{2\,{\rm L}}\,$ on $\,\widetilde\txG_2$,\ expressible as a (dual) linear combination 
\qq\nn
\widetilde\theta^{\widetilde{\unl A}}_{2\,{\rm L}}=V^{\widetilde{\unl A}}_{\ \widetilde{\unl\a}}\,\widetilde\theta^{\widetilde{\unl\a}}_{2\,{\rm L}}\,.
\qqq
Assuming that an extension $\,\widetilde\ggt_2\,$ thus constrained has been found \emph{and} that there exist collections $\,\{\mu^2_{\widetilde{\unl A}_1\widetilde{\unl A}_2\ldots\widetilde{\unl A}_{p+1}}\}_{\widetilde{\unl A_k}\in\ovl{1,\dim\,\widetilde\tgt_1},\ k\in\ovl{1,p+1}}\,$ of ($\vec R$-)scalable constants which form $\txH_2$-invariant tensors
\qq\nn
\widetilde\mu^2_{\widetilde{\unl\a}_1\widetilde{\unl\a}_2\ldots\widetilde{\unl\a}_{p+1}}:=\mu^2_{\widetilde{\unl A}_1\widetilde{\unl A}_2\ldots\widetilde{\unl A}_{p+1}}\,V^{\widetilde{\unl A}_1}_{\ \widetilde{\unl\a}_1}\,V^{\widetilde{\unl A}_2}_{\ \widetilde{\unl\a}_2}\cdots V^{\widetilde{\unl A}_{p+1}}_{\ \widetilde{\unl\a}_{p+1}}
\qqq
and go over to the $\,\mu^1_{\widetilde{\unl A}_1\widetilde{\unl A}_2\ldots\widetilde{\unl A}_{p+1}}\,$ in the limit\footnote{This may involve a restriction on the representations of $\,\txH_2\,$ surviving in the limit, {\it cp} \Rcite{Metsaev:1998it}.} \eqref{eq:IWlim},
\qq\nn
\mu^2_{\widetilde{\unl A}_1\widetilde{\unl A}_2\ldots\widetilde{\unl A}_{p+1}}\xrightarrow{\ \vec R\to\vec R_*\ }\mu^1_{\widetilde{\unl A}_1\widetilde{\unl A}_2\ldots\widetilde{\unl A}_{p+1}}\,,
\qqq
we define 
\qq\nn
\underset{\tx{\ciut{(p+1)}}}{\widetilde\b^2}=\mu^2_{\widetilde{\unl A}_1\widetilde{\unl A}_2\ldots\widetilde{\unl A}_{p+1}}\,\widetilde\theta^{\widetilde{\unl A}_1}_{2\,{\rm L}}\wedge\widetilde\theta^{\widetilde{\unl A}_2}_{2\,{\rm L}}\wedge\cdots\wedge\widetilde\theta^{\widetilde{\unl A}_{p+1}}_{2\,{\rm L}}\,.
\qqq 
At this stage, there are two possibilities. Either the exterior derivative of the super-$(p+1)$-form $\,\underset{\tx{\ciut{(p+1)}}}{\widetilde\b^2}\,$ engineered above \emph{does} reproduce the pullback of the GS super-$(p+2)$-cocycle $\,\underset{\tx{\ciut{(p+2)}}}{\chi^2}\,$ along $\,\widetilde\pi_2$,\ in which case we have attained the cohomological goal of trivialising the latter (in the CaE cohomology) in a manner compatible with the \.In\"on\"u--Wigner contraction, or it \emph{does not}, that is -- we obtain 
\qq\nn
\underset{\tx{\ciut{(p+2)}}}{\widetilde\chi^2}:=\sfd\underset{\tx{\ciut{(p+1)}}}{\widetilde\b^2}\neq\widetilde\pi_2^*\underset{\tx{\ciut{(p+2)}}}{\chi^2}
\qqq
but even then we are right to speak of having realised an important field-theoretic goal, conceived along the lines of and motivated by the original argument due to Metsaev and Tseytlin, {\it cp} \Rcite{Metsaev:1998it}, invoked again, among others, in \Rcite{Zhou:1999sm} and recently recalled in \Rcite{Suszek:2018bvx}, to wit: We have found \emph{a} super-$(p+2)$-cocycle $\,\underset{\tx{\ciut{(p+2)}}}{\widetilde\chi^2}\,$ representing a class in the CaE cohomology of \emph{an} extension of the body Lie group $\,\vert\txG_2\vert\,$ asymptoting to the reference super-$(p+2)$-cocycle $\,\underset{\tx{\ciut{(p+2)}}}{\chi^1}\,$ on the Lie supergroup $\,\txG_1$,\ with the thus understood contractibility of the geometric objects over $\,\vert\txG_2\vert\,$ inherited by the CaE primitive $\,\underset{\tx{\ciut{(p+1)}}}{\widetilde\b^2}\,$ of $\,\underset{\tx{\ciut{(p+2)}}}{\widetilde\chi^2}\,$ -- a {\it sine qua non} condition for the contractibility of a full-fledged geometrisation of $\,\underset{\tx{\ciut{(p+2)}}}{\widetilde\chi^2}\,$ based on $\,\underset{\tx{\ciut{(p+1)}}}{\widetilde\b^2}$,\ should some such exist. 

Thus equipped, we are ready to dive into a detailed study of a simple model of the general mechanism described above.

\section{Integrable super-0-brane extensions of supersymmetry algebras}\label{sec:ints0bext}

In this section, we perform a parallel analysis of cohomologically motivated wrapping-charge deformations of the Lie superalgebras $\,\gt{siso}(3,1\,\vert\,2\cdot 4)\,$ and $\,\gt{su}(1,1\,\vert\,2)_2\,$ associated with a pair of homogeneous spaces: $\,{\rm sMink}^{3,1\,\vert\,2\cdot 4}\,$ and $\,{\rm s}({\rm AdS}_2\x\bS^2)$.\ Along the way, we enforce associativity of the deformations as well as a correspondence between them that extends the \.In\"on\"u--Wigner contraction relating $\,\gt{su}(1,1\,\vert\,2)_2\,$ and $\,\gt{smink}^{3,1\,\vert\,2\cdot 4}\subset\gt{siso}(3,1\,\vert\,2\cdot 4)$.

\subsection{The $\,N=2\,$ super-Minkowskian algebra}\label{sec:sMinkext}

We begin by reconsidering the super-Minkowski space with $N$ supersymmetries (essentially in the notation of Refs.\,\cite{Suszek:2017xlw,Suszek:2018bvx})
\qq\nn
{\rm sMink}^{d,1\,\vert\,N\cdot D_{d,1}}\equiv{\rm sISO}(d,1\,\vert\,N\cdot D_{d,1})/{\rm SO}(d,1)\,,
\qqq
a homogeneous space of the super-Poincar\'e supergroup
\qq\nn
{\rm sISO}(d,1\,\vert\,N\cdot D_{d,1})\equiv\bR^{d,1\,\vert\,N\cdot D_{d,1}}\rx{\rm SO}(d,1)
\qqq
with the (complexified) Lie superalgebra
\qq\nn
\gt{siso}(d,1\,\vert\,N\cdot D_{d,1})=\bigg(\bigoplus_{I=1}^N\,\bigoplus_{\widehat\a=1}^{D_{d,1}}\,\corr{Q_{\widehat\a I}}_\bC\oplus\bigoplus_{\widehat a=0}^d\,\corr{P_{\widehat a}}_\bC\bigg)\oplus\bigoplus_{\widehat a,\widehat b=0}^d\,\corr{J_{\widehat a\widehat b}\equiv-J_{\widehat b\widehat a}}_\bC
\qqq
defined by the structure equations (here, $\,(\eta_{\widehat a\widehat b})\equiv\diag(-1,\underbrace{+1,+1,\ldots,+1}_{d\ {\rm times}})$)
\qq
&[P_{\widehat a},P_{\widehat b}]=0\,,\qquad\qquad[Q_{\widehat\a I},Q_{\widehat\b J}]=2\d_{IJ}\,\ovl\G^{\widehat a}_{\widehat\a\widehat\b}\,P_{\widehat a}\,,\qquad\qquad[P_{\widehat a},Q_{\widehat\a I}]=0\,,&\cr\cr
&[J_{\widehat a\widehat b},J_{\widehat c\widehat d}]=\eta_{\widehat a\widehat d}\,J_{\widehat b\widehat c}-\eta_{\widehat a\widehat c}\,J_{\widehat b\widehat d}+\eta_{\widehat b\widehat c}\,J_{\widehat a\widehat d}-\eta_{\widehat b\widehat d}\,J_{\widehat a\widehat c}\,,&\label{eq:sisosalg}\\ \cr
&[J_{\widehat a\widehat b},P_{\widehat c}]=\eta_{\widehat b\widehat c}\,P_{\widehat a}-\eta_{\widehat a\widehat c}\,P_{\widehat b}\,,\qquad\qquad[J_{\widehat a\widehat b},Q_{\widehat\a I}]=\tfrac{1}{2}\,\bigl(\G_{\widehat a\widehat b}\bigr)^{\widehat\b}_{\ \widehat\a}\,Q_{\widehat\b\,I}\,.&\nn
\qqq
The generators $\,\{\G^{\widehat a}\}^{\widehat a\in\ovl{0,d}}\,$ of the Clifford algebra $\,\Cliff\,(\bR^{d,1})\,$ are taken in a Majorana representation of dimension $\,D_{d,1}\,$ in which their products $\,\ovl\G^{\widehat a}\equiv\cC\,\G^{\widehat a}\,$ with the skew-symmetric charge-conjugation matrix 
\qq\nn
\cC^{\rm T}=-\cC
\qqq
are symmetric,
\qq\nn
\bigl(\cC\,\G^{\widehat a}\bigr)^{\rm T}=\cC\,\G^{\widehat a}\,.
\qqq
The Lie superbracket manifestly closes on the sub-superspace spanned on the $\,Q_{\widehat\a I}\,$ and the $\,P_{\widehat a}$,\ defining the (complexified) $N$-extended supertranslations Lie superalgebra
\qq\nn
\gt{smink}^{d,1\,\vert\,N\cdot D_{d,1}}:=\bigoplus_{I=1}^N\,\bigoplus_{\widehat\a=1}^{D_{d,1}}\,\corr{Q_{\widehat\a I}}_\bC\oplus\bigoplus_{\widehat a=0}^d\,\corr{P_{\widehat a}}_\bC\,.
\qqq
The homogeneous space is embedded in the supersymmetry group $\,{\rm sISO}(d,1\,\vert\,N\cdot D_{d,1})\,$ by a \emph{single} section of the principal ${\rm SO}(d,1)$-bundle 
\qq\label{eq:sMinkasbas}
\alxydim{@C=1cm@R=1cm}{{\rm SO}(d,1) \ar[r] & {\rm sISO}\bigl(d,1\,\vert\,N\cdot D_{d,1}\bigr) \ar[d]^{\pi_1\equiv\pi_{{\rm sISO}(d,1\,\vert\,N\cdot D_{d,1})/{\rm SO}(d,1)}} \\ & {\rm sISO}(d,1\,\vert\,N\cdot D_{d,1})/{\rm SO}(d,1)}\,,
\qqq
using the standard (global) coordinates $\,\{\theta^{\widehat\a I},x^{\widehat a}\}^{(\widehat\a,I)\in\ovl{1,D_{d,1}}\x\ovl{1,N},\ \widehat a\in\ovl{0,d}}\,$ on $\,{\rm sMink}^{d,1\,\vert\,N\cdot D_{d,1}}$,
\qq\nn
\si^1\ :\ {\rm sMink}^{d,1\,\vert\,N\cdot D_{d,1}}\too{\rm sISO}(d,1\,\vert\,N\cdot D_{d,1})\ :\ \bigl(\theta^{\widehat\a I},x^{\widehat a}\bigr)\longmapsto\ee^{x^{\widehat a}\,P_{\widehat a}}\cdot\ee^{\theta^{\widehat\a I}\,Q_{\widehat\a I}}\,.
\qqq
Here, the right-hand side is to be understood as the unital(-time) flow of the group unit first along the integral lines of the left-invariant vector field engendered by the Lie-superalgebra element $\,x^{\widehat a}\,P_{\widehat a}\,$ and subsequently along the one associated with $\,\theta^{\widehat\a I}\,Q_{\widehat\a I}$.\ The quotient is itself a Lie supergroup, namely the supertranslation group $\,\bR^{d,1\,\vert\,N\cdot D_{d,1}}$.\ Its binary operation takes the familiar form
\qq\nn
(\theta_1^{\widehat\a I},x_1^{\widehat a})\cdot(\theta_2^{\widehat\b J},x_2^{\widehat b})=\bigl(\theta_1^{\widehat\a I}+\theta_2^{\widehat\a I},x_1^{\widehat a}+x_2^{\widehat a}-\d_{IJ}\,\ovl\theta^I_1\,\G^{\widehat a}\,\theta^J_2\bigr)\,.
\qqq
We also have a natural (left) action of the supersymmetry group $\,{\rm sISO}(d,1\,\vert\,N\cdot D_{d,1})\,$ on the homogeneous space $\,{\rm sMink}^{d,1\,\vert\,N\cdot D_{d,1}}$,\ defined along the lines of \cite[Sec.\,4.1.]{Suszek:2017xlw} as
\qq
[\ell^1]_\cdot\ &:&\ {\rm sISO}(d,1\,\vert\,N\cdot D_{d,1})\x{\rm sMink}^{d,1\,\vert\,N\cdot D_{d,1}}\too{\rm sMink}^{d,1\,\vert\,N\cdot D_{d,1}}\cr\cr 
&:&\ \bigl(\bigl(\vep^{\widehat\a I},y^{\widehat a},h\bigr),\bigl(\theta^{\widehat\b J},x^{\widehat b}\bigr)\bigr)\longmapsto\bigl(S_I(h)^{\widehat\a}_{\ \widehat\b}\,\theta^{\widehat\b\,I}+\vep^{\widehat\a I},L(h)^{\widehat a}_{\ \widehat b}\,x^{\widehat b}+y^{\widehat a}-\d_{IJ}\,\ovl\vep^I\,\G^{\widehat a}\,S_J(h)\,\theta^J\bigr)\,, \label{eq:susyonsMink}
\qqq
where the isotropy group $\,{\rm SO}(d,1)\,$ acts linearly through its representations: the vectorial one $\,L(\cdot)\,$ and the spinorial ones $\,S_I(\cdot),\ I\in\ovl{1,N}$.\ The action is realised on the section $\,\si^1\,$ in a natural manner:
\qq\label{eq:susyonsisMink}
\bigl(\bigl(\vep^{\widehat\a I},y^{\widehat a},h\bigr),\si^1\bigl(\theta^{\widehat\b J},x^{\widehat b}\bigr)\bigr)\\ \cr
\longmapsto\bigl(\vep^{\widehat\a I},y^{\widehat a},h\bigr)\cdot\si\bigl(\theta^{\widehat\b J},x^{\widehat b}\bigr)\cdot(0,0,h)^{-1}\equiv\si^1\bigl(S_I(h)^{\widehat\a}_{\ \widehat\b}\,\theta^{\widehat\b\,I}+\vep^{\widehat\a I},L(h)^{\widehat a}_{\ \widehat b}\,x^{\widehat b}+y^{\widehat a}-\d_{IJ}\,\ovl\vep^I\,\G^{\widehat a}\,S_J(h)\,\theta^J\bigr)\,,\nn
\qqq
{\it cp} Eq.\,(2.4) of \Rcite{Suszek:2018bvx}. The triviality of the principal bundle \eqref{eq:sMinkasbas} in conjunction with the said Lie-supergroup structure on its base allow for a significant simplification of the ensuing supersymmetry-equivariant differential calculus -- indeed, we may define it independently of that on the total space of the bundle, only to find out that it can be obtained through a global restriction and pullback resp.\ pushforward from $\,{\rm sISO}(d,1\,\vert\,N\cdot D_{d,1})$.

Next, we specialise our discussion to the case $\,(d,N)=(3,2)\,$ and take a closer look at the relevant Green--Schwarz super-2-cocycle over the super-Minkowski space (left-)invariant under the action of the super-Poincar\'e supergroup $\,{\rm sISO}(3,1\,\vert\,2\cdot 4)$.\ When written in terms of the components of the $\gt{siso}(3,1\,\vert\,2\cdot 4)$-valued left-invariant Maurer--Cartan super-1-form $\,\theta_{1\,{\rm L}}\,$ on $\,{\rm sISO}(3,1\,\vert\,2\cdot 4)\,$ associated with the generators $\,Q_{\widehat\a I}\,$ and $\,P_{\widehat a}$,\ readily calculated, upon restriction to $\,\si^1({\rm sMink}^{3,1\,\vert\,2\cdot 4})$,\ from \Reqref{eq:sisosalg} with the help of the Schur--Poincar\'e formula 
\qq\nn
\si^1{}^*\theta_{1\,{\rm L}}(\theta,x)&=&\sum_{k=0}^\infty\,\tfrac{(-1)^k}{(k+1)!}\,\bigl(\sfd x^{\widehat a}\ox\sum_{l=0}^\infty\,\tfrac{(-1)^l}{l!}\,\ad^l_{\theta^{\widehat\a I}\,Q_{\widehat\a I}}\circ\ad^k_{x^{\widehat b}\,P_{\widehat b}}(P_{\widehat a})+\sfd\theta^{\widehat\a I}\ox\ad^k_{\theta^{\widehat\b J}\,Q_{\widehat\b J}}(Q_{\widehat\a I})\bigr)\cr\cr
&=&\sfd\theta^{\widehat\a I}\ox Q_{\widehat\a I}+\bigl(\sfd x^{\widehat a}+\d_{IJ}\,\ovl\theta^I\,\G^{\widehat a}\,\sfd\theta^J\bigr)\ox P_{\widehat a}
\qqq 
as
\qq\nn
\si^1{}^*\theta_{1\,{\rm L}}^{\widehat\a I}(\theta,x)=\sfd\theta^{\widehat\a I}\,,\qquad\qquad\si^1{}^*\theta_{1\,{\rm L}}^{\widehat a}(\theta,x)=\sfd x^{\widehat a}+\d_{IJ}\,\ovl\theta^I\,\G^{\widehat a}\,\sfd\theta^J\,,
\qqq
the super-2-cocycle takes the form 
\qq\label{eq:GSs2csMink}
\underset{\tx{\ciut{(2)}}}{\txH^1}(\theta,x)\equiv\si^1{}^*\underset{\tx{\ciut{(2)}}}{\chi}^1(\theta,x)=\si^1{}^*\bigl(\ep_{IJ}\,\theta_{1\,{\rm L}}^{\widehat\a I}\wedge\cC_{\widehat\a\widehat\b}\,\theta^{\widehat\b J}_{1\,{\rm L}}\bigr)(\theta,x)\equiv\ep_{IJ}\,\sfd\ovl\theta^I\wedge\sfd\theta^J\,,
\qqq
Here, $\,\ep_{IJ}=-\ep_{JI}\,$ is the standard Levi-Civita symbol, with $\,\ep_{12}=1$.\ The super-2-cocycle admits a primitive
\qq\label{eq:GSs2cprimsMink}
\underset{\tx{\ciut{(1)}}}{\txB^1}(\theta,x)=\ep_{IJ}\,\ovl{\theta}^I\,\sfd\theta^J\,.
\qqq
The latter is pseudo-invariant, in the sense made precise in \cite[Sec.\,3]{Suszek:2018bvx}, under supertranslations and we may compute the corresponding supersymmetry-variation super-0-forms $\,\underset{\tx{\ciut{(0)}}}{\G_X}$, defined {\it ibidem} through 
\qq\nn
\pLie{\cR_X}\underset{\tx{\ciut{(1)}}}{\txB^1}=:\sfd\underset{\tx{\ciut{(0)}}}{\G_X}\,,
\qqq
using the coordinate form of the relevant right-invariant vector fields\footnote{These are right-invariant vector fields on the Lie supergroup $\,\bR^{d,1\,\vert\,N\cdot D_{d,1}}$,\ obtained from their counterparts on $\,{\rm sISO}(d,1\,\vert\,N\cdot D_{d,1})\,$ through restriction to $\,\si^1({\rm sMink}^{d,1\,\vert\,N\cdot D_{d,1}})\,$ and pushforward along $\,\pi_1$.}
\qq\nn
\cR_{(\vep,0)}(\theta,x)=\vep^{\widehat\a I}\,\tfrac{\vec\p\ }{\p\theta^{\widehat\a I}}-\d_{IJ}\,\ovl\vep^I\,\G^{\widehat a}\,\theta^J\,\tfrac{\p\ }{\p x^{\widehat a}}\,,\qquad\qquad\cR_{(0,y)}(\theta,x)=y^{\widehat a}\,\tfrac{\p\ }{\p x^{\widehat a}}\,.
\qqq
Putting the pieces together, we obtain
\qq\nn
\pLie{\cR_{(\vep,0)}}\underset{\tx{\ciut{(1)}}}{\txB^1}(\theta,x)=\ep_{IJ}\,\ovl\vep^I\,\sfd\theta^J\,,\qquad\qquad\pLie{\cR_{(0,y)}}\underset{\tx{\ciut{(1)}}}{\txB^1}(\theta,x)=0\,,
\qqq
and so also
\qq\nn
\underset{\tx{\ciut{(0)}}}{\G_{(\vep,0)}}(\theta,x)=\ep_{IJ}\,\ovl\vep^I\,\theta^J\,,\qquad\qquad\underset{\tx{\ciut{(0)}}}{\G_{(0,y)}}(\theta,x)=0\,.
\qqq
These give us the Noether charges (written in terms of the Cauchy data $\,(\theta,x,p)\,$ of a field configuration, defining a state of the classical field theory, with the momentum component $\,p_{\widehat a}$) 
\qq\nn
h_X[\theta,x,p]=p_{\widehat a}\,\bigl(\cR_X\con\theta^{\widehat a}_{1\,{\rm L}}\bigr)\bigl(\si^1(\theta,x)\bigr)+\bigl(\cR_X\con\underset{\tx{\ciut{(1)}}}{\txB}-\underset{\tx{\ciut{(0)}}}{\G_X}\bigr)\bigl(\si^1(\theta,x)\bigr)
\qqq
of the supersymmetry of the associated 1-dimensional Green--Schwarz super-$\si$-model, defined in \cite[Eq.\,(2.15)]{Suszek:2018bvx},
\qq\nn
h_{(\vep,0)}[\theta,x,p]=-2\ovl\vep^I\,\bigl(\d_{IJ}\,p_{\widehat a}\,\G^{\widehat a}+\ep_{IJ}\bigr)\,\theta^J\,,\qquad\qquad h_{(0,y)}[\theta,x,p]=y^{\widehat a}\,p_{\widehat a}\,.
\qqq
Their Poisson bracket, determined by the (pre-)symplectic 2-form 
\qq\nn
\Om_{{\rm GS},0}[\theta,x,p]=\d\bigl(p_{\widehat a}\,\theta^{\widehat a}_{1\,{\rm L}}\bigl(\si^1(\theta,x)\bigr)\bigr)+\underset{\tx{\ciut{(2)}}}{\txH^1}(\theta,x)\,,
\qqq
given in \cite[Eq.\,(3.2)]{Suszek:2018bvx}, exhibits an anomaly,
\qq\nn
&\{h_{X_1},h_{X_2}\}_{\Om_{{\rm GS},0}}=h_{-[X_1,X_2]}+\xcW_{X_1,X_2}\,,&\cr\cr
&\xcW_{X_1,X_2}[\theta,x,p]=\bigl(\cR_{X_2}\con\cR_{X_1}\con\underset{\tx{\ciut{(2)}}}{\txH^1}+\cR_{[X_1,X_2]}\con\underset{\tx{\ciut{(1)}}}{\txB}-\underset{\tx{\ciut{(0)}}}{\G_{[X_1,X_2]}}\bigr)(\theta,x)&
\qqq
that reads
\qq\label{eq:wrapanosMink}\qquad\qquad
\xcW_{(\vep_1,0),(\vep_2,0)}[\theta,x,p]=2\ep_{IJ}\,\ovl\vep_1^I\,\vep_2^J\,,\qquad\qquad\xcW_{(0,y_1),(0,y_2)}[\theta,x,p]=0=\xcW_{(\vep,0),(0,y)}[\theta,x,p]\,.
\qqq
The ensuing deformation of the super-Minkowski superalgebra (obtained, {\it e.g.}, through canonical quantisation of the super-centrally extended Poisson--Lie algebra of Noether charges, after an obvious sign flip) is the Lie superalgebra
\qq\label{eq:s0bextsiso1}
\{Q_{\widehat\a I},Q_{\widehat\b J}\}^\sim=2\d_{IJ}\,\ovl\G^{\widehat a}_{\widehat\a\widehat\b}\,P_{\widehat a}+2\ep_{IJ}\,\cC_{\widehat\a\widehat\b}\,Z\,,\qquad\qquad[P_{\widehat a},P_{\widehat b}]^\sim=0=[Q_{\widehat\a I},P_{\widehat a}]^\sim\,,
\qqq
with $\,Z\,$ central,
\qq\label{eq:s0bextsiso2}
[Q_{\widehat\a I},Z]^\sim=0\,,\qquad\qquad[P_{\widehat a},Z]^\sim=0\,,\qquad\qquad[Z,Z]^\sim=0\,,
\qqq
The deformation is readily seen to extend to the entire supersymmetry algebra. This we achieve by taking the wrapping charge to be a Lorentz scalar, 
\qq\label{eq:s0bextsiso3}
[J_{\widehat a\widehat b},Z]^\sim=0\,,
\qqq
and transplanting the remaining structure relations in an unchanged form. We shall denote the resultant Lie superalgebra as
\qq\label{eq:s0bextsiso}\qquad\qquad
\widetilde{\gt{siso}(3,1\,\vert\,2\cdot 4)}=\bigl(\corr{Z}_\bC\oplus\bigl(\bigoplus_{I=1}^N\,\bigoplus_{\widehat\a=1}^{D_{d,1}}\,\corr{Q_{\widehat\a I}}_\bC\oplus\bigoplus_{\widehat a=0}^d\,\corr{P_{\widehat a}}_\bC\bigr)\bigr)\oplus\bigoplus_{\widehat a,\widehat b=0}^d\,\corr{J_{\widehat a\widehat b}\equiv-J_{\widehat b\widehat a}}_\bC\,.
\qqq
It fits into the short exact sequence of Lie superalgebras
\qq\label{eq:SESsiso}
\brd0\too\corr{Z}_\bC\cong\bC\too\widetilde{\gt{siso}(3,1\,\vert\,2\cdot 4)}\too\gt{siso}(3,1\,\vert\,2\cdot 4)\too\brd0\,.
\qqq
Within it, we find a Lie sub-superalgebra
\qq\nn
\widetilde{\gt{smink}}^{3,1\,\vert\,2\cdot 4}:=\corr{Z}_\bC\oplus\bigoplus_{I=1}^N\,\bigoplus_{\widehat\a=1}^{D_{d,1}}\,\corr{Q_{\widehat\a I}}_\bC\oplus\bigoplus_{\widehat a=0}^d\,\corr{P_{\widehat a}}_\bC\,,
\qqq
described by the corresponding short exact sequence of Lie superalgebras
\qq\nn
\brd0\too\bC\too\widetilde{\gt{smink}}^{3,1\,\vert\,2\cdot 4}\too\gt{smink}^{3,1\,\vert\,2\cdot 4}\too\brd0\,.
\qqq

Based on the above supercommutation relations, we define the corresponding Lie supergroup as the set
\qq\label{eq:sISOext}
\widetilde{{\rm sISO}(3,1\,\vert\,2\cdot 4)}={\rm sISO}(3,1\,\vert\,2\cdot 4)\x\bC^\x
\qqq
equipped with the binary operation
\qq\nn
&&\widetilde{{\rm sISO}(3,1\,\vert\,2\cdot 4)}^{\x 2}\too\widetilde{{\rm sISO}(3,1\,\vert\,2\cdot 4)}\cr\cr 
&:&\ \bigl(\bigl(\bigl(\theta^{\widehat\a I}_1,x^{\widehat a}_1,h_1\bigr),z_1\bigr),\bigl(\bigl(\theta^{\widehat\b J}_2,x^{\widehat b}_2,h_2\bigr),z_2\bigr)\bigr)\cr\cr
&&\hspace{1.5cm}\longmapsto\bigl(\bigl(S_I(h_1)^{\widehat\a}_{\ \widehat\b}\,\theta^{\widehat\b\,I}_2+\theta^{\widehat\a I}_1,L(h_1)^{\widehat a}_{\ \widehat b}\,x^{\widehat b}_2+x^{\widehat a}_1-\d_{IJ}\,\ovl\theta^I_1\,\G^{\widehat a}\,S_J(h_1)\,\theta^J_2,h_1\cdot h_2\bigr),\cr\cr
&&\hspace{2.5cm}\ee^{-\ep_{IJ}\,\ovl\theta^I_1\,S_J(h_1)\,\theta^J_2}\cdot z_1\cdot z_2\bigr)\,,
\qqq
with the neutral element
\qq\nn
\bigl((0,0,\bd1),1\bigr)\,,
\qqq
and the inverse
\qq\nn
\widetilde{{\rm sISO}(3,1\,\vert\,2\cdot 4)}\too\widetilde{{\rm sISO}(3,1\,\vert\,2\cdot 4)}\ :\ \bigl(\bigl(\theta^{\widehat\a I},x^{\widehat a},h\bigr),z\bigr)\longmapsto\bigl(-S_I\bigl(h^{-1}\bigr)^{\widehat\a}_{\ \widehat\b}\,\theta^{\widehat\b\,I},L\bigl(h^{-1}\bigr)^{\widehat a}_{\ \widehat b}\,x^{\widehat b},h^{-1}\bigr),z^{-1}\bigr)\,.
\qqq
It is easy to verify the associativity of the binary operation and thus convince ourselves that it defines a Lie-supergroup structure on $\,\widetilde{{\rm sISO}(3,1\,\vert\,2\cdot 4)}\,$ that lifts the short exact sequence \eqref{eq:SESsiso} to the Lie-supergroup level,
\qq\label{eq:sMinkextgrp}
\bd1\too\bC^\x\xrightarrow{\ \jmath_{\bC^\x}\ }\widetilde{{\rm sISO}(3,1\,\vert\,2\cdot 4)}\xrightarrow{\ \widetilde\pi_1\equiv\pr_1\ }{\rm sISO}(3,1\,\vert\,2\cdot 4)\too\bd1\,,
\qqq
with
\qq\nn
\jmath_{\bC^\x}\ :\ \bC^\x\too\widetilde{{\rm sISO}(3,1\,\vert\,2\cdot 4)}\ :\ z\longmapsto\bigl((0,0,\bd1),z\bigr)\,.
\qqq
Instead of checking that, we may invoke Thm.\,\ref{thm:TWextgrp} and verify that the conditions listed in it are satisfied by the GS super-2-cocycle. Indeed, the latter is exact, and hence 
\qq\nn
{\rm Per}\bigl(\underset{\tx{\ciut{(2)}}}{\chi^1}\bigr)=\{0\}
\qqq 
is the trivial group. Furthermore, in consequence of the left-invariance (and closedness) of $\,\underset{\tx{\ciut{(2)}}}{\chi^1}$,\ we obtain
\qq\nn
\sfd\bigl(\cR_X\con\underset{\tx{\ciut{(2)}}}{\chi^1}\bigr)\equiv\pLie{\cR_X}\underset{\tx{\ciut{(2)}}}{\chi^1}=0\,,
\qqq
and so, given the triviality of the de Rham cohomology of the supertranslation Lie supergroup to which this identity projects, there exists a smooth moment map for the left regular action of $\,{\rm sISO}(3,1\,\vert\,2\cdot 4)\,$ on $\,({\rm sISO}(3,1\,\vert\,2\cdot 4),\underset{\tx{\ciut{(2)}}}{\chi^1})$.\ In its full version presented in \Rcite{Tuynman:1987ij}, the theorem also ensures uniqueness of the extension, whence the result verified by hand.
Obviously, the extended supersymmetry group fibres over the original one as a trivial principal $\bC^\x$-bundle
\qq\label{eq:extsISOCx}
\alxydim{@C=1cm@R=1cm}{\bC^\x \ar[r] & {\rm sISO}(3,1\,\vert\,2\cdot 4)\x\bC^\x\equiv\widetilde{{\rm sISO}(3,1\,\vert\,2\cdot 4)} \ar[d]^{\widetilde\pi_1\equiv\pr_1} \\ & {\rm sISO}(3,1\,\vert\,2\cdot 4)}\,,
\qqq
with the bundle projection
\qq\nn
\widetilde\pi_1\equiv\pr_1\ :\ \widetilde{{\rm sISO}(3,1\,\vert\,2\cdot 4)}\too{\rm sISO}(3,1\,\vert\,2\cdot 4)\ :\ \bigl(\bigl(\theta^{\widehat\a I},x^{\widehat a},h\bigr),z\bigr)\longmapsto\bigl(\theta^{\widehat\a I},x^{\widehat a},h\bigr)\,,
\qqq
and the Lie-supergroup structure on its total space endows this bundle with a natural ${\rm sISO}(3,1\,\vert\,2\cdot 4)$-equivariant structure -- the (left) action of $\,{\rm sISO}(3,1\,\vert\,2\cdot 4)\,$ induced by the binary operation is realised by a principal $\bC^\x$-bundle automorphism.

\subsection{A contractible super-0-brane deformation of the $\,N=2\,$ super-${\rm AdS}_2\x\bS^2\,$ algebra}\label{sec:IWextaAdSS}

At the focus of our interest, we find the homogeneous space
\qq\nn
{\rm SU}(1,1\,\vert\,2)_2/({\rm SO}(1,1)\x{\rm SO}(2))\equiv{\rm s}\bigl({\rm AdS}_2\x\bS^2\bigr)
\qqq
of the Lie supergroup $\,{\rm SU}(1,1\,\vert\,2)_2$,\ with the body given by the homogeneous space
\qq\nn
{\rm SO}(1,2)/{\rm SO}(1,1)\x{\rm SO}(3)/{\rm SO}(2)\equiv{\rm AdS}_2\x\bS^2
\qqq
of the Lie group $\,{\rm SO}(1,2)\x{\rm SO}(3)$.\ The supersymmetry algebra $\,\gt{su}(1,1\,\vert\,2)_2\,$ and the supergeometry of the model (and so also its field-theoretic content) are readily cast in a form compatible with the decomposition of the body into its independent constituents: $\,{\rm AdS}_2\,$ and $\,\bS^2$.\ To this end, we work with the Majorana spinors of the product $\,{\rm Spin}\,$ group $\,{\rm Spin}(1,1)\x{\rm Spin}(2)\x{\rm Spin}(2,1)\,$ (the last factor accounts for the two species of spinors entering the construction), whence the presence of the multi-indices $\,\widehat\a\equiv\a\a' I\,$ on them, with $\,\a,\a'\in\{1,2\}\,$ and $\,I\in\{1,2\}\,$ (and that of the tensor products of elements of the Clifford algebras of the quadratic spaces $\,\bR^{1,1},\bR^{2,0}\,$ and $\,\bR^{2,1}$), and with tensors of the product isotropy group $\,{\rm SO}(1,1)\x{\rm SO}(2)$,\ whence the two subsets of vector indices: $\,a\in\ovl{0,1}\,$ and $\,a'\in\ovl{2,3}$.\ Important properties of the distinguished representations of the said Clifford algebras entering the construction have been recapitulated in App.\,\ref{app:CliffAdSS}. The (complexified) Lie superalgebra $\,\gt{su}(1,1\,\vert\,2)_2\,$ of the supersymmetry group has generators
\qq\nn
\gt{su}(1,1\,\vert\,2)_2&\equiv&\bigl(\tgt^{(0)}\oplus\tgt^{(1)}\bigr)\oplus\bigl(\gt{so}(1,1)\oplus\gt{so}(2)\bigr)\cr\cr
&=&\bigg(\bigl(\bigoplus_{a\in\{0,1\}}\,\corr{P_a}_\bC\oplus\bigoplus_{a'\in\{2,3\}}\,\corr{P_{a'}}_\bC\bigr)\oplus\bigoplus_{(\a,\a', I)\in\{1,2\}^{\x 3}}\,\corr{Q_{\a\a' I}}_\bC\bigg)\cr\cr
&&\oplus\bigg(\corr{J_{01}=-J_{10}}_\bC\oplus\corr{J_{23}=-J_{32}}_\bC\bigg)
\qqq
subject to the structure relations
\qq
\{Q_{\a\a' I},Q_{\b\b' J}\}=2\bigl(\bigl(\unl C\,\unl\g^{\widehat a}\ox\bd1\bigr)_{\a\a'I\b\b'J}\,P_{\widehat a}-\sfi\,\bigl(\unl C\ox\si_2\bigr)_{\a\a'I\b\b'J}\,J_{01}+\sfi\,\bigl(\unl C\,\unl\g_5\ox\si_2\bigr)_{\a\a'I\b\b'J}\,J_{23}\bigr)\,,\cr\cr\cr
[P_0,P_1]=J_{01}\,,\qquad\qquad[P_2,P_3]=-J_{23}\,,\qquad\qquad[P_a,P_{a'}]=0\,,\cr\cr\cr
[J_{01},J_{23}]=0\,,\label{eq:AdSS}\\ \cr\cr
[P_a,J_{01}]=\eta_{a0}\,P_1-\eta_{a1}\,P_0\,,\qquad\qquad[P_{a'},J_{23}]=\d_{a'2}\,P_3-\d_{a'3}\,P_2\,,\qquad\qquad[P_a,J_{23}]=0=[P_{a'},J_{01}]\,,\cr\cr\cr
[Q_{\a\a' I},P_{\widehat a}]=-\tfrac{\sfi}{2}\,\bigl(\widetilde\g_3\,\unl\g_{\widehat a}\ox\si_2\bigr)^{\b\b'J}_{\ \ \a\a'I}\,Q_{\b\b' J}\,,\qquad\qquad[Q_{\a\a' I},J_{\widehat a\widehat b}]=-\tfrac{1}{2}\,\bigl(\unl\g_{\widehat a\widehat b}\ox\bd1\bigr)^{\b\b'J}_{\ \ \a\a'I}\,Q_{\b\b'J}\,.\nn
\qqq
From the above, we readily recover the Lie superalgebra $\,\gt{smink}^{3,1\,\vert\,2\cdot 4}\,$ upon rescaling the generators as
\qq\label{eq:IWrescale}
\varsigma_R\ :\ \bigl(Q_{\a\a'I},P_{\widehat a},J_{01},J_{23}\bigr)\longmapsto\bigl(R^{\frac{1}{2}}\,Q_{\a\a'I},R\,P_{\widehat a},J_{01},J_{23}\bigr)\,,\quad R\in\bR
\qqq
and subsequently passing to the limit 
\qq\label{eq:sAdSS2sMinklim}
R\to\infty\,,
\qqq
which jointly defines the \.In\"on\"u--Wigner contraction
\qq\label{eq:IWcontrsAdS}
\gt{su}(1,1\,\vert\,2)_2\xrightarrow{\ \varsigma_R\ }\gt{su}(1,1\,\vert\,2)_{2,R}\xrightarrow{\ R\to\infty\ }\gt{smink}^{3,1\,\vert\,2\cdot 4}\,.
\qqq

The Lie supergroup $\,{\rm SU}(1,1\,\vert\,2)_2\,$ forms a principal ${\rm SO}(1,1)\x{\rm SO}(2)$-bundle 
\qq\label{eq:princSOSO}
\alxydim{@C=1cm@R=1cm}{{\rm SO}(1,1)\x{\rm SO}(2) \ar[r] & {\rm SU}(1,1\,\vert\,2)_2 \ar[d]^{\pi_2\equiv\pi_{{\rm SU}(1,1\,\vert\,2)_2/({\rm SO}(1,1)\x{\rm SO}(2))}} \\ & {\rm SU}(1,1\,\vert\,2)_2/\bigl({\rm SO}(1,1)\x{\rm SO}(2)\bigr)}\,,
\qqq
over the homogeneous space $\,{\rm SU}(1,1\,\vert\,2)_2/({\rm SO}(1,1)\x{\rm SO}(2))$,\ and the latter may be locally realised within the total space $\,{\rm SU}(1,1\,\vert\,2)_2\,$ of the bundle by a collection of sections 
\qq
\si^2_i\ :\ \cO^2_i\too{\rm SU}(1,1\,\vert\,2)\ :\ Z_i\equiv\bigl(\theta_i^{\a\a' I},x_i^a,x_i'{}^{a'}\bigr)\equiv\bigl(\theta_i^{\a\a'I},X_i^{\widehat a}\bigr)\longmapsto\unl{g_i}\cdot\ee^{X_i^{\widehat a}\,P_{\widehat a}}\cdot\ee^{\theta_i^{\a\a' I}\,Q_{\a\a' I}}\,,\qquad i\in I_2\cr \label{eq:MTsect}
\qqq
defined on the respective elements $\,\cO^2_i\,$ of a trivialising cover $\,\{\cO^2_i\}_{i\in I_2}\,$ of (the base of) the bundle in terms of local coordinates $\,\{\theta_i^{\a\a'I},X_i^{\widehat a}\}\,$ centred on some $\,\unl{g_i}\,({\rm SO}(1,1)\x{\rm SO}(2))\in\cO^2_i$.\ The formula for the sections is to be understood as in the super-Minkowskian case. There is, this time, no obvious Lie-supergroup structure on the homogeneous space $\,{\rm s}({\rm AdS}_2\x\bS^2)$.

The supersymmetry group acts on $\,{\rm s}\bigl({\rm AdS}_2\x\bS^2\bigr)\,$ in a natural manner,
\qq\nn
[\ell^2]_\cdot\ &:&\ {\rm SU}(1,1\,\vert\,2)_2\x{\rm s}\bigl({\rm AdS}_2\x\bS^2\bigr)\too{\rm s}\bigl({\rm AdS}_2\x\bS^2\bigr)\cr\cr 
&:&\ \bigl(g',g\,\bigl({\rm SO}(1,1)\x{\rm SO}(2)\bigr)\bigr)\longmapsto (g'\cdot g)\,\bigl({\rm SO}(1,1)\x{\rm SO}(2)\bigr)\,,
\qqq
the action becoming linear\footnote{The isotropy group $\,{\rm SO}(1,1)\x{\rm SO}(2)\,$ is realised in the spinor representations (indexed by $\,\{1,2\}\ni I$) on the Gra\ss mann-odd local coordinates $\,\theta^{\a\a'I}\,$ and in the vector representation on the Gra\ss mann-even local coordinate $\,X^{\widehat a}$,\ with the $\,x^a\,$ (resp.\ the $\,x'{}^{a'}$) behaving as scalars with respect to the component $\,{\rm SO}(2)\,$ (resp.\ $\,{\rm SO}(1,1)$)} upon restriction to the isotropy group $\,{\rm SO}(1,1)\x{\rm SO}(2)$.\ This is modelled locally on the space
\qq\nn
\si^2\bigl({\rm s}\bigl({\rm AdS}_2\x\bS^2\bigr)\bigr):=\bigsqcup_{i\in I_2}\,\si^2_i(\cO_i)
\qqq
as in \Reqref{eq:cosetactal}.

The very last piece of data that we need prior to launching a detailed analysis of the Cartan--Eilenberg cohomology behind Zhou's super-0-brane model with the supertarget $\,{\rm s}({\rm AdS}_2\x\bS^2)$,\ introduced in \Rcite{Zhou:1999sm}, is the asymptotic expansion of the various components of the $\gt{su}(1,1\,\vert\,2)_2$-valued left-invariant Maurer--Cartan super-1-form $\,\theta_{2\,{\rm L}}\,$ on $\,{\rm SU}(1,1\,\vert\,2)_2\,$ restricted to $\,\si^2({\rm s}({\rm AdS}_2\x\bS^2))\,$ in the r\'egime of large scaling parameter $\,R>>1$,\ to be identified with the common radius of the generating 1-cycle in $\,{\rm AdS}_2\cong\bS^1\x\bR\,$ and that of the 2-sphere $\,\bS^2\,$ in the body $\,{\rm AdS}_2\x\bS^2\,$ of the supertarget. We use the parameter to redefine the local coordinates on the latter in such a manner (dual to \eqref{eq:IWrescale})
\qq\nn
\bigl(\theta_i^{\a\a'I},X_i^{\widehat a}\bigr)\longmapsto\bigl(R^{-\frac{1}{2}}\,\theta_i^{\a\a'I},R^{-1}\,X_i^{\widehat a}\bigr)
\qqq
as to obtain the supertarget $\,{\rm sMink}^{3,1\,\vert\,2\cdot 4}\,$ in the limit \eqref{eq:sAdSS2sMinklim}. Clearly, the rescaling can be extended (trivially) to the fibre coordinates (generalised angles) on the total space $\,{\rm SU}(1,1\,\vert\,2)\,$ of the principal ${\rm SO}(1,1)\x{\rm SO}(2)$-bundle over $\,{\rm s}({\rm AdS}_2\x\bS^2)$.\ In the said r\'egime, we obtain, with the help of the same Schur--Poincar\'e formula as before, the result
\qq\nn
\si_i^*\theta_{2\,{\rm L}}(\theta,X)=\si_i^2{}^*\theta^{\a\a'I}_{2\,{\rm L}}(\theta,X)\ox Q_{\a\a'I}+\si_i^2{}^*\theta^{\widehat a}_{2\,{\rm L}}(\theta,X)\ox P_{\widehat a}+\si_i^2{}^*\theta^{01}_{2\,{\rm L}}(\theta,X)\ox J_{01}+\si_i^2{}^*\theta^{23}_{2\,{\rm L}}(\theta,X)\ox J_{23}\,.
\qqq
with the components
\qq\nn
\si_i^2{}^*\theta^{\a\a'I}_{2\,{\rm L}}(\theta_i,X_i)&=&\tfrac{1}{R^{\frac{1}{2}}}\,\bigl[\sfd\theta^{\a\a'I}_i+\tfrac{\sfi}{2R}\,\bigl(\sfd X^{\widehat a}_i+\tfrac{1}{3}\,\ovl\theta_i\,\bigl(\unl\g^{\widehat a}\ox\bd1\bigr)\,\sfd\theta_i\bigr)\,\bigl(\widetilde\g_3\,\unl\g^{\widehat a}\ox\si_2\bigr)^{\a\a'I}_{\ \b\b'J}\,\theta_i^{\b\b'J}\cr\cr
&&-\tfrac{\sfi}{6R}\,\bigl(\ovl\theta_i\,\bigl(\bd1\ox\si_2\bigr)\,\sfd\theta_i\,\bigl(\widetilde\g_3\ox\bd1\bigr)^{\a\a'I}_{\ \b\b'J}-\ovl\theta_i\,\bigl(\unl\g_5\ox\si_2\bigr)\,\sfd\theta_i\,\bigl(\widetilde\g_3'\ox\bd1\bigr)^{\a\a'I}_{\ \b\b'J}\bigr)\,\theta_i^{\b\b'J}\bigr]+O\bigl(R^{-\frac{5}{2}}\bigr)\,,\cr\cr
\si_i^2{}^*\theta^{\widehat a}_{2\,{\rm L}}(\theta_i,X_i)&=&\tfrac{1}{R}\,\bigl[\sfd X_i^{\widehat a}+\ovl\theta_i\,\bigl(\unl\g^{\widehat a}\ox\bd1\bigr)\,\sfd\theta_i+\tfrac{\sfi}{2R}\,\bigl(\sfd X_i^{\widehat b}+\tfrac{1}{6}\,\ovl\theta_i\,\bigl(\unl\g^{\widehat b}\ox\bd1\bigr)\,\sfd\theta_i\bigr)\,\ovl\theta_i\,\bigl(\unl\g^{\widehat a}\,\widetilde\g_3\,\unl\g_{\widehat b}\ox\si_2\bigr)\,\theta\cr\cr
&&-\tfrac{\sfi}{12R}\,\bigl(\ovl\theta_i\,\bigl(\bd1\ox\si_2\bigr)\,\sfd\theta_i\cdot\unl\theta_i\,\bigl(\unl\g^{\widehat a}\,\widetilde\g_3\ox\bd1\bigr)\,\theta_i\bigr)-\ovl\theta_i\,\bigl(\unl\g_5\ox\si_2\bigr)\,\sfd\theta_i\cdot\unl\theta_i\,\bigl(\unl\g^{\widehat a}\,\widetilde\g_3'\ox\bd1\bigr)\,\theta_i\bigr)\bigr]+O\bigl(R^{-3}\bigr)\,,\cr\cr
\si_i^2{}^*\theta^{01}_{2\,{\rm L}}(\theta_i,X_i)&=&\tfrac{1}{R}\,\bigl[-\sfi\,\ovl\theta_i\,\bigl(\bd1\ox\si_2\bigr)\,\sfd\theta_i+\tfrac{1}{2R}\,\bigl(x_i^1\,\sfd x_i^0-x_i^0\,\sfd x_i^1\bigr)\cr\cr
&&+\tfrac{1}{2R}\,\bigl(\sfd x'{}^{a'}+\tfrac{1}{6}\,\ovl\theta_i\,\bigl(\unl\g^{a'}\ox\bd1\bigr)\,\sfd\theta_i\bigr)\,\ovl\theta_i\,\bigl(\widetilde\g_3\,\unl\g_{a'}\ox\bd1\bigr)\,\theta_i\cr\cr
&&-\tfrac{1}{12R}\,\bigl(\ovl\theta_i\,\bigl(\bd1\ox\si_2\bigr)\,\sfd\theta_i\cdot\ovl\theta_i\,\bigl(\widetilde\g_3\ox\si_2\bigr)\,\theta_i\bigr)-\ovl\theta_i\,\bigl(\unl\g_5\ox\si_2\bigr)\,\sfd\theta_i\cdot\ovl\theta_i\,\bigl(\widetilde\g_3'\ox\si_2\bigr)\,\theta_i\bigr)\bigr]+O\bigl(R^{-3}\bigr)\,,\cr\cr
\si_i^2{}^*\theta^{23}_{2\,{\rm L}}(\theta_i,X_i)&=&\tfrac{1}{R}\,\bigl[\sfi\,\ovl\theta_i\,\bigl(\unl\g_5\ox\si_2\bigr)\,\sfd\theta_i+\tfrac{1}{2R}\,\bigl(x_i^2\,\sfd x_i^3-x_i^3\,\sfd x_i^2\bigr)\cr\cr
&&-\tfrac{1}{2R}\,\bigl(\sfd x^a+\tfrac{1}{6}\,\ovl\theta_i\,\bigl(\unl\g^a\ox\bd1\bigr)\,\sfd\theta_i\bigr)\,\ovl\theta_i\,\bigl(\widetilde\g_3'\,\unl\g_a\ox\bd1\bigr)\,\theta_i\cr\cr
&&+\tfrac{1}{12R}\,\bigl(\ovl\theta_i\,\bigl(\bd1\ox\si_2\bigr)\,\sfd\theta_i\cdot\ovl\theta_i\,\bigl(\widetilde\g_3'\ox\si_2\bigr)\,\theta_i\bigr)-\ovl\theta_i\,\bigl(\unl\g_5\ox\si_2\bigr)\,\sfd\theta_i\cdot\ovl\theta_i\,\bigl(\widetilde\g_3\ox\si_2\bigr)\,\theta_i\bigr)\bigr]+O\bigl(R^{-3}\bigr)\,.
\qqq

The point of departure of our supergerbe analysis is the supersymmetric de Rham super-2-cocycle
\qq\label{eq:Zhs0c}
\underset{\tx{\ciut{(2)}}}{\chi^2}=\sfi\,\ovl\theta_{2\,{\rm L}}\wedge\bigl(\bd1\ox\si_2\bigr)\,\theta_{2\,{\rm L}}+\theta_{2\,{\rm L}}^0\wedge\theta_{2\,{\rm L}}^1\,,
\qqq
corresponding to the choice $\,(A,B)=(1,0)\,$ of the parameters introduced in \Rcite{Zhou:1999sm}. The super-2-cocycle descends to a supersymmetric Green--Schwarz super-2-cocycle on the homogeneous space $\,{\rm s}({\rm AdS}_2\x\bS^2)\,$ with restrictions 
\qq\label{eq:GSs2csAdSS}
\underset{\tx{\ciut{(2)}}}{\txH^2}\rstr_{\cO_i}\equiv\si_i^2{}^*\underset{\tx{\ciut{(2)}}}{\chi^2}
\qqq
and admits a global primitive
\qq\nn
\underset{\tx{\ciut{(2)}}}{\chi^2}=\sfd\underset{\tx{\ciut{(1)}}}{\b^2}\,,\qquad\qquad\underset{\tx{\ciut{(1)}}}{\b^2}=-\theta^{01}_{2\,{\rm L}}
\qqq
that can be restricted to $\,\si^2({\rm s}({\rm AdS}_2\x\bS^2))\,$ but does \emph{not} descend to a smooth super-1-form on the homogeneous space. Locally, we find the asymptotics
\qq\nn
\underset{\tx{\ciut{(2)}}}{\txH^2}(\theta_i,X_i)&=&\tfrac{\sfi}{R}\,\sfd\ovl\theta^I_i\wedge\bigl(\bd1\ox\si_2\bigr)\,\sfd\theta^J_i+O\bigl(R^{-2}\bigr)\,,\cr\cr
\si_i^2{}^*\underset{\tx{\ciut{(1)}}}{\b^2}(\theta_i,X_i)&=&\tfrac{\sfi}{R}\,\ovl\theta^I_i\,\bigl(\bd1\ox\si_2\bigr)\,\sfd\theta^J_i+O\bigl(R^{-2}\bigr)\,,
\qqq
and so, clearly, the super-2-cocycle and its (local) primitive go over to their flat-superspace counterparts \eqref{eq:GSs2csMink} and \eqref{eq:GSs2cprimsMink}, respectively, after an overall rescaling by $\,R$.

According to the general scheme laid out in Sec.\,\ref{sec:scheme}, we shall now look for an extension $\,\widetilde{\gt{su}(1,1\,\vert\,2)_2}\,$ of the supersymmetry algebra $\,\gt{su}(1,1\,\vert\,2)_2\,$ that contracts to the formerly described extension $\,\widetilde{\gt{smink}}^{3,1\,\vert\,2\cdot 4}\,$ of the $\,N=2\,$ super-Minkowski Lie superalgebra $\,\gt{smink}^{3,1\,\vert\,2\cdot 4}\,$ and integrates to an extended supersymmetry group 
\qq\nn
\widetilde\pi_2\ :\ \widetilde{{\rm SU}(1,1\,\vert\,2)_2}\too{\rm SU}(1,1\,\vert\,2)_2
\qqq
such that $\,\widetilde\pi_2^*\underset{\tx{\ciut{(2)}}}{\chi^2}\,$ admits a primitive descending to a supersymmetric super-1-form on the homogeneous space $\,\widetilde{{\rm SU}(1,1\,\vert\,2)_2}/({\rm SO}(1,1)\x{\rm SO}(2))\,$ with the asymptotics (of the restriction of) $\,\widetilde\theta^Z_{1\,{\rm L}}$.\ Taking into account the flat-superspace result \eqref{eq:s0bextsiso1}-\eqref{eq:s0bextsiso3}, we consider a generic \emph{scalar} deformation of $\,\gt{su}(1,1\,\vert\,2)_2\,$ with the germ
\qq\nn
\{Q_{\a\a' I},Q_{\b\b' J}\}^\sim&=&2\bigl(\bigl(\unl C\,\unl\g^{\widehat a}\ox\bd1\bigr)_{\a\a'I\b\b'J}\,P_{\widehat a}-\sfi\,\bigl(\unl C\ox\si_2\bigr)_{\a\a'I\b\b'J}\,(J_{01}+\a\,Z_{01})\cr\cr
&&+\sfi\,\bigl(\unl C\,\unl\g_5\ox\si_2\bigr)_{\a\a'I\b\b'J}\,J_{23}\bigr)\,,\qquad\a\in\bC^\x\cr\cr\cr
[P_0,P_1]^\sim&=&J_{01}+\b\,Z_{01}\,,\qquad\b\in\bC^\x\,,
\qqq
and all the remaining (super)commutation relations of the mother Lie superalgebra (within the new one) undeformed\footnote{We only have a meaningful choice in the case of the commutators $\,[P_2,P_3]^\sim\,$and $\,[P_a,P_{a'}]^\sim$,\ and we minimalistically \emph{choose} them, with hindsight, to be undeformed.}. This leaves us with the following commutation relations involving the new generator
\qq\nn
&[Q_{\a\a'I},Z_{01}]^\sim=\G_{\a\a'I}^{\ \b\b'J}\,Q_{\b\b'J}\,,\qquad\qquad[P_{\widehat a},Z_{01}]^\sim=0\,,\qquad\qquad[J_{\widehat a\widehat b},Z_{01}]^\sim=0\,,&\cr\cr
&[Z_{01},Z_{01}]^\sim=\la\,Z_{01}\,,\qquad\la\in\bC\,,&
\qqq 
where $\,\G_{\a\a'I}^{\ \b\b'J}\,$ is (the matrix element of) an element of the (product) Clifford algebra to be determined in the analysis to follow. The requirement that the above-defined deformation contract to the previously derived $\,\widetilde{\gt{smink}}^{3,1\,\vert\,2\cdot 4}\,$ determines the scaling of the new generator uniquely, and we have, altogether,
\qq\nn
\widetilde\varsigma_R\ :\ \bigl(Q_{\a\a'I},P_{\widehat a},J_{01},J_{23},Z_{01}\bigr)\longmapsto\bigl(R^{\frac{1}{2}}\,Q_{\a\a'I},R\,P_{\widehat a},J_{01},J_{23},R\,Z_{01}\bigr)\,.
\qqq

As we want to remain in the category of Lie superalgebras, we next impose the super-Jacobi identities as constraints upon the deformation parameters $\,\a,\b,\la\,$ and $\,\G_{\a\a'I}^{\ \b\b'J}$.\ We commence with the identity
\qq\nn
{\rm sJac}(Q_{\a\a'I},P_{\widehat a},P_{\widehat b})=0
\qqq
which yields, upon invoking its counterpart for $\,\gt{su}(1,1\,\vert\,2)_2$,\ the condition
\qq\nn
\b\,\bigl(\d_{\widehat a 0}\,\d_{\widehat b 1}-\d_{\widehat a 1}\,\d_{\widehat b 0}\bigr)\,\G_{\a\a'I}^{\ \b\b'J}\,Q_{\b\b'J}=0\,,
\qqq
whence
\qq\nn
\G_{\a\a'I}^{\ \b\b'J}=0\,.
\qqq
Next, we consider the super-Jacobi identity
\qq\nn
{\rm sJac}(P_0,P_1,Z_{01})=0\,,
\qqq
whereby we obtain the condition
\qq\nn
\b\,\la\,Z_{01}=0\,,
\qqq
leading to 
\qq\nn
\la=0\,.
\qqq
Finally, we impose the super-Jacobi identity
\qq\nn
{\rm sJac}(Q_{\a\a'I},Q_{\b\b'J},P_a)=0\,,
\qqq
which produces the condition
\qq\nn
0&=&2\b\,\ep_{ba}\,\bigl(\unl C\,\unl\g^b\ox\bd1\bigr)_{\a\a'I\b\b'J}+\a\,\bigl(\unl C\,\widetilde\g_3\,\unl\g_a\ox\bd1\bigr)_{\b\b'J\a\a'I}+\a\,\bigl(\unl C\,\widetilde\g_3\,\unl\g_a\ox\bd1\bigr)_{\a\a'I\b\b'J}\cr\cr
&=&2\b\,\ep_{ba}\,\bigl(\unl C\,\unl\g^b\ox\bd1\bigr)_{\a\a'I\b\b'J}+2\a\,\bigl(\unl C\,\widetilde\g_3\,\unl\g_a\ox\bd1\bigr)_{\a\a'I\b\b'J}\equiv 2(\b-\a)\,\ep_{ba}\,\bigl(\unl C\,\unl\g^b\ox\bd1\bigr)_{\a\a'I\b\b'J}\,,
\qqq
{\it cp} \Reqref{eq:gam3act}, giving the relation
\qq\nn
\b=\a\,.
\qqq
It is now easy to check that no further constraints follow from the imposition of the super-Jacobi identities, and so we end up with the extension $\,\widetilde{\gt{su}(1,1\,\vert\,2)_2}\,$ generated by the $\,Q_{\a\a'I},P_{\widehat a},J_{01},J_{23}\,$ and $\,Z_{01}\,$ subject to the structure relations
\qq
\{Q_{\a\a' I},Q_{\b\b' J}\}^\sim=2\bigl(\bigl(\unl C\,\unl\g^{\widehat a}\ox\bd1\bigr)_{\a\a'I\b\b'J}\,P_{\widehat a}-\sfi\,\bigl(\unl C\ox\si_2\bigr)_{\a\a'I\b\b'J}\,(J_{01}+\a\,Z_{01})+\sfi\,\bigl(\unl C\,\unl\g_5\ox\si_2\bigr)_{\a\a'I\b\b'J}\,J_{23}\bigr)\,,\cr\cr\cr
[P_0,P_1]^\sim=J_{01}+\a\,Z_{01}\,,\qquad\qquad[P_2,P_3]^\sim=-J_{23}\,,\qquad\qquad[P_a,P_{a'}]^\sim=0\,,\cr\cr\cr
[J_{01},J_{23}]^\sim=0\,,\label{eq:s0bextsu112}\\ \cr\cr
[P_a,J_{01}]^\sim=\eta_{a0}\,P_1-\eta_{a1}\,P_0\,,\qquad\qquad[P_{a'},J_{23}]^\sim=\d_{a'2}\,P_3-\d_{a'3}\,P_2\,,\qquad\qquad[P_a,J_{23}]^\sim=0=[P_{a'},J_{01}]^\sim\,,\cr\cr\cr
[Q_{\a\a' I},P_{\widehat a}]^\sim=-\tfrac{\sfi}{2}\,\bigl(\widetilde\g_3\,\unl\g_{\widehat a}\ox\si_2\bigr)^{\b\b'J}_{\ \ \a\a'I}\,Q_{\b\b' J}\,,\qquad\qquad[Q_{\a\a' I},J_{\widehat a\widehat b}]^\sim=-\tfrac{1}{2}\,\bigl(\unl\g_{\widehat a\widehat b}\ox\bd1\bigr)^{\b\b'J}_{\ \ \a\a'I}\,Q_{\b\b'J}\,,\cr\cr\cr
[Q_{\a\a'I},Z_{01}]^\sim=0\,,\qquad\qquad[P_{\widehat a},Z_{01}]^\sim=0\,,\qquad\qquad[J_{\widehat a\widehat b},Z_{01}]^\sim=0\,,\qquad\qquad[Z_{01},Z_{01}]^\sim=0\,.\nn
\qqq
A change of the basis in the supervector space $\,\widetilde{\gt{su}(1,1\,\vert\,2)_2}\,$ affecting only the pair $\,(J_{01},Z_{01})\,$ as
\qq\nn
(J_{01},Z_{01})\longmapsto(J_{01}+\a\,Z_{01},Z_{01})
\qqq
gives us a Lie-superalgebra isomorphism 
\qq\nn
\widetilde{\gt{su}(1,1\,\vert\,2)_2}\cong\bC\oplus\gt{su}(1,1\,\vert\,2)_2\,,
\qqq
with the right-hand side given by a direct sum of decoupled Lie superalgebras, of which the first has a trivial commutator. Moreover, and more importantly, the extended supersymmetry algebra contracts as desired,
\qq\label{eq:IWcontrsAdSMink}
\widetilde{\gt{su}(1,1\,\vert\,2)_2}\xrightarrow{\ \widetilde\varsigma_R\ }\widetilde{\gt{su}(1,1\,\vert\,2)_{2,R}}\xrightarrow{\ R\to\infty\ }\widetilde{\gt{smink}}^{3,1\,\vert\,2\cdot 4}\,.
\qqq
Thus, at the level of the Lie superalgebras, the goal has been achieved. It remains to check if the extension
\qq\label{eq:sAdSSextalg}
\brd0\too\corr{Z_{01}}_\bC\cong\bC\too\widetilde{\gt{su}(1,1\,\vert\,2)_2}\too\gt{su}(1,1\,\vert\,2)_2\too\brd0
\qqq
integrates to \emph{a} Lie-supergroup extension
\qq\label{eq:sAdSSextgrp}
\bd1\too\bC^\x\too\widetilde{{\rm SU}(1,1\,\vert\,2)_2}\too{\rm SU}(1,1\,\vert\,2)_2\too\bd1\,.
\qqq 
To this end, we note that 
\qq\nn
{\rm Per}\bigl(\underset{\tx{\ciut{(2)}}}{\chi^2}\bigr)=\{0\}
\qqq
owing to the exactness of $\,\underset{\tx{\ciut{(2)}}}{\chi^2}$,\ and the left-invariance of the global primitive $\,\underset{\tx{\ciut{(1)}}}{\b^2}\,$ of the super-2-cocycle on $\,{\rm SU}(1,1\,\vert\,2)_2\,$ (or, in other words, the triviality of the class $\,[\underset{\tx{\ciut{(2)}}}{\chi^2}]\in{\rm CaE}^2({\rm SU}(1,1\,\vert\,2)_2)$) gives us the identity
\qq\nn
\cR_X\con\underset{\tx{\ciut{(2)}}}{\chi^2}\equiv-\cR_X\con\sfd\theta^{01}_{2\,{\rm L}}=-\pLie{\cR_X}\theta^{01}_{2\,{\rm L}}+\sfd\bigl(\cR_X\con\theta^{01}_{2\,{\rm L}}\bigr)=\sfd\bigl(\cR_X\con\theta^{01}_{2\,{\rm L}}\bigr)
\qqq
for an arbitrary element $\,X\in\gt{su}(1,1\,\vert\,2)_2$,\ whence the existence of the moment map
\qq\nn
\mu_\cdot^2\ :\ \gt{su}(1,1\,\vert\,2)_2\too C^\infty\bigl({\rm SU}(1,1\,\vert\,2)_2,\bR\bigr)\ :\ X\longmapsto-\cR_X\con\theta^{01}_{2\,{\rm L}}\,.
\qqq
This concludes the proof of existence of a Lie-supergroup extension sought-after.

\section{Geometrisations of the GS super-2-cocycles}\label{sec:geometrise}

We shall, next, use the superalgebraic findings from the previous section towards a full-fledged geometrisation of the GS super-2-cocycles \eqref{eq:GSs2csMink} and \eqref{eq:GSs2csAdSS}, subject to the additional requirements: equivariance with respect to the supersymmetry present and compatibility with the contraction \eqref{eq:IWcontrsAdSMink}. In so doing, we shall, first, refer solely to the topology and differential supergeometry of the respective supertargets: $\,{\rm sMink}^{3,1\,\vert\,2\cdot 4}\,$ and $\,{\rm s}({\rm AdS}_2\x\bS^2)$,\ without invoking the supergeometry of the Lie supergroups $\,{\rm sISO}(3,1\,\vert\,2\cdot 4)\,$ and $\,{\rm SU}(1,1\,\vert\,2)_2\,$ or their extensions. Only in the final stage of our discussion shall we identify the objects found as 
restrictions, to the realisations of the respective homogeneous spaces within the Lie supergroup, of the structures engendered by the extensions \eqref{eq:sMinkextgrp} and \eqref{eq:sAdSSextgrp}. The point of this exercise is to work out a path towards a geometrisation that circumnavigates the overarching Lie-supergroup structure, apparently superfluous at this stage, and yields directly a supermanifold endowed with an action of the supersymmetry group and a super-1-form invariant under an action of the latter.

\subsection{The Green--Schwarz super-0-gerbe over $\,{\rm sMink}^{3,1\,\vert\,2\cdot 4}$}\label{sec:spgsMink}

The sought-after geometrisation of the Green--Schwarz super-2-cocycle $\,\underset{\tx{\ciut{(2)}}}{\txH^1}\,$ can be deduced directly from the transformation law for its primitive $\,\underset{\tx{\ciut{(1)}}}{\txB^1}\,$ under the quotient action \eqref{eq:susyonsMink} of the supersymmetry group. Indeed, the latter reads
\qq\nn
[\ell^1]_{(\vep^{\widehat\a I},y^{\widehat a},h)}^*\underset{\tx{\ciut{(1)}}}{\txB^1}(\theta,X)=\underset{\tx{\ciut{(1)}}}{\txB^1}(\theta,X)+\sfd\bigl(\ep_{IJ}\,\ovl\vep^I\,S_J(h)\,\theta^J\bigr)\,,
\qqq
and so we are led to erect a trivial principal $\bC^\x$-bundle 
\qq\label{eq:s0gsMink}
\alxydim{@C=1cm@R=1cm}{\bC^\x \ar[r] & {\rm sMink}^{3,1\,\vert\,2\cdot 4}\x\bC^\x\equiv\xcL_1 \ar[d]^{[\widetilde\pi_1]\equiv\pi_{\xcL_1}} \\ & {\rm sMink}^{3,1\,\vert\,2\cdot 4}}\,.
\qqq
over $\,{\rm sMink}^{3,1\,\vert\,2\cdot 4}$,\ with the obvious projection to the base
\qq\nn
[\widetilde\pi_1]\equiv\pr_1\ :\ {\rm sMink}^{3,1\,\vert\,2\cdot 4}\x\bC^\x\too{\rm sMink}^{3,1\,\vert\,2\cdot 4}\ :\ \bigl(\bigl(\theta^{\widehat\a I},X^{\widehat a}\bigr),z\bigr)\longmapsto\bigl(\theta^{\widehat\a I},X^{\widehat a}\bigr)\,,
\qqq
and endow it with a \emph{projective} action of the supersymmetry group $\,{\rm sISO}(3,1\,\vert\,2\cdot 4)\,$ by principal $\bC^\x$-bundle automorphisms that preserve the principal $\bC^\x$-connection 1-form sewn from $\,\underset{\tx{\ciut{(1)}}}{\txB^1}\,$ as 
\qq\label{eq:princConnsMink}
\underset{\tx{\ciut{(1)}}}{\cA^1}(\theta,X,z)=\tfrac{\sfd z}{z}+\underset{\tx{\ciut{(1)}}}{\txB^1}(\theta,X)\,.
\qqq
The latter condition fixes the action in the form
\qq
\widetilde{[\unl{\ell^1}]}_\cdot\ &:&\ {\rm sISO}(3,1\,\vert\,2\cdot 4)\x\xcL_1\too\xcL_1\cr\cr 
&:&\ \bigl(\bigl(\vep^{\widehat\a I},y^{\widehat a},h\bigr),\bigl(\bigl(\theta^{\widehat\b J},X^{\widehat b}\bigr),z\bigr)\bigr)\cr\cr
&&\hspace{2cm}\longmapsto\bigl(\bigl(S_I(h)^{\widehat\a I}_{\ \widehat\b}\,\theta^{\widehat\b\,I}+\vep^{\widehat\a I},L(h)^{\widehat a}_{\ \widehat b}\,X^{\widehat b}+y^{\widehat a}-\d_{IJ}\,\ovl\vep^I\,\G^{\widehat a}\,S_J(h)\,\theta^J\bigr),\ee^{-\ep_{IJ}\,\ovl\vep^I\,S_J(h)\,\theta^J}\cdot z\bigr)\,, \label{eq:extsusyacts09sMink}
\qqq
with the homomorphicity super-2-cocycle
\qq\nn
d^{(0)}_{(\vep_1^{\widehat\a I},y_1^{\widehat a},h_1),(\vep_2^{\widehat\b J},y_2^{\widehat b},h_2)}=\ee^{-\ep_{IJ}\,\ovl\vep^I_1\,S_J(h_1)\,\vep^J_2}\in Z^2\bigl({\rm sISO}(3,1\,\vert\,2\cdot 4),\bC^\x\bigr)
\qqq
entering through
\qq\nn
\widetilde{[\unl{\ell^1}]}_{(\vep_1^{\widehat\a I},y_1^{\widehat a},h_1)}\circ\widetilde{[\unl{\ell^1}]}_{(\vep_2^{\widehat\b J},y_2^{\widehat b},h_2)}=\bigl(\id_{{\rm sMink}^{3,1\,\vert\,2\cdot 4}}\x\widehat\ell_{d^{(0)}_{(\vep_1^{\widehat\a I},y_1^{\widehat a},h_1),(\vep_2^{\widehat\b J},y_2^{\widehat b},h_2)}}\bigr)\circ\widetilde{[\unl{\ell^1}]}_{(\vep_1^{\widehat\a I},y_1^{\widehat a},h_1)\cdot(\vep_2^{\widehat\b J},y_2^{\widehat b},h_2)}\,,
\qqq
where
\qq\nn
\widehat\ell_\cdot\ :\ \bC^\x\x\bC^\x\too\bC^\x\ :\ (\z,z)\longmapsto\z\cdot z
\qqq
is the regular action of the structure group $\,\bC^\x\,$ upon itself (and so also on the fibre). The fact that we are dealing with a super-2-cocycle on $\,{\rm sISO}(3,1\,\vert\,2\cdot 4)\,$ (with values in the trivial ${\rm sISO}(3,1\,\vert\,2\cdot 4)$-module $\,\bC^\x$),
\qq\nn
\bigl(\d_{{\rm sISO}(3,1\,\vert\,2\cdot 4)}d^{(0)}\bigr)_{(\vep_1^{\widehat\a I},y_1^{\widehat a},h_1),(\vep_2^{\widehat\b J},y_2^{\widehat b},h_2),(\vep_3^{\widehat\g K},y_3^{\widehat c},h_3)}\equiv d^{(0)}_{(\vep_1^{\widehat\a I},y_1^{\widehat a},h_1),(\vep_2^{\widehat\b J},y_2^{\widehat b},h_2)}\cr\cr
\cdot\bigl(d^{(0)}_{(\vep_1^{\widehat\a I},y_1^{\widehat a},h_1),(\vep_2^{\widehat\b J},y_2^{\widehat b},h_2)\cdot(\vep_3^{\widehat\g K},y_3^{\widehat c},h_3)}\bigr)^{-1}\cdot d^{(0)}_{(\vep_1^{\widehat\a I},y_1^{\widehat a},h_1)\cdot(\vep_2^{\widehat\b J},y_2^{\widehat b},h_2),(\vep_3^{\widehat\g K},y_3^{\widehat c},h_3)}\cdot\bigl(d^{(0)}_{(\vep_2^{\widehat\b J},y_2^{\widehat b},h_2),(\vep_3^{\widehat\g K},y_3^{\widehat c},h_3)}\bigr)^{-1}=1\,,
\qqq
implies that the projective action of the original supersymmetry group $\,{\rm sISO}(3,1\,\vert\,2\cdot 4)\,$ can be lifted to a standard action of a central extension of that group, and we immediately see that the relevant extension is precisely the Lie supergroup $\,\widetilde{{\rm sISO}(3,1\,\vert\,2\cdot 4)}\,$ of \Reqref{eq:sISOext} that integrates the central extension $\,\widetilde{\gt{siso}(3,1\,\vert\,2\cdot 4)}\,$ defined by the wrapping anomaly \eqref{eq:wrapanosMink}, with the action on $\,\xcL_1\,$ induced by the binary operation on $\,\widetilde{{\rm sISO}(3,1\,\vert\,2\cdot 4)}\,$ in the form
\qq\nn
\widetilde{[\ell^1]}_\cdot\ &:&\ \widetilde{{\rm sISO}(3,1\,\vert\,2\cdot 4)}\x\xcL_1\too\xcL_1\cr\cr 
&:&\ \bigl(\bigl(\bigl(\vep^{\widehat\a I},y^{\widehat a},h\bigr),\z\bigr),\bigl(\bigl(\theta^{\widehat\b J},X^{\widehat b}\bigr),z\bigr)\bigr)\cr\cr
&&\hspace{2cm}\longmapsto\bigl(\bigl(S_I(h)^{\widehat\a I}_{\ \widehat\b}\,\theta^{\widehat\b\,I}+\vep^{\widehat\a I},L(h)^{\widehat a}_{\ \widehat b}\,X^{\widehat b}+y^{\widehat a}-\d_{IJ}\,\ovl\vep^I\,\G^{\widehat a}\,S_J(h)\,\theta^J\bigr),\ee^{-\ep_{IJ}\,\ovl\vep^I\,S_J(h)\,\theta^J}\cdot\z\cdot z\bigr)\,.
\qqq

The extended supersymmetry group fibres (trivially) over the original one, {\it cp} \Reqref{eq:extsISOCx}, just as the extended superspacetime $\,\xcL_1\equiv\widetilde{{\rm sMink}}^{3,1\,\vert\,2\cdot 4}\,$ does over the original spacetime $\,{\rm sMink}^{3,1\,\vert\,2\cdot 4}\,$ in \Reqref{eq:s0gsMink}. Furthermore, $\,\widetilde{{\rm sISO}(3,1\,\vert\,2\cdot 4)}\,$ fibres over $\,\widetilde{{\rm sMink}}^{3,1\,\vert\,2\cdot 4}\,$ just as $\,{\rm sISO}(3,1\,\vert\,2\cdot 4)\,$ does over $\,{\rm sMink}^{3,1\,\vert\,2\cdot 4}$,\ and, by the end of the day, we arrive at the commutative diagram of principal bundles (with structure groups $\,\bC^\x\,$ and $\,{\rm SO}(3,1)$,\ respectively)
\qq\label{diag:princbunsMink}\qquad\qquad
\alxydim{@C=2cm@R=2cm}{ & \bC^\x \ar[d] \ar@{=}[rr] & & \bC^\x \ar[d] \\ {\rm SO}(3,1) \ar[r] \ar@{=}[d] & \widetilde{{\rm sISO}(3,1\,\vert\,2\cdot 4)} \ar[rr]^{\pi_{\widetilde{{\rm sISO}(3,1\,\vert\,2\cdot 4)}/{\rm SO}(3,1)}} \ar[d]^{\widetilde\pi_1} & & \widetilde{{\rm sMink}^{3,1\,\vert\,2\cdot 4}} \ar[d]^{[\widetilde\pi_1]} \\ {\rm SO}(3,1) \ar[r] & {\rm sISO}(3,1\,\vert\,2\cdot 4) \ar[rr]_{\pi_1} & & {\rm sMink}^{3,1\,\vert\,2\cdot 4} }\,,
\qqq
which incidentally happens to be a commutative diagram of Lie supergroups as well. In fact, we may readily put even more structure in the above diagram. Indeed, the principal $\bC^\x$-bundle $\,\bC^\x\too\widetilde{{\rm sISO}(3,1\,\vert\,2\cdot 4)}\too{\rm sISO}(3,1\,\vert\,2\cdot 4)\,$ admits a natural principal $\bC^\x$-connection 1-form, given by the component of the $\widetilde{\gt{siso}(3,1\,\vert\,2\cdot 4)}$-valued Maurer--Cartan super-1-form on $\,\widetilde{{\rm sISO}(3,1\,\vert\,2\cdot 4)}\,$ along the central generator $\,Z$,
\qq\nn
\widetilde{\underset{\tx{\ciut{(1)}}}{\cA^1}}=\widetilde\theta^Z_{1\,{\rm L}}\,.
\qqq
Upon pullback along the global section
\qq\nn
\widetilde\si^1\ :\ \widetilde{{\rm sMink}^{3,1\,\vert\,2\cdot 4}}\too\widetilde{{\rm sISO}(3,1\,\vert\,2\cdot 4)}\ :\ \bigl(\bigl(\theta^{\widehat\a I},X^{\widehat a}\bigr),z\bigr)\longmapsto\bigl(\si^1\bigl(\theta^{\widehat\a I},X^{\widehat a}\bigr),z\bigr)\,,
\qqq
the latter reproduces ({\it i.e.}, descends to) the principal $\bC^\x$-connection 1-form on $\,\bC^\x\too\widetilde{{\rm sMink}^{3,1\,\vert\,2\cdot 4}}\too{\rm sMink}^{3,1\,\vert\,2\cdot 4}$,
\qq\nn
\widetilde\si^1{}^*\widetilde{\underset{\tx{\ciut{(1)}}}{\cA^1}}\equiv\underset{\tx{\ciut{(1)}}}{\cA^1}\,.
\qqq

Our discussion leads us naturally to
\bedef\label{def:s0gerbesMink}
The \textbf{Green--Schwarz super-0-gerbe} of curvature $\,\underset{\tx{\ciut{(2)}}}{\txH^1}\,$ over $\,{\rm sMink}^{3,1\,\vert\,2\cdot 4}\,$ is the triple
\qq\nn
\sG^{(0)}_{\rm GS}\bigl({\rm sMink}^{3,1\,\vert\,2\cdot 4}\bigr):=\bigl(\xcL_1,\pi_{\xcL_1},\underset{\tx{\ciut{(1)}}}{\cA^1}\bigr)
\qqq
constructed in the preceding paragraphs. It admits the action $\,\widetilde{[\ell^1]}_\cdot\,$ of the extended supersymmetry group $\,\widetilde{{\rm sISO}(3,1\,\vert\,2\cdot 4)}\,$ by connection-preserving principal $\bC^\x$-bundle automorphisms.
\exdef

\subsection{The contractible Green--Schwarz super-0-gerbe over $\,{\rm s}({\rm AdS}_2\x\bS^2)$}\label{sec:spgsAdSS}

Drawing inspiration from our earlier treatment of its super-Minkowskian counterpart, we read off the local structure of the super-0-gerbe to be constructed from the supersymmetry transformations of the locally smooth primitive 
\qq\nn
\underset{\tx{\ciut{(1)}}}{\txB_i^2}:=\si_i^2{}^*\underset{\tx{\ciut{(1)}}}{\b^2}\equiv-\si_i^2{}^*\theta^{01}_{2\,{\rm L}}
\qqq 
of the Green--Schwarz super-2-cocycle $\,\underset{\tx{\ciut{(2)}}}{\txH^2}\,$ of \Reqref{eq:GSs2csAdSS}. Their effect on the primitive can be extracted from their local model \eqref{eq:cosetactal} (written for arbitrary $\,g'\in{\rm SU}(1,1\,\vert\,2)_2\,$ and $\,x\in\cO^2_i,\ i\in I_2$)
\qq
g'\cdot\si^2_i(x)=\si^2_j\bigl(\widetilde x(x;g')\bigr)\cdot\unl h^2_{ij}(x;g')^{-1}\,,\qquad\qquad\unl h^2_{ij}(x;g')=\bigl(\unl h^{01}_{ij}(x;g'),\unl h^{23}_{ij}(x;g')\bigr)\in{\rm SO}(1,1)\x{\rm SO}(2)\,,\cr \label{eq:locmodAdSact}
\qqq
upon recalling that under a \emph{right} regular translation by an element $\,h=(h^{01},h^{23})\in{\rm SO}(1,1)\x{\rm SO}(2)\,$ of the isotropy group, the component $\,\theta^{01}_{2\,{\rm L}}\,$ of the Maurer--Cartan super-1-form $\,\theta_{2\,{\rm L}}\,$ on $\,{\rm SU}(1,1\,\vert\,2)_2\,$ undergoes an affine tranformation 
\qq\nn
r_h^*\theta^{01}_{2\,{\rm L}}=\theta^{01}_{2\,{\rm L}}+h^{01}{}^{-1}\,\sfd h^{01}\,.
\qqq
These imply the transformation law
\qq\nn
[\ell^2]_{g'}^*\underset{\tx{\ciut{(1)}}}{\txB^2_j}(x)=\underset{\tx{\ciut{(1)}}}{\txB^2_i}(x)-\unl h^{01}_{ij}{}^{-1}\,\sfd\unl h^{01}_{ij}(x;g')\,,
\qqq
valid in a neighbourhood of $\,(x,g')\in\cO^2_i\x{\rm SU}(1,1\,\vert\,2)_2$.\ The above immediately suggests that the supertarget $\,{\rm s}({\rm AdS}_2\x\bS^2)\,$ be locally extended by $\,\bC^\x$,\ and that in the adapted local coordinates $\,((\theta_i^{\a\a'I},X_i^{\widehat a}),z_i)\in\cO^2_i\x\bC^\x\,$ the primitive $\,\underset{\tx{\ciut{(1)}}}{\txB_i^2}\,$ be replaced by the super-1-form
\qq\label{eq:locconsAdS}
\underset{\tx{\ciut{(1)}}}{\txA_i^2}\bigl(\theta_i^{\a\a'I},X_i^{\widehat a},z_i\bigr):=\tfrac{\sfd z_i}{z_i}+\underset{\tx{\ciut{(1)}}}{\txB_i^2}\bigl(\theta_i^{\a\a'I},X_i^{\widehat a}\bigr)\,,
\qqq
assumed invariant under a suitable realisation of \eqref{eq:locmodAdSact} on a global structure glued from the local trivialisations $\,\cO^2_i\x\bC^\x$.\ The obvious way to proceed is to reconstruct the total space of the sought-after geometrisation of $\,\underset{\tx{\ciut{(2)}}}{\txH^2}\,$ with the help of The Clutching Theorem, using the ${\rm SO}(1,1)$-component\footnote{Strictly speaking, we are working with the connected component $\,\uj\,$ of the group unit in $\,{\rm SO}(1,1)$,\ {\it cp} \Rcite{Berkovits:1999zq}.} $\,h^{01}_{ij}\ :\ \cO^2_{ij}\too{\rm SO}(1,1)\,$ of the transition maps 
\qq\nn
h^2_{ij}=(h^{01}_{ij},h^{23}_{ij})\ :\ \cO^2_{ij}\too{\rm SO}(1,1)\x{\rm SO}(2)
\qqq
of the principal ${\rm SO}(1,1)\x{\rm SO}(2)$-bundle \eqref{eq:princSOSO}, as the quotient (super)manifold
\qq\nn
\bigl(\bigsqcup_{i\in I_2}\,\cO^2_i\x\bC^\x\bigr)/_{\sim_{h^{01}_{\cdot\cdot}}}
\qqq
obtained from the disjoint union $\,\bigsqcup_{i\in I_2}\,\cO^2_i\x\bC^\x\,$ of the local trivialisations through the identification, over $\,\cO^2_{ij}\ni x$,\ engendered by the equivalence relation  
\qq\nn
(x,z_i,i)\sim_{h^{01}_{\cdot\cdot}}\bigl(x,z_i\cdot h^{01}_{ij}(x),j\bigr)\,.
\qqq
Taking into account the gluing law for the local sections $\,\si^2_i\,$ over $\,\cO^2_{ij}\ni x$, 
\qq\nn
\si^2_j(x)=\si^2_i(x)\cdot h^2_{ij}(x)\,,
\qqq
we readily convince ourselves that the locally smooth super-1-forms \eqref{eq:locconsAdS} compose a global super-1-form on the quotient as
\qq\nn
\underset{\tx{\ciut{(1)}}}{\txA_j^2}\bigl(\theta_j^{\a\a'I}(x),X_j^{\widehat a}(x),z_i\cdot h^{01}_{ij}(x)\bigr)\equiv\tfrac{\sfd\bigl(z_i\cdot h^{01}_{ij}(x)\bigr)}{z_i\cdot h^{01}_{ij}(x)}-\bigl(\si_i^2{}^*r_{h^2_{ij}(\cdot)}^*\theta^{01}_{2\,{\rm L}}\bigr)(x)\equiv\underset{\tx{\ciut{(1)}}}{\txA_i^2}\bigl(\theta_i^{\a\a'I}(x),X_i^{\widehat a}(x),z_i\bigr)\,,
\qqq
so that we may define
\qq\nn
\underset{\tx{\ciut{(1)}}}{\txA^2}\bigl(\bigl[(x,z_i,i)\bigr]\bigr):=\underset{\tx{\ciut{(1)}}}{\txA_i^2}\bigl(\theta_i^{\a\a'I}(x),X_i^{\widehat a}(x),z_i\bigr)\,.
\qqq
This super-1-form is invariant under the supersymmetry transformations
\qq\nn
\bigl(g',\bigl[(x,z_i,i)\bigr]\bigr)\longmapsto\bigl[\bigl([\ell^2]_{g'}(x),\unl h^{01}_{ij}(x;g')\cdot z_i,j\bigr)\bigr]\,.
\qqq 
These are, indeed, well-defined as in view of relation \eqref{eq:cosetactglue}, we obtain, for $\,\cO^2_{jk}\ni[\ell^2]_{g'}(x)$,
\qq\nn
\bigl[\bigl([\ell^2]_{g'}(x),\unl h^{01}_{ik}(x;g')\cdot z_i,k\bigr)\bigr]&\equiv&\bigl[\bigl([\ell^2]_{g'}(x),\unl h^{01}_{ij}(x;g')\cdot z_i\cdot h^{01}_{jk}\bigl([\ell^2]_{g'}(x)\bigr),k\bigr)\bigr]\cr\cr
&=&\bigl[\bigl([\ell^2]_{g'}(x),\unl h^{01}_{ij}(x;g')\cdot z_i,j\bigr)\bigr]\,.
\qqq
We also establish the desired behaviour: 
\qq\nn
\underset{\tx{\ciut{(1)}}}{\txA^2}\bigl(\bigl[\bigl([\ell]_{g'}(x),\unl h^{01}_{ij}(x;g')\cdot z_i,j\bigr)\bigr]\bigr)&\equiv&\underset{\tx{\ciut{(1)}}}{\txA_i^2}\bigl(\theta_j^{\a\a'I}\bigl(\bigl([\ell]_{g'}(x)\bigr),X_j^{\widehat a}\bigl(\bigl([\ell]_{g'}(x)\bigr),\unl h^{01}_{ij}(x;g')\cdot z_i\bigr)\cr\cr
&=&\tfrac{\sfd\bigl(\unl h^{01}_{ij}(x;g')\cdot z_i\bigr)}{\unl h^{01}_{ij}(x;g')\cdot z_i}+[\ell]_{g'}^*\underset{\tx{\ciut{(1)}}}{\txB_j^2}(x)=\tfrac{\sfd z_i}{z_i}+\underset{\tx{\ciut{(1)}}}{\txB_i^2}(x)\cr\cr
&\equiv&\underset{\tx{\ciut{(1)}}}{\txA^2}\bigl(\bigl[(x,z_i,i)\bigr]\bigr)\,.
\qqq
We shall now recover the local description from global structures erected over the supersymmetry group $\,{\rm SU}(1,1\,\vert\,2)_2\,$ just as in the super-Minkowskian setting.

We begin by descending the principal $\bC^\x$-bundle 
\qq\label{eq:princbunextAdS}
\alxydim{@C=1cm@R=1cm}{\bC^\x\cong\exp\bigl(\corr{Z_{01}}_\bC\bigr) \ar[r] & \widetilde{{\rm SU}(1,1\,\vert\,2)_2} \ar[d]^{\widetilde\pi_2} \\ & \widetilde{{\rm SU}(1,1\,\vert\,2)_2}/\exp\bigl(\corr{Z_{01}}_\bC\bigr)\cong{\rm SU}(1,1\,\vert\,2)_2}
\qqq
obtained, in virtue of Thm.\,\ref{thm:TWextgrp}, through integration of the Lie-superalgebra extension \eqref{eq:sAdSSextalg} to the Lie-supergroup extension \eqref{eq:sAdSSextgrp}, to the homogeneous space $\,{\rm s}({\rm AdS}_2\x\bS^2)$.\ In other words, we seek to derive a commutative diagram of principal bundles 
\qq\nn
\alxydim{@C=2cm@R=2cm}{ & \bC^\x \ar[d] \ar@{=}[rr] & & \bC^\x \ar[d] \\ {\rm SO}(1,1)\x{\rm SO}(2) \ar[r] \ar@{=}[d] & \widetilde{{\rm SU}(1,1\,\vert\,2)_2} \ar[rr]^{\pi_{\widetilde{{\rm SU}(1,1\,\vert\,2)_2}/({\rm SO}(1,1)\x{\rm SO}(2))}\qquad\qquad} \ar[d]^{\widetilde\pi_2} & & \widetilde{{\rm SU}(1,1\,\vert\,2)_2}/({\rm SO}(1,1)\x{\rm SO}(2)) \ar[d]^{[\widetilde\pi_2]} \\ {\rm SO}(1,1)\x{\rm SO}(2) \ar[r] & {\rm SU}(1,1\,\vert\,2)_2 \ar[rr]_{\pi_2} & & {\rm s}\bigl({\rm AdS}_2\x\bS^2\bigr) }\,,
\qqq
in full structural analogy with the one for the Lie supergroup $\,{\rm SO}(3,1\,\vert\,2\cdot 4)\,$ and its homogeneous space $\,{\rm sMink}^{3,1\,\vert\,2\cdot 4}$,\ {\it cp} Diag.\,\eqref{diag:princbunsMink}. For that, we need to define the projection to the base $\,[\widetilde\pi_2]\,$ and identify an action of the structure group $\,\bC^\x\,$ on the total space $\,\widetilde{{\rm SU}(1,1\,\vert\,2)_2}/({\rm SO}(1,1)\x{\rm SO}(2))\,$ with the standard properties. The existence of the latter follows directly from the relative commutativity of the relevant subgroups: $\,{\rm SO}(1,1)\x{\rm SO}(2)\,$ and $\,\bC^\x\,$ within the extended supersymmetry group $\,\widetilde{{\rm SU}(1,1\,\vert\,2)_2}$,\ a consequence of the relation 
\qq\nn
[J_{\widehat a\widehat b},Z_{01}]^\sim=0\,,
\qqq 
{\it cp} \Reqref{eq:s0bextsu112}. It implies equivariance of $\,\widetilde\pi_2\,$ under the right (regular) action of the subgroup $\,{\rm SO}(1,1)\x{\rm SO}(2)\,$ on the Lie supergroups $\,\widetilde{{\rm SU}(1,1\,\vert\,2)_2}\,$ and $\,{\rm SU}(1,1\,\vert\,2)_2$,\ whence the existence of a unique smooth map $\,[\widetilde\pi_2]\,$ that closes the commutative square in the diagram. More concretely, let $\,\{\widetilde\cO^2_i\}_{i\in\widetilde I_2}\,$ be a trivialising open cover for the principal ${\rm SO}(1,1)\x{\rm SO}(2)$-bundle  
\qq\nn
\alxydim{@C=1cm@R=1cm}{{\rm SO}(1,1)\x{\rm SO}(2) \ar[r] & \widetilde{{\rm SU}(1,1\,\vert\,2)_2} \ar[d]^{\pi_{\widetilde{{\rm SU}(1,1\,\vert\,2)_2}/({\rm SO}(1,1)\x{\rm SO}(2))}\equiv\widetilde\pi} \\ & \widetilde{{\rm SU}(1,1\,\vert\,2)_2}/({\rm SO}(1,1)\x{\rm SO}(2))}\,,
\qqq
and let 
\qq\nn
\widetilde\si^2_i\ :\ \widetilde\cO^2_i\too\widetilde{{\rm SU}(1,1\,\vert\,2)_2}\,,\qquad i\in\widetilde I_2
\qqq
be the corresponding local sections, related over double intersections $\,\widetilde\cO^2_{ij}\equiv\widetilde\cO^2_i\cap\widetilde\cO^2_j\,$ by the transition maps 
\qq\nn
\widetilde h^2_{ij}\ :\ \widetilde\cO^2_{ij}\too{\rm SO}(1,1)\x{\rm SO}(2)
\qqq
of the bundle as 
\qq\nn
\widetilde\si^2_j(\widetilde x)=\widetilde\si^2_i(\widetilde x)\cdot\widetilde h^2_{ij}(\widetilde x)\,,\qquad\widetilde x\in\widetilde\cO^2_{ij}\,.
\qqq
We may then define (manifestly) locally smooth maps
\qq\nn
[\widetilde\pi_2]\rstr_{\widetilde\cO^2_i}:=\pi_2\circ\widetilde\pi_2\circ\widetilde\si^2_i\equiv[\widetilde\pi_2]_i\,,
\qqq
and check that they glue to a globally smooth one,
\qq\nn
[\widetilde\pi_2]_j(\widetilde x)\equiv\pi_2\bigl(\widetilde\si^2_j(\widetilde x)\,\bC^\x\bigr)=\pi_2\bigl(\widetilde\si^2_i(\widetilde x)\,\bC^\x\,\widetilde h^2_{ij}(\widetilde x)\bigr)=\pi_2\bigl(\widetilde\si^2_i(\widetilde x)\,\bC^\x\bigr)\equiv[\widetilde\pi_2]_i(\widetilde x)\,,
\qqq
that closes the commutative square -- indeed, for any $\,\widetilde g\in\widetilde{{\rm SU}(1,1\,\vert\,2)_2}\,$ such that $\,\widetilde g\,({\rm SO}(1,1)\x{\rm SO}(2))\in\widetilde\cO^2_i$,\ we have
\qq\nn
\widetilde\si^2_i\bigl(\widetilde\pi(\widetilde g)\bigr)=\widetilde g\cdot\widetilde h_i(\widetilde g)
\qqq
for some $\,\widetilde h_i(\widetilde g)\in{\rm SO}(1,1)\x{\rm SO}(2)$,\ and so
\qq\nn
[\widetilde\pi_2]\circ\widetilde\pi(\widetilde g)\equiv\pi_2\bigl(\widetilde g\cdot\widetilde h_i(\widetilde g)\,\bC^\x\bigr)=\pi_2\bigl(\widetilde g\,\bC^\x\,\widetilde h_i(\widetilde g)\bigr)=\pi_2\bigl(\widetilde g\,\bC^\x\bigr)\equiv\pi_2\circ\widetilde\pi_2(\widetilde g)\,.
\qqq
Clearly, $\,[\widetilde\pi_2]\,$ is a surjective submersion. In the next step, we use the ${\rm SO}(1,1)\x{\rm SO}(2)$-equivariant right (defining) action 
\qq\nn
r_\cdot\ :\ \widetilde{{\rm SU}(1,1\,\vert\,2)_2}\x\bC^\x\too\widetilde{{\rm SU}(1,1\,\vert\,2)_2}\ :\ (\widetilde g,\z)\longmapsto\widetilde g\cdot\z
\qqq
to write a smooth action 
\qq\nn
[r]_\cdot\ &:&\ \widetilde{{\rm SU}(1,1\,\vert\,2)_2}/({\rm SO}(1,1)\x{\rm SO}(2))\x\bC^\x\too\widetilde{{\rm SU}(1,1\,\vert\,2)_2}/({\rm SO}(1,1)\x{\rm SO}(2))\cr\cr 
&:&\ \bigl(\widetilde g\,({\rm SO}(1,1)\x{\rm SO}(2)),\z\bigr)\longmapsto(\widetilde g\cdot\z)\,({\rm SO}(1,1)\x{\rm SO}(2))\,.
\qqq
The action is readily checked to be free,
\qq\nn
[r]_\z\bigl(\widetilde\pi(\widetilde g)\bigr)=\widetilde\pi(\widetilde g)\qquad
\Longleftrightarrow\qquad\exists_{h_1,h_2\in{\rm SO}(1,1)\x{\rm SO}(2)}\ :\ \z\cdot h_1=h_2\qquad\Longrightarrow\qquad\z=e\,.
\qqq
It preserves level sets of $\,[\widetilde\pi_2]$,
\qq\nn
[\widetilde\pi_2]\circ[r]_\z\bigl(\widetilde\pi\bigl(\widetilde g\bigr)\bigr)&\equiv&\pi_2\circ\widetilde\pi_2\circ\widetilde\si^2_i\circ\widetilde\pi\bigl(\widetilde g\cdot\z\bigr)=\pi_2\bigl(\widetilde g\cdot\z\cdot\widetilde h_i(\widetilde g\cdot\z)\,\bC^\x\bigr)=\pi_2\bigl(\widetilde g\cdot\z\,\bC^\x\,\widetilde h_i(\widetilde g\cdot\z)\bigr)=\pi_2\bigl(\widetilde g\,\bC^\x\bigr)\cr\cr
&\equiv&[\widetilde\pi_2]\bigl(\widetilde\pi\bigl(\widetilde g\bigr)\bigr)\,,
\qqq
and, conversely, any two points $\,\widetilde x_\a,\widetilde x_2\in[\widetilde\pi_2]^{-1}(\{x\})\cap\widetilde\cO^2_{i_\a},\ \a\in\{1,2\}\,$ and in a given level set (of $\,x\in{\rm s}({\rm AdS}_2\x\bS^2)$) belong to the same ${\rm SO}(1,1)\x{\rm SO}(2)$-orbit as the equality
\qq\nn
[\widetilde\pi_2](\widetilde x_1)=[\widetilde\pi_2](\widetilde x_2)
\qqq
implies the existence of a pair $\,(h,\z)\in({\rm SO}(1,1)\x{\rm SO}(2))\x\bC^\x\,$ satisfying the relation
\qq\nn
\widetilde\si^2_{i_2}(\widetilde x_2)=\widetilde\si^2_{i_1}(\widetilde x_1)\cdot\z\cdot h\,,
\qqq
and hence
\qq\nn
\widetilde x_2\equiv\widetilde\pi\circ\widetilde\si^2_{i_2}(\widetilde x_2)=\widetilde\pi\bigl(\widetilde\si^2_{i_1}(\widetilde x_1)\cdot h\cdot\z\bigr)\equiv[r]_\z\circ\widetilde\pi\circ\widetilde\si^2_{i_1}(\widetilde x_1)=[r]_\z(\widetilde x_1)\,.
\qqq
Thus, altogether, the structure
\qq\nn
\alxydim{@C=1cm@R=1cm}{\bC^\x \ar[r] & \widetilde{{\rm SU}(1,1\,\vert\,2)_2}/({\rm SO}(1,1)\x{\rm SO}(2))\equiv\xcL_2 \ar[d]^{[\widetilde\pi_2]\equiv\pi_{\xcL_2}} \\ & {\rm s}\bigl({\rm AdS}_2\x\bS^2\bigr)}
\qqq
is a principal $\bC^\x$-bundle, as claimed.

We conclude the present section by inducing a principal $\bC^\x$-connection 1-form on the above principal $\bC^\x$-bundle in a manner structurally identical as in the super-Minkowskian setting. That is, we descend the principal $\bC^\x$-connection 1-form on the bundle \eqref{eq:princbunextAdS}, given by the component of the $\widetilde{\gt{su}(1,1\,\vert\,2)_2}$-valued Maurer--Cartan super-1-form on $\,\widetilde{{\rm SU}(1,1\,\vert\,2)_2}\,$ along the central generator $\,Z_{01}$,
\qq\nn
\underset{\tx{\ciut{(1)}}}{\widetilde\cA^2}:=\widetilde\theta^{Z_{01}}_{2\,{\rm L}}\,,
\qqq
upon noting that the latter is ${\rm SO}(1,1)\x{\rm SO}(2)$-basic owing to the centrality of $\,Z_{01}$.\ The descent gives us a globally smooth super-1-form with restrictions
\qq\nn
\underset{\tx{\ciut{(1)}}}{\cA^2}\rstr_{\widetilde\cO^2_i}\equiv\widetilde\si^2_i{}^*\underset{\tx{\ciut{(1)}}}{\widetilde\cA^2}\,.
\qqq
Inspection of the super-commutation relations \eqref{eq:s0bextsu112} reveals that, locally, we recover the $\,\underset{\tx{\ciut{(1)}}}{\txA_i^2}$.

The natural left action of the extended supersymmetry group on the total space of the above principal $\bC^\x$-bundle
\qq\nn
\widetilde{[\ell^2]}_\cdot\ &:&\ \widetilde{{\rm SU}(1,1\,\vert\,2)_2}\x\xcL_2\too\xcL_2\cr\cr 
&:&\ \bigl(\widetilde g',\widetilde g\,({\rm SO}(1,1)\x{\rm SO}(2))\bigr)\longmapsto\bigl(\widetilde g'\cdot\widetilde g\bigr)\,({\rm SO}(1,1)\x{\rm SO}(2))
\qqq 
is realised by its automorphisms. The action is modelled locally on the space
\qq\nn
\widetilde\si^2(\xcL_2):=\bigsqcup_{i\in\widetilde I_2}\,\widetilde\si^2_i\bigl(\widetilde\cO_i\bigr)
\qqq
in the standard manner described in Sec.\,\ref{sec:scheme} ({\it cf} Eqs.\,\eqref{eq:cosetactal} and \eqref{eq:cosetactglue}), and as the restrictions of the connection 1-form $\,\underset{\tx{\ciut{(1)}}}{\widetilde\cA^2}\,$ are invariant under this modelling action, the automorphisms are promoted to the rank of connection-reserving ones.

Thus, after the dust has cleared, we arrive at
\bedef\label{def:s0gerbeAdS}
The \textbf{Green--Schwarz super-0-gerbe} of curvature $\,\underset{\tx{\ciut{(2)}}}{\txH^2}\,$ over $\,{\rm s}({\rm AdS}_2\x\bS^2)\,$ is the triple
\qq\nn
\sG^{(0)}_{\rm GS}\bigl({\rm s}({\rm AdS}_2\x\bS^2)\bigr):=\bigl(\xcL_2,\pi_{\xcL_2},\underset{\tx{\ciut{(1)}}}{\cA^2}\bigr)
\qqq
constructed in the preceding paragraphs. It admits the action $\,\widetilde{[\ell^2]}_\cdot\,$ of the extended supersymmetry group $\,\widetilde{{\rm SU}(1,1\,\vert\,2)_2}\,$ by connection-preserving principal $\bC^\x$-bundle automorphisms.
\exdef
\noindent Crucially, the GS super-0-gerbe $\,\sG^{(0)}_{\rm GS}({\rm s}({\rm AdS}_2\x\bS^2))\,$ is -- by construction -- compatible with the \.In\"on\"u--Wigner contraction \eqref{eq:IWcontrsAdS}, which we symbolically write as
\qq\nn
\sG^{(0)}_{\rm GS}\bigl({\rm s}({\rm AdS}_2\x\bS^2)\bigr)\xrightarrow{\ \varsigma_R\ }\sG^{(0)}_{\rm GS}\bigl({\rm s}({\rm AdS}_2(R)\x\bS^2(R))\bigr)\xrightarrow{\ R\to\infty\ }\sG^{(0)}_{\rm GS}\bigl({\rm sMink}^{3,1\,\vert\,2\cdot 4}\bigr)\,.
\qqq

\section{The weak $\k$-equivariance of the extended super-0-gerbes}\label{sec:kappa}

The findings discussed in the foregoing sections provide a clear indication that the objects constructed in Sec.\,\ref{sec:geometrise} are not merely mathematically consistent but also physically relevant. A clinching argument comes from an analysis of the behaviour of their extensions, extracted directly from the Hughes--Polchinski reformulation of the original Nambu--Goto super-$\si$-model, under $\k$-symmetry, recalled towards the end of Sec.\,\ref{sec:Carthomsp}. In what follows, we examine the Green--Schwarz super-0-gerbes of Defs.\,\ref{def:s0gerbesMink} and \ref{def:s0gerbeAdS} from this perspective. In both cases, $\,p=0\,$ and we choose 
\qq\label{eq:t0vacspec}
\tgt^{(0)}_{\rm vac}\equiv\corr{P_0}_\bC\,,
\qqq
so that the $\k$-symmetry constraint reduces to 
\qq\nn
f_{\a\b}^{\ \ 0}\,\d^\b_{\unl A}+\chi_{\a\unl A}\must\bigl(\id_{\tgt^{(1)}}-\txP^{\tgt^{(1)}}_{\ \tgt^{(1)}_{\rm vac}}\bigr)^\b_{\ \a}\,\la_{\b\,\unl A}
\qqq
for \emph{some} idempotent
\qq\nn
\txP^{\tgt^{(1)}}_{\ \tgt^{(1)}_{\rm vac}}\in\bC(8)
\qqq
with the proprety 
\qq\nn
\tr_{\bC^{\x 8}}\,\txP^{\tgt^{(1)}}_{\ \tgt^{(1)}_{\rm vac}}=4\,.
\qqq
That the Hughes--Polchinski model can be used (equivalently) in the supergeometric settings of interest is a consequence of the following observations ({\it cp} the assumptions of Thm.\,\ref{prop:IHCartMink}):
\bit
\item in both cases, there exist non-degenerate bilinear forms on $\,\tgt^{(0)}_{\rm vac}\,$ of \Reqref{eq:t0vacspec} and on 
\qq\nn
\egt^{(0)}\equiv\corr{P_1,P_2,P_3}_\bC\,,
\qqq
to wit,
\qq\nn
\unl\g=-\t^0\ox\t^0\,,\qquad\qquad\widehat\g=\t^1\ox\t^1+\t^2\ox\t^2+\t^3\ox\t^3\,;
\qqq
\item using the above, we readily verify condition \eqref{eq:f2fIH} -- on $\,{\rm sMink}^{3,1\,\vert\,2\cdot 4}$,\ it boils down to the identity, written for $\,\widehat a,\widehat b\in\{1,2,3\}$,
\qq\nn
-f_{0\widehat a\,0}^{\ \ \ \widehat b}=-\d_{\widehat a}^{\ \widehat b}\equiv-\d_{\widehat a\widehat b}=-f_{0\widehat a\,\widehat b}^{\ \ \ 0}\,,
\qqq
whereas on $\,{\rm s}({\rm AdS}_2\x\bS^2)$,\ it reads, for $\,\widehat a\in\{1,2,3\}$,
\qq\nn
-f_{01\,0}^{\ \ \ \widehat a}=-\d_1^{\ \widehat a}\equiv-\d_{1\widehat a}=-f_{01\,\widehat a}^{\ \ \ 0}\,.
\qqq
\eit
The reformulation of the original super-$\si$-models leads, along the lines of \Rcite{Suszek:2018bvx}, to the emergence of new supergeometric objects. We have 
\bedef\label{def:exts0gsMink}
The \textbf{extended Green--Schwarz super-0-gerbe} over $\,\widehat\cM^1:={\rm sISO}(3,1\,\vert\,2\cdot 4)/{\rm SO}(3)\,$ of curvature $\,\underset{\tx{\ciut{(2)}}}{\widehat\txH^1}\,$ fixed by the condition
\qq\nn
\pi_{{\rm sISO}(3,1\,\vert\,2\cdot 4)/{\rm SO}(3)}^*\underset{\tx{\ciut{(2)}}}{\widehat\txH^1}=(\cC\ox\sfi\,\si_2)_{\widehat\a I\widehat\b J}\,\theta_{\rm L}^{\widehat\a I}\wedge\theta^{\widehat\b J}_{1\,{\rm L}}+\sfd\theta_{1\,{\rm L}}^0 
\qqq
is the triple
\qq\nn
\sG^{(0)}_{\rm GS}\bigl(\widehat\cM^1\bigr):=\bigl(\widetilde{{\rm sISO}(3,1\,\vert\,2\cdot 4)}/{\rm SO}(3),\pr_1,\underset{\tx{\ciut{(1)}}}{\widehat\cA^1}\bigr)\,,
\qqq
with the principal $\bC^\x$-connection $\,\underset{\tx{\ciut{(1)}}}{\widehat\cA^1}\,$ determined by the condition
\qq\nn
\pi_{\widetilde{{\rm sISO}(3,1\,\vert\,2\cdot 4)}/{\rm SO}(3)}^*\underset{\tx{\ciut{(1)}}}{\widehat\cA^1}=\widetilde\theta_{1\,{\rm L}}^Z+\widetilde\theta_{1\,{\rm L}}^0\,.
\qqq
\exdef
\noindent and
\bedef\label{def:exts0gsAdSS}
The \textbf{extended Green--Schwarz super-0-gerbe} over $\,\widehat\cM^2:={\rm SU}(1,1\,\vert\,2)_2/{\rm SO}(2)\,$ of curvature $\,\underset{\tx{\ciut{(2)}}}{\widehat\txH^2}\,$ fixed by the condition
\qq\nn
\pi_{{\rm SU}(1,1\,\vert\,2)_2/{\rm SO}(2)}^*\underset{\tx{\ciut{(2)}}}{\widehat\txH^2}=(\unl C\ox\sfi\,\si_2)_{\a\a'I\b\b'J}\,\theta_{2\,{\rm L}}^{\a\a'I}\wedge\theta^{\b\b'J}_{2\,{\rm L}}+\theta_{2\,{\rm L}}^0\wedge\theta_{2\,{\rm L}}^1+\sfd\theta_{2\,{\rm L}}^0 
\qqq
is the triple
\qq\nn
\sG^{(0)}_{\rm GS}\bigl(\widehat\cM^2\bigr):=\bigl(\widetilde{{\rm SU}(1,1\,\vert\,2)_2}/{\rm SO}(2),\pr_1,\underset{\tx{\ciut{(1)}}}{\widehat\cA^2}\bigr)\,,
\qqq
with the principal $\bC^\x$-connection $\,\underset{\tx{\ciut{(1)}}}{\widehat\cA^2}\,$ determined by the condition
\qq\nn
\pi_{\widetilde{{\rm SU}(1,1\,\vert\,2)_2}/{\rm SO}(2)}^*\underset{\tx{\ciut{(1)}}}{\widehat\cA^2}=\widetilde\theta_{2\,{\rm L}}^{Z_{01}}+\widetilde\theta_{2\,{\rm L}}^0\,.
\qqq
\exdef
\noindent We are now ready to examine $\k$-equivariance of the above extended geometrisations.

\subsection{The $\k$-symmetry of the super-$\si$-model and super-0-gerbe over $\,{\rm sMink}^{3,1\,\vert\,2\cdot 4}$}

The non-zero components of the relevant super-2-cocycle on $\,{\rm sISO}(3,1\,\vert\,2\cdot 4)\,$ are  
\qq\nn
\chi_{\widehat\a I\widehat\b J}=2\cC_{\widehat\a\widehat\b}\,\ep_{IJ}\,,
\qqq
and so the $\k$-symmetry constraint reads
\qq\nn
\left\{\barr{l}
\bigl(\id_{\tgt^{(1)}}-\txP^{\tgt^{(1)}}_{\ \tgt^{(1)}_{\rm vac}}\bigr)^{\widehat\b J}_{\ \widehat\a I}\,\la_{\widehat\b J\,\widehat\g K}\must2\bigl(\d^{\widehat\b}_{\ \widehat\a}\,\d^J_{\ I}-\sfi\,\G^0{}^{\widehat\b}_{\ \widehat\a}\,\si_2{}^J_{\ I}\bigr)\,\cC_{\widehat\b\widehat\g}\,\ep_{JK}\cr\cr
\bigl(\id_{\tgt^{(1)}}-\txP^{\tgt^{(1)}}_{\ \tgt^{(1)}_{\rm vac}}\bigr)^{\widehat\b J}_{\ \widehat\a I}\,\la_{\widehat\b J\,\widehat a}\must0 \earr\right.\,.
\qqq
This is solved by
\qq\nn
\la_{\widehat\a I\,\unl A}=4\sfi\,\d_{\unl A}^{\ \ \widehat\b J}\,(\cC\ox\si_2)_{\widehat\a I\widehat\b J}
\qqq
and
\qq\nn
\txP^{\tgt^{(1)}}_{\ \tgt^{(1)}_{\rm vac}}=\tfrac{1}{2}\,\bigl(\bd1+\sfi\,\G^0\ox\si_2\bigr)\,,
\qqq
with -- as desired --
\qq\nn
\tr_{\bC^{\x 8}}\,\txP^{\tgt^{(1)}}_{\ \tgt^{(1)}_{\rm vac}}=\tfrac{1}{2}\,\tr_{\bC^{\x 8}}\,\bd1=4\,.
\qqq
Having identified the symmetry of interest, we may next examine the behaviour of the extension of the super-0-gerbe $\,\sG^{(0)}_{\rm GS}({\rm sMink}^{3,1\,\vert\,2\cdot 4})\,$ determined by the Hughes--Polchinski model under the corresponding transformations. Given that 
\qq\nn
\ceL_{Q_{\widehat\a I}}\con\bigl(\widetilde\theta_{1\,{\rm L}}^Z+\widetilde\theta_{1\,{\rm L}}^0\bigr)=0\,,
\qqq
we are led, once more, to the $\k$-symmetry constraint,
\qq\nn
\d_\k\bigl(\widetilde\theta_{1\,{\rm L}}^Z+\widetilde\theta_{1\,{\rm L}}^0\bigr)\equiv\kappa^{\widehat\a I}\,\ceL_{Q_{\widehat\a I}}\con\sfd\bigl(\widetilde\theta_{1\,{\rm L}}^Z+\widetilde\theta_{1\,{\rm L}}^0\bigr)\equiv 0\,.
\qqq
Thus, we conclude that the extended Green--Schwarz super-0-gerbe $\,\sG^{(0)}_{\rm GS}(\widehat\cM^1)\,$ of Def.\,\ref{def:exts0gsMink} is endowed with a weak $\k$-equivariant structure.

\subsection{The $\k$-symmetry of the super-$\si$-model and super-0-gerbe over $\,{\rm s}({\rm AdS}_2\x\bS^2)$}

Zhou's super-2-cocycle on $\,{\rm SU}(1,1\,\vert\,2)_2\,$ has non-zero components 
\qq\nn
\chi_{\a\a'I\b\b'J}=2\unl C_{\widehat\a\widehat\b}\,\ep_{IJ}\,,\qquad\qquad\chi_{ab}=-\ep_{ab}\,,
\qqq
whence the $\k$-symmetry constraint in the form
\qq\nn
\left\{\barr{l}
\bigl(\id_{\tgt^{(1)}}-\txP^{\tgt^{(1)}}_{\ \tgt^{(1)}_{\rm vac}}\bigr)^{\b\b'J}_{\ \a\a'I}\,\la_{\b\b'J\,\g\g'K}\must2\bigl(\d^{\b\b'}_{\ \a\a'}\,\d^J_{\ I}-\sfi\,\unl\g^0{}^{\b\b'}_{\ \a\a'}\,\si_2{}^J_{\ I}\bigr)\,(\unl C\ox\sfi\,\si_2)_{\b\b'J\g\g'K}\cr\cr
\bigl(\id_{\tgt^{(1)}}-\txP^{\tgt^{(1)}}_{\ \tgt^{(1)}_{\rm vac}}\bigr)^{\b\b'J}_{\ \a\a'I}\,\la_{\b\b'J\,\widehat a}\must0 \earr\right.\,.
\qqq
This is solved by
\qq\nn
\la_{\a\a'I\,\unl A}=4\sfi\,\d_{\unl A}^{\ \ \b\b'J}\,(\unl C\ox\si_2)_{\a\a'I\b\b'J}
\qqq
and
\qq\nn
\txP^{\tgt^{(1)}}_{\ \tgt^{(1)}_{\rm vac}}=\tfrac{1}{2}\,\bigl(\bd1+\sfi\,\unl\g^0\ox\si_2\bigr)\,,
\qqq
with -- as desired, once again --
\qq\nn
\tr_{\bC^{\x 8}}\,\txP^{\tgt^{(1)}}_{\ \tgt^{(1)}_{\rm vac}}=\tfrac{1}{2}\,\tr_{\bC^{\x 8}}\,\bd1=4\,.
\qqq
Just as in the super-Minkowskian case, weak $\k$-equivariance of the extended Green--Schwarz super-0-gerbe $\,\sG^{(0)}_{\rm GS}(\widehat\cM^2)\,$ of Def.\,\ref{def:exts0gsAdSS} is now a direct consequence of the identity
\qq\nn
\ceL_{Q_{\widehat\a I}}\con\bigl(\widetilde\theta_{1\,{\rm L}}^{Z_{01}}+\widetilde\theta_{1\,{\rm L}}^0\bigr)=0\,.
\qqq

\section{Conclusions \& Outlook}\label{sec:C&O}

In the present paper, we have laid out a general scheme of a \emph{correlated} supersymmetry-equivariant geometrisation of physically distinguished Green--Schwarz super-$(p+2)$-cocycles representing classes in the supersymmetry-invariant refinement of the de Rham cohomology of a \emph{pair} of homogeneous spaces of Lie supergroups set in correspondence by a blow-up transformartion dual to the \.In\"on\"u--Wigner contraction relating the respective Lie superalgebras. The geometrisation over each of the homogeneous spaces is realised through a universal Cartan-geometric mechanism worked out and illustrated on a number of concrete examples in a series of papers \cite{Suszek:2017xlw,Suszek:2018bvx}. It associates a supergeometric object, termed a super-$p$-gerbe in \Rcite{Suszek:2017xlw}, with the super-$(p+2)$-cocycle in a manner structurally similar to that known from the works of Murray {\it et al.} in standard geometry and (de Rham) cohomology. The mechanism is based essentially on the classical equivalence between the Cartan--Eilenberg cohomology of a Lie supergroup and the Chevalley--Eilenberg cohomology of its Lie superalgebra with values in $\,\bR\,$ as well as on the one-to-one correspondence, discussed in \Rcite{Suszek:2017xlw} in analogy with its non-graded counterpart, between equivalence classes of central extensions of the Lie superalgebra and classes in its 2nd cohomology group with values in the extending (trivial) abelian module. Compatibility of the geometrisation with the contraction, reflected in the existence of a consistent extension of the blow-up transformation to the super-$p$-gerbe, has been promoted to the status of a defining property of the scheme, in conformity with the original proposal formulated in \Rcite{Suszek:2018bvx}. The validity of the scheme has been tested in an in-depth case study of the pair of physically motivated $\,N=2\,$ super-0-brane backgrounds with four-dimensional bodies: the super-Minkowski space $\,{\rm sMink}^{3,1\,\vert\,2\cdot 4}\equiv{\rm sISO}(3,1\,\vert\,2\cdot 4)/{\rm SO}(3,1)\,$ with the Green--Schwarz super-2-cocycle \eqref{eq:GSs2csMink} and the super-${\rm AdS}_2\x\bS^2\,$ space $\,{\rm s}({\rm AdS}_2\x\bS^2)\equiv{\rm SU}(1,1\,\vert\,2)_2/({\rm SO}(1,1)\x{\rm SO}(2))\,$ with the Green--Schwarz super-2-cocycle \eqref{eq:GSs2csAdSS}, the latter asymptoting to the former in the limit of an infinite common radius of the generating 1-cycle in $\,{\rm AdS}_2\cong\bS^1\x\bR\,$ and of the 2-sphere $\,\bS^2\,$ in the body $\,{\rm AdS}_2\x\bS^2\,$ of the supertarget $\,{\rm s}({\rm AdS}_2\x\bS^2)$.\ The super-0-gerbes obtained as a result of the study are none other than the supersymmetry-equivariant prequantisation bundles for the respective super-$\si$-models of the Green--Schwarz type in the Nambu--Goto formulation. Further corroboration of our proposal comes from the explicit verification of the existence of a weak $\k$-equivariant structure on each of the super-0-gerbes, manifest in the purely geometric setting of the Hughes--Polchinski formulation of the super-$\si$-models in which the super-0-gerbes are replaced, in a manner first put forward in \Rcite{Suszek:2017xlw} and further elaborated in Sec.\,\ref{sec:Carthomsp} of the present paper, by their extended variants over the larger homogeneous spaces: $\,{\rm sISO}(3,1\,\vert\,2\cdot 4)/{\rm SO}(3)\,$ and $\,{\rm SU}(1,1\,\vert\,2)_2/{\rm SO}(2)$,\ respectively. 

The above pair of superbackgrounds, picked up for the case study because of its tractability, the nontrivial topology and the non-vanishing metric curvature of the supermanifold $\,{\rm s}({\rm AdS}_2\x\bS^2)\equiv\txG_2/\txH_2\,$ as well as the genericness of the attendant principal ${\rm SO}(1,1)\x{\rm SO}(2)$-bundle \eqref{eq:princSOSO} notwithstanding, and because of its relevance for the overarching project of elucidating the higher (super)geometry of the ${\rm AdS}$/CFT correspondence, realises a particular cohomological scenario. In it, the super-$(p+2)$-cocycle on the supertarget $\,\txG_2/\txH_2\,$ to be blown up admits, upon pullback to the corresponding supersymmetry group $\,\txG_2$,\ a global left-invariant primitive that does not descend to the homogeneous space $\,\txG_2/\txH_2\,$ but locally goes over, under the blow up, to the non-supersymmetric primitive of the super-$(p+2)$-cocycle on the asymptotic supertarget $\,\txG_1/\txH_1$.\ Accordingly, the task at hand, accomplished for the distinguished pair of superbackgrounds in the present paper, consists in finding a pair of Lie-superalgebra extensions $\,\widetilde\ggt_\a,\ \a\in\{1,2\}\,$ of the respective supersymmetry algebras $\,\ggt_\a\,$ that integrate to Lie-supergroup extensions $\,\widetilde\txG_\a\,$ of the respective supersymmetry groups $\,\txG_\a\,$ inducing, along the way, surjective submersions $\,\widetilde\txG_\a/\txH_\a\,$ over the two homogeneous spaces $\,\txG_\a/\txH_\a\,$ in (asymptotic) correspondence with the structure of homogeneous spaces of the respective extended supersymmetry groups on which pullbacks of the original super-$(p+2)$-cocycles admit global supersymmetric primitives, and in such a way that the contraction resp.\ the blow-up transformation can be prolonged to the extended supersymmetry algebras resp.\ the new homogeneous spaces, whereupon it maps the supersymmetric primitives to one another. Furthermore, the dimensionality of the super-$\si$-models under consideration ($p=0$) implies that the task effectively reduces to the construction of the said surjective submersions and the attendant supersymmetric primitives of the Green--Schwarz super-$(p+2)$-cocycles -- a considerable simplification of the general situation which emphasises its status of a toy model in the geometrisation project. Passing to Zhou's superstring model with the same curved supertarget $\,{\rm s}({\rm AdS}_2\x\bS^2)$,\ we encounter a different scenario, familiar from the recent study reported in \Rcite{Suszek:2018bvx}, in which the super-$(p+2)$-cocycle on the supertarget to be blown up admits, upon pullback to the supersymmetry group, a supersymmetric primitive that \emph{does} descend to the homogeneous space but \emph{does not} asymptote to its counterpart on the asymptotic supertarget, and so it seems likely that the logic advanced in \cite[Sec.\,9]{Suszek:2018bvx} has to be pursued, that is we ought to look for a Lie-superalgebra extension $\,\widetilde\ggt_2\,$ of the supersymmetry algebra $\,\ggt_2\,$ of the homogeneous space $\,\txG_2/\txH_2\,$ to be blown up that contracts to the known extension $\,\widetilde\ggt_1\,$ of the asymptotic supersymmetry algebra $\,\ggt_1\,$ and yields a trivialisation of the pullback of a \emph{potentially deformed} super-3-cocycle on the associated Lie supergroup $\,\widetilde\txG_2\,$ that subsequently descends to the extended homogeneous space $\,\widetilde\txG_2/\txH_2\,$ and asymptotes to its counterpart on $\,\txG_1/\txH_1\,$ (or rather the pullback of the latter to $\,\widetilde\txG_1/\txH_1$). Moreover, we are bound to confront a host of novel challenges entailed by the appearance of higher-order structures in the geometrisation of a super-3-cocycle, such as a coherent definition of fibred products of the previously obtained surjective submersions over the homogeneous space supporting the \emph{effective} super-$(p+2)$-cocycle just mentioned, endowed with an action of the extended supersymmetry group, and a super-0-gerbe thereover. The study of the two-dimensional super-$\si$-model with the supertarget $\,{\rm s}({\rm AdS}_2\x\bS^2)\,$ inevitably raises further physically motivated and mathematically intriguing questions concerning, {\it e.g.}, the supergeometry behind the supersymmetric bi-branes (and, in particular, D-branes) asociated with this curved backround. Answers obtained in the tractable four-dimensional supergeometric setting are expected to give us insights necessary to reconsider and better understand the critical superstring background over the super-${\rm AdS}_5\x\bS^5\,$ space whose in-depth analysis was initiated in \Rcite{Suszek:2018bvx}. We hope to return to these ideas and challenges in a future work.
\newpage

\appendix

\section{Conventions for \& facts about the $\,{\rm AdS}_5\x\bS^5\,$ Clifford algebras}\label{app:CliffAdSS}

In the present paper, we are dealing with a ditignuished, geometrically/physically motivated realisation of the Clifford algebra $\,\Cliff(\bR^{3,1})\,$ in terms of the generators of the Clifford algebras $\,\Cliff(\bR^{1,1})\,$ and $\,\Cliff(\bR^{2,0})$.\ Let us denote the generators of $\,\Cliff(\bR^{1,1})\,$ (in the 2-dimensional spinor representation) as
\qq\nn
\{\G^a\}_{a\in\ovl{0,1}}
\qqq
and those of $\,\Cliff(\bR^{2,0})\,$ (also in the 2-dimensional spinor representation) as
\qq\nn
\{\G^{a'}\}_{a'\in\ovl{2,3}}\,.
\qqq
The standard generators of $\,\Cliff(\bR^{3,1})\,$ are now given by
\qq\nn
\{\unl\g^a\equiv\G^a\ox\bd1,\unl\g^{a'}\equiv\G_0\,\G_1\ox\G^{a'}\}_{a\in\ovl{0,1},a'\in\ovl{2,3}}\,,
\qqq
{\it cp} \Rcite{Zhou:1999sm}. Out of these, we form bi-vectorial objects
\qq\nn
\G^{ab}:=\tfrac{1}{2}\,[\G^a,\G^b]\,,\qquad\qquad\G^{a'b'}:=\tfrac{1}{2}\,[\G^{a'},\G^{b'}]\,,
\qqq
and
\qq\nn
\unl\g^{\widehat a\widehat b}:=\tfrac{1}{2}\,[\unl\g^{\widehat a},\unl\g^{\widehat b}]\,,\qquad\widehat a,\widehat b\in\ovl{0,3}\,.
\qqq
linearly independent of the generators. Define
\qq\nn
\g_3:=\G_0\,\G_1\,,\qquad\qquad\g_3':=\G_2\,\G_3\,.
\qqq
The chirality operator of $\,\Cliff(\bR^{3,1})\,$ takes the form
\qq\nn
\unl\g_5\equiv\unl\g_0\,\unl\g_1\,\unl\g_2\,\unl\g_3=\g_3\ox\g_3'\,.
\qqq
We shall also use the shorthand notation
\qq\nn
\widetilde\g_3\equiv\g_3\ox\bd1\,,\qquad\qquad\widetilde\g_3'\equiv\bd1\ox\g_3'\,.
\qqq
Note the identity
\qq\label{eq:gam3act}
\widetilde\g_3\cdot\unl\g^a=-\eta^{ab}\,\ep_{bc}\,\unl\g^c\,,
\qqq
expressed in terms of the totally antisymmetric tensor $\,\ep_{ab}=-\ep_{ba}\,$ with $\,\ep_{01}=-1$.

The sets of generators of the Clifford algebras of interest are augmented with the respective charge-conjugation matrices -- for $\,\Cliff(\bR^{1,1})$:
\qq\nn
C=-C^{\rm T}\,,
\qqq
for $\,\Cliff(\bR^{2,0})$:
\qq\nn
C'=C'{}^{\rm T}\,,
\qqq
and -- in the end -- also for $\,\Cliff(\bR^{3,1})$:
\qq\label{eq:CCc}
\unl C=C\ox C'=-\unl C^{\rm T}\,.
\qqq
These enable us to describe the basic symmetry properties of the said generators,
\qq\nn
\bigl(\G^a\bigr)^{\rm T}=-C\,\G^a\,C^{-1}\,,\qquad\qquad\bigl(\G^{a'}\bigr)^{\rm T}=C'\,\G^{a'}\,C'{}^{-1}\,,
\qqq
which we may rewrite equivalently as
\qq\nn
\bigl(C\,\G^a\bigr)^{\rm T}=C\,\G^a\,,\qquad\qquad\bigl(C'\,\G^{a'}\bigr)^{\rm T}=C'\,\G^{a'}\,.
\qqq

\end{document}